\documentclass[phd, titlesmallcaps, examinerscopy, copyrightpage, foronline]{mqthesis}

\usepackage{lscape}
\usepackage{rotating}  
\usepackage{theorem}   
\usepackage{bm}  
\usepackage{amsmath,amsfonts,amssymb}
\usepackage{booktabs}  
\usepackage{pdfsync}   
\usepackage{graphicx}
\usepackage{latexsym}
\usepackage{amsmath}
\usepackage{amssymb}
\usepackage{fancyvrb}
\usepackage{subfigure}
\usepackage{colortbl}
\usepackage{enumerate}
\usepackage{pifont}
\usepackage{stmaryrd}
\usepackage{textcomp}
\usepackage{fncylab}
\usepackage{multirow}
\usepackage{paralist}
\usepackage{wrapfig}
\usepackage{colortbl}
\usepackage{longtable}
\usepackage[table]{xcolor}
\usepackage[protrusion=true,expansion=true]{microtype}
\usepackage[T1]{fontenc}
\usepackage{lscape}
\usepackage{lipsum}
\usepackage{pgfplots}
\usepackage{algpseudocode}
\usepackage{algorithm}
\usepackage{graphicx}
\usepackage{pgfplots}
\usepackage{pgfplotstable}
\usepgfplotslibrary{groupplots}
\usepackage{pgf,tikz}
\usetikzlibrary{patterns}
\usetikzlibrary{positioning}
\usepackage{longtable}
\usepackage{diagbox}

\begin{document}

\frontmatter

\title{Towards Time-Aware Context-Aware Deep Trust Prediction in Online Social Networks}

\ifthenelse{\boolean{foronline}}{
  \author{\href{mailto:seyed-mohssen.ghafari@hdr.mq.edu.au}{Seyed Mohssen Ghafari}}
  \department{Computing}
}{
  \author{Seyed Mohssen Ghafari}
  \department{Computing}
}
\degrees{}

 \submitdate{December 2019}

\renewcommand{\degreetext}%
{Doctorate of Philosophy}

\titlepage

\chapter{Dedication}

This work is dedicated to my lovely wife for her understanding of my commitment to my research and for always encouraging me to be the best version of myself, and to my mother and father, without whose their inspiration, drive and support I might not be the person I am today.
It is also dedicated to my supervisor, Dr Amin Beheshti, for his support and his strong commitment to teaching. He believed in me, helped me to be a better researcher and shed light on the issues I faced during my PhD studies. I also thank my associate supervisor Dr Aditya Joshi, for always being supportive and sharing his knowledge with me. 

Mark Twain once said, `Kindness is a language which the deaf can hear and the blind can see'. I feel so blessed to be surrounded by many kind people. Finally, I dedicate this work to Macquarie University and the community that provided me the chance to conduct my research over the past three years.

\chapter{Acknowledgements}

Working at the Data Analytics Research Lab\footnote{https://data-science-group.github.io/}, Department of Computing, Macquarie University (MQU\footnote{https://www.mq.edu.au/}) has been a great pleasure and a wonderful privilege.

In the first place, I would like to express my sincere appreciation and deep gratitude to my supervisor, Dr Amin Beheshti, for his exceptional support, encouragement and guidance during the PhD program. Amin taught me how to do high-quality research and helped me think creatively. His truly incredible academic excellence and beautiful mind have made him a constant oasis of ideas and passions in science, which has inspired and enriched my growth as a student, a researcher and a scientist. Moreover, I thank him for providing me with the opportunity to work with a talented team of researchers.

I would like to express my gratitude to my associate supervisors, Prof. Mehmet Orgun, Dr Aditya Joshi and Dr Cecile Paris who have always supported my research.

I am thankful to everyone in the Data Analytics Research Lab at MQU for their friendship, support and helpful comments. In addition, I would like to thank the review panels and the anonymous reviewers who provided suggestions and helpful feedback on my publications.

I acknowledge Macquarie University and CSIRO's Data61\footnote{https://data61.csiro.au/} for providing the scholarships (iRTP, PGRF, Data61-Topup) that allowed me to pursue my doctoral studies.

I would like to thank the administrative and technical staff members of the Department of Computing at MQU, who have been kind enough to advise and help me in their respective roles.

Last, but not the least, I would like to dedicate this thesis to my family, for their love, patience and understanding. They allowed me to spend much of my time on this thesis. They are my source of strength and without their endless support this thesis would have never been started nor completed.
\\
\\
\noindent Seyed Mohssen Ghafari\\
\noindent Sydney, Australia\\
\noindent December 2019

\chapter{Dissertation Examiners}

\begin{itemize}
\item[$\bullet$] Professor Abdul Sattar, Griffith University, Australia
\item[$\bullet$] Associate Professor Flora Salim, RMIT University, Australia
\item[$\bullet$] Associate Professor Ashkan Sami, Shiraz University, Iran
\end{itemize}

\chapter{Publications}
This thesis is based on my research during my PhD program at the Department of Computing,
Macquarie University, between 2017 and 2019. Some parts of my research have been
published in the following venues:

\begin{itemize}
\item[$\bullet$] \textbf{Seyed Mohssen Ghafari}, Shahpar Yakhchi, Amin Beheshti and Mehmet Orgun, `Social context-aware trust Prediction: Methods for Identifying fake news', Published in Web Information Systems Engineering (WISE), pp. 161--177, 2018. (\textbf{Core Rank: A}).
\item[$\bullet$] \textbf{Seyed Mohssen Ghafari}, Shahpar Yakhchi, Amin Beheshti and Mehmet Orgun, `SETTRUST: Social Exchange Theory Based Context-Aware Trust Prediction in Online Social Networks', Published in Web Information Systems Engineering (WISE) workshop QUAT'18, pp. 46-61, 2018. (\textbf{Core Rank: A})
\item[$\bullet$] Amin Beheshti, Vahid Moraveji Hashemi, Shahpar Yakhchi, Hamid Reza Motahari-Nezhad, \textbf{Seyed Mohssen Ghafari} and Jian Yang, `personality2vec: Enabling the Analysis of behavioural Disorders in Social Networks', The $13^{th}$ ACM International WSDM Conference, Houston, USA, Texas, pp. 1-4, 2020. (\textbf{Core Rank: A*})
\item[$\bullet$] \textbf{Seyed Mohssen Ghafari}, Aditya Joshi, Amin Beheshti, Cecile Paris, Shahpar Yakhchi and Mehmet Orgun, `DCAT: A Deep Context-Aware Trust Prediction Approach for Online Social Networks', Accepted in the $17^{th}$ International Conference on Advances in MobileComputing and Multimedia (MoMM'19), Munich, Germany, December, pp. 1-8, 2019. (\textbf{Core Rank: B})
\item[$\bullet$]Shahpar Yakhchi, Amin Beheshti, \textbf{Seyed Mohssen Ghafari} and Mehmet Orgun, `Enabling the Analysis of Personality Aspects in Recommender Systems', Published in $26^{th}$ Pacific Asia Conference on Information Systems (PACIS), Xian, China, pp. 1-15, 2019 . (\textbf{Core Rank: A})
\item[$\bullet$] Shahpar Yakhchi, \textbf{Seyed Mohssen Ghafari}, Amin Beheshti and Mehmet Orgun, `CNR: Cross-network Recommendation Embedding User's Personality', Published in Web Information Systems Engineering (WISE) workshop QUAT'18, pp. 62-77, 2018. (\textbf{Core Rank: A})
\item[$\bullet$] \textbf{Seyed Mohssen Ghafari}, Amin Beheshti, Aditya Joshi, Cecile Paris, Shahpar Yakhchi and Mehmet Orgun, `Intelligent Trust Prediction: Methods for Identifying Fake News', is Accepted in Cyber Defence Next Generation Technology and science Conference (CDNG), Brisbane, Australia, 2020.
\end{itemize}

\chapter{Abstract}

Trust can be defined as a measure to determine which source of information is reliable and with whom we should share or from whom we should accept information. There are several applications for trust in Online Social Networks (OSNs), including social spammer detection, fake news detection, retweet behaviour detection and recommender systems. 
Trust prediction is the process of predicting a new trust relation between two users who are not currently connected. In applications of trust, trust relations among users need to be predicted. This process faces many challenges, such as the sparsity of user-specified trust relations, the context-awareness of trust and changes in trust values over time.

In this dissertation, we analyse the state-of-the-art in pair-wise trust prediction models in OSNs. We discuss three main challenges in this domain and present novel trust prediction approaches to address them. We first focus on proposing a low-rank representation of users that incorporates users' personality traits as additional information. Then, we propose a set of context-aware trust prediction models. Finally, by considering the time-dependency of trust relations, we propose a dynamic deep trust prediction approach. We design and implement five pair-wise trust prediction approaches and evaluate them with real-world datasets collected from OSNs. The experimental results demonstrate the effectiveness of our approaches compared to other state-of-the-art pair-wise trust prediction models.

\tableofcontents
\listoffigures
\listoftables

\mainmatter

\chapter{Introduction}
\label{chap:introduction}

In early human societies, (hunter-gatherer) people realised that to fulfil their needs, they had to interact with each other. Quickly, they found that not all interactions were beneficial for them. For instance, their experiences in trading with other people (traders) were not always satisfactory, and sometimes they were deceived by the traders. At that point, they learned to interact with trustworthy people.   
Trust can be defined as the `willingness of a party to be vulnerable to actions of another party based on the expectation that the other will perform a particular action important to the trustor, irrespective of the ability to monitor or control that other party'~\cite{Mayer}. 
`Trust is necessary in order to face the unknown, whether that unknown is another human being, or simply the future and its contingent events'~\footnote{https://reviews.history.ac.uk/review/287a}. Sociologically speaking, `a complete absence of trust would prevent [one] even getting up in the morning'~\cite{TrustDefinationSociology3}.

\section{Definition of Trust}
With the development of human societies, trust has played an important role in people's lives, including in their relationships, families and their businesses and in social management systems. With the  development of science and scientific knowledge, different branches of science that focused on human behavioural analysis and human interaction analysis started to study the concept of trust. 
Trust has different definitions in different scientific fields. Here, a brief overview is provided on the definition of trust in psychology, sociology, economics and, of particular relevance to the subject of this thesis, computer science.

\subsection{Trust in Psychology}

Schlenker et al.~\cite{TrustDefinationPsychology3} provided a definition for trust: 
being confident about received information from another party in an uncertain environmental
state.
Psychologists also define trust as `the
subjective probability by which an individual expects that another performs a given
action on which its welfare depends'~\cite{TrustDefinationPsychology}.
Psychologically speaking, 
an inclination towards trusting others can be considered a personality trait~\cite{TangBook}.
Moreover, `trusting behaviour takes place when an individual confronts an ambiguous path leading to a perceived either beneficial or harmful result
contingent on the action of another person'~\cite{PayanName}.

\subsection{Trust in Sociology}

Although in sociology studies, the main focus is on the trust in the society or social relations, some research has also focused on trust at the individual level. At this level, the definition of trust is similar to that in psychology~\cite{PayanName}; for example, Sztompka stated that `trust is a bet about the future contingent actions of others'~\cite{TrustDefinationSociology1}. At the society or social relations level, sociologists consider trust as a properties of social groups~\cite{PayanName} and define it as `a set of expectations shared by
all those involved in an exchange'~\cite{TrustDefinationSociology2}. Another sociologist defined trust as `a means for reducing the complexity of society' ~\cite{TrustDefinationSociology3}. A different definition of trust was provided by Seligman~\cite{TrustDefinationSociology4}:
`trust enters into social interaction in the interstices of systems, when for
one reason or another systematically defined role expectations are no longer viable'.
Hence, according to Seligman, if people play their expected roles, we can safely have our own transactions~\cite{TrustDefinationSociology4}.

\subsection{Trust in Economics}

In economics, trust is defined as `the property of a business relationship, such that reliance can be
placed on the business partners and the business transactions developed with them'~\cite{TrustDefinationEconomics}.
Economists also conceptualise trust as `existing when one party has confidence in an exchange partner's reliability and integrity'~\cite{TrustDefinationEconomics2}.
Moreover, in online trading environments, where there is a lack of direct interaction with customer and products, `trust can reduce transaction risks, mitigate information asymmetry and generate price premiums for reputable vendors'~\cite{PayanName}~\cite{TrustDefinationEconomics3}~\cite{TrustDefinationEconomics4}. 

\subsection{Trust in Computer Science}

The concept of trust is widely used in computer science. 
Artz and Gil~\cite{TrustDefinationComputer} classified trust related research domains in computer science into four major categories: i) policy-based trust, which covers studies in topics related to network security credentials, security policies and trust languages; ii) reputation-based trust, which includes research on trust in peer-to-peer networks, and grids and trust metrics in a web of trust;
iii) general models of trust, encompassing research addressing general considerations and properties of trust and software engineering; and
iv) trust in information resources, which focus on trust concerns on the Web, the semantic Web and information filtering based on trust. 

Trust also plays a significant role in the online activities of users of platforms such as Online Social Networks (OSNs). Tang et al.~\cite{TangGHL13} provided a popular definition for trust in OSNs: `Trust provides information about with whom we should share information, from whom we should accept information and what considerations to give to information from people when aggregating or filtering data'. There are many applications for trust in OSNs, including:
social spammer detection~\cite{Li2015}, fake news detection~\cite{GhafariWise}, retweet behaviour detection~\cite{RetweetBehavior1}~\cite{RetweetBehavior2}
and recommender systems~\cite{Recom1}~\cite{Recomm4}. All these applications require predicting the trust relations among users. 

\section{Preliminaries}

This section briefly introduces the main concepts of this dissertation, namely
OSNs and trust prediction in OSNs.

\subsection{Online Social Networks}
Garton et al.'s~\cite{OSNDefination} widely accepted definition concerning OSNs holds that `when a computer network connects people or organisations, it is a social network. Just as a computer network is a set of machines connected by a set of cables, a social network is a set of people (or organisations or other social entities) connected by a set of social relationships, such as friendship, co-working or information exchange'. 
OSNs are relatively new and evolving phenomena on the Web. Users of these online platforms can communicate with others and present themselves through their profiles~\cite{OSNDefination3}~\cite{OSNDefination2}.

Social network analysis is an area of study focusing on OSNs that looks for patterns of relations among people~\cite{OSNDefination}. In OSNs, relations can be characterised by their content (i.e., resources exchanged, such as information), direction and strength~\cite{OSNDefination}. One of the relations on which social network analysis focuses is the trust among people in OSNs. These studies aim to understand why people trust each other and establish trust relations in OSNs, with a view to predicting trust relations among people in OSNs.

\subsection{Trust Prediction in Online Social Networks}

Trust prediction
is defined as `the process of estimating a new pair-wise trust relation
between two users who are not directly connected based on
existing observations'~\cite{ZhengWOZL14}.
The literature of trust prediction can be divided into two main categories~\cite{TangGHL13}: supervised approaches~\cite{LiuLLLSSK08}~\cite{Nguyen2009}~\cite{TangandLiu} and unsupervised approaches~\cite{TangGHL13}~\cite{WangWTZC15}~\cite{Borzymek2009}~\cite{IjCAI2018}~\cite{Guha2004}. 
Supervised trust prediction approaches treat the trust prediction problem as a classification problem. They create a feature set for their classifiers and consider the existence of trust as labels to train a binary classifier~\cite{TangGHL13}. Unsupervised approaches can also identify the trust relations among users, even if they are not directly connected. These approaches use methods like trust propagation or low-rank representation~\cite{WangWTZC15}.

\section{Challenges of Trust Prediction in Online Social Networks}
This thesis will focus on the following three significant problems in OSNs:  sparsity of user-specified trust relations, context-aware pair-wise trust relations and time-aware pair-wise trust relations.

\subsection{Sparsity of User-Specified Trust Relations}

User-specified trust relations are extremely rare~\cite{IjCAI2018}. For instance, `the density of a typical trust network
in social media is less than 0:01'~\cite{TangBook}~\cite{SocialNetwork}.
As another example, `the sparsity of Advogato, Ciao,
and Epinions [frequently used datasets in trust prediction related research], i.e., the ratio of the observed trust relations
to all the possible relations, is 0.1195\%, 0.2055\% and
0.4135\%, respectively. It is challenging to predict the trust
relations well with so limited observed links'~\cite{IjCAI2018}.
Moreover, trust relations
follow the rules of the power law distribution: many trust relations can be
accounted for a small number of users and a large number of users participate in only a
few trust relations~\cite{TangGHL13}. For any trust prediction approach in OSNs, this sparsity of user-specified trust relations compared to all possible relations among users is low. 
This makes the pair-wise trust prediction problem
in OSNs a challenging task; any trust prediction approach should be able to deal with this data sparsity problem.

\subsection{Context-Aware Pair-wise Trust Relations}
The notion of trust is context-dependent~\cite{GhafariWise}~\cite{TangBook}: Trusting someone in one context does not guarantee trusting them in another context~\cite{TangBook}.
As an example, the context dependency of trust has been investigated by~\cite{trustcontext} in the collected data from a real-world product review website~\footnote{http://www.Epinions.com}. In this website there is an option for users to explicitly indicate which users are trustworthy. Tang et al.~\cite{TangGHL13} used this information as the ground truth of their analysis. They considered items' categories (e.g., electronics, sports and entertainment) as the context of trust and reported that: `less than 1\% of users, trust their friends in all categories' and `on average, people trust only 35.4\% of their trust networks for a specific category'. 

Hence, people trust each other
in certain contexts. Context is the information about the condition of an entity~\cite{Zheng/ShortPaper/2014}. As
an illustration of a single context (focusing on the domain of the trust), consider David who is a PhD student at the Computing Department. He trusts his supervisor in the computer science
field; however, he does not necessarily trust him in sports. As a result, predicting pair-wise trust relations with respect to the different context of trust can be a daunting task.

  
  
\subsection{Time-Aware Pair-Wise Trust Relations}
Trust values can also change over time. Users can establish new trust relations or eliminate their existing trust relations after a period of time. For instance, if Jack trusts David (as two users in an OSN) at time $T_1$, this does not necessary mean that he trusts him at time $T_2$ (where $T_2$=$T_1 +h$ and h is a fraction of time). As another example, David may not trust Sarah at time $T_1$, but he could trust her at time $T_2$.  

Hence, predicting the pair-wise trust relations statically may not be a realistic approach for OSNs. Trust is time-sensitive: if John trusts David at time $T_1$, this trust relation may change at time $T_2$. This can be affected by many factors, such as some new behaviour on David's part or a change in John's interests. Hence, predicting pair-wise trust relations in OSNs dynamically can be a challenging task.    
  
\section{Dissertation Contributions}
This thesis makes the following contributions:
  
\begin{itemize}
  \item \textbf{Proposing a Low Rank Representation of Users that Incorporates Users' Personality Traits as Additional Information:}
  
Many social studies have attempted to explore the reasons behind the
establishing of trust relations among people. Although many of
these studies consider trust a situational construct, some investigate individual
characteristics in their trusting behaviour predictions~\cite{EVANS20081585}. One
of these characteristics is people's personality. Alarcon et al.~\cite{ALARCON201869}
stated that `personality can assist researchers in understanding
the processes underlying trust interactions'. Studies in social
science consider people's personality as a part of developing trust
relations in their face-to-face interactions. However, this important
attribute remains unexplored for pair-wise trust relations prediction
in OSNs. Based on a well-known theory from psychology, the Big Five personality model~\cite{BigFive}, people's personalities can be
characterised by five personality traits: Openness, Conscientiousness,
Extraversion, Agreeableness and Neuroticism. Augmenting
these personality traits (as implicit or additional information) in our trust prediction approach can help us improve its prediction performance (see \textbf{Chapter 4}). 

  \item  \textbf{Considering the Context of Trust and Proposing a Matrix Factorisation Based Trust Prediction Approach}:
  
 We apply a well-known psychological theory, Social Exchange Theory
(SET), to evaluate the potential trust relations among users in OSNs. Based on
SET, one person may start a relationship with another person, if and only if the cost
of that relationship is less than its benefit. To evaluate potential trust relations
in OSNs based on SET, we first propose some factors to capture the costs and
benefits of a relationship. Then, based on these factors, we propose a trust metric
called trust degree. At this point, we also propose a trust prediction method, based
on matrix factorisation (MF) and apply the context of trust in a mathematical model (see \textbf{Chapter 5}).

 \item \textbf{Proposing a Tensor Decomposition Based Trust Prediction Approach that Directly Considers the Context of Trust}:
 
We present another context-aware trust prediction approach, which improves the previous method by directly considering the context of trust in its model. It considers the notion of a context (i.e., any knowledge
to specify the condition of an entity) as well as the social actor's behaviour
(supported by theories from social psychology) as first-class citizens. We present
novel algorithms that employ social context factors inspired by social psychology
theories and mathematically model our approach based on tensor decomposition (TD) (see \textbf{Chapter 5}).

\item \textbf{Proposing a Deep Classifier for Context-Aware Trust Prediction in OSNs}:
 
We propose another novel approach for considering the context of trust using a supervised approach, Deep Context-Aware Trust prediction approach (DCAT). The proposed model is based on a deep structure, which makes it one of the first deep context-aware trust
predictors for OSNs. $DCAT$ has a higher prediction performance compared to our two previous context-aware trust prediction approaches (see \textbf{Chapter 5}).

\item \textbf{Focusing on the Dynamic Nature of Pair-Wise Trust Relations:}
  
Most of the existing trust prediction approaches assume that trust relations are fixed over time. Thus, they fail to capture the dynamic behaviour of users in OSNs. We propose a dynamic deep trust prediction model and a novel deep structure that incorporates users' emotions (psychology studies have proven that incidental emotions have a significant effect on trust) and the textual contents provided by users in OSNs.
We also consider different time windows to dynamically predict pair-wise trust relations in OSNs (see \textbf{Chapter 6}).
\end{itemize}

\section{Dissertation Organisation}
The remainder of this dissertation is organised as follows. We start with a discussion
of the current state-of-the-art in trust prediction approaches in OSNs in Chapter~\ref{chap:chapterRelatedWorks}. We
explain in more depth what trust relations and trust prediction processes are. We analyse existing trust prediction approaches based on factors such as whether they consider the context of trust or the dynamic nature of trust and the types of algorithms they use. In Chapter~\ref{chap:DataandMetrics}, we thoroughly discuss the datasets, evaluation metrics and baseline methods that we use in the evaluations presented in this dissertation. 

In Chapter~\ref{chap:chapter4}, we present the details of our novel pair-wise trust prediction approach, which can predict trust relations among users in the presence of the sparsity of user-specified trust relations. This approach focuses on the properties of the users and tries to use the users' personality as additional information for its trust prediction model. This model is based on low-rank representation of users and employs a three-dimensional $TD$.

Chapter~\ref{chap:context} discusses our proposed context-aware trust prediction models and explains how our models facilitate the representation and analysis of
trust relations in different contexts of trust. Two of them are based on low-rank representation of users ($MF$ and $TD$) and one is a deep classifier. In these approaches, we first propose some new context factor to capture trust relations between users. Then, with the help of these context factors, we predict trust among users in OSNs.

In Chapter~\ref{chap:time}, we propose a novel time-aware trust prediction approach by focusing on the dynamic nature of pair-wise trust relations. This approach has a deep structure and focuses on analysing users' demographic features (e.g., the number of followers and followees) and their textual content features in OSNs. More specifically, it investigates the relation between users' emotions and their trust relations in OSNs. 

Finally, Chapter~\ref{chap:conclusion} presents the concluding remarks of this dissertation and discusses possible directions for future work.


\chapter{Related Work}
\label{chap:chapterRelatedWorks}

\begin{figure*}
\centering
  \includegraphics[ width=0.50\textwidth]{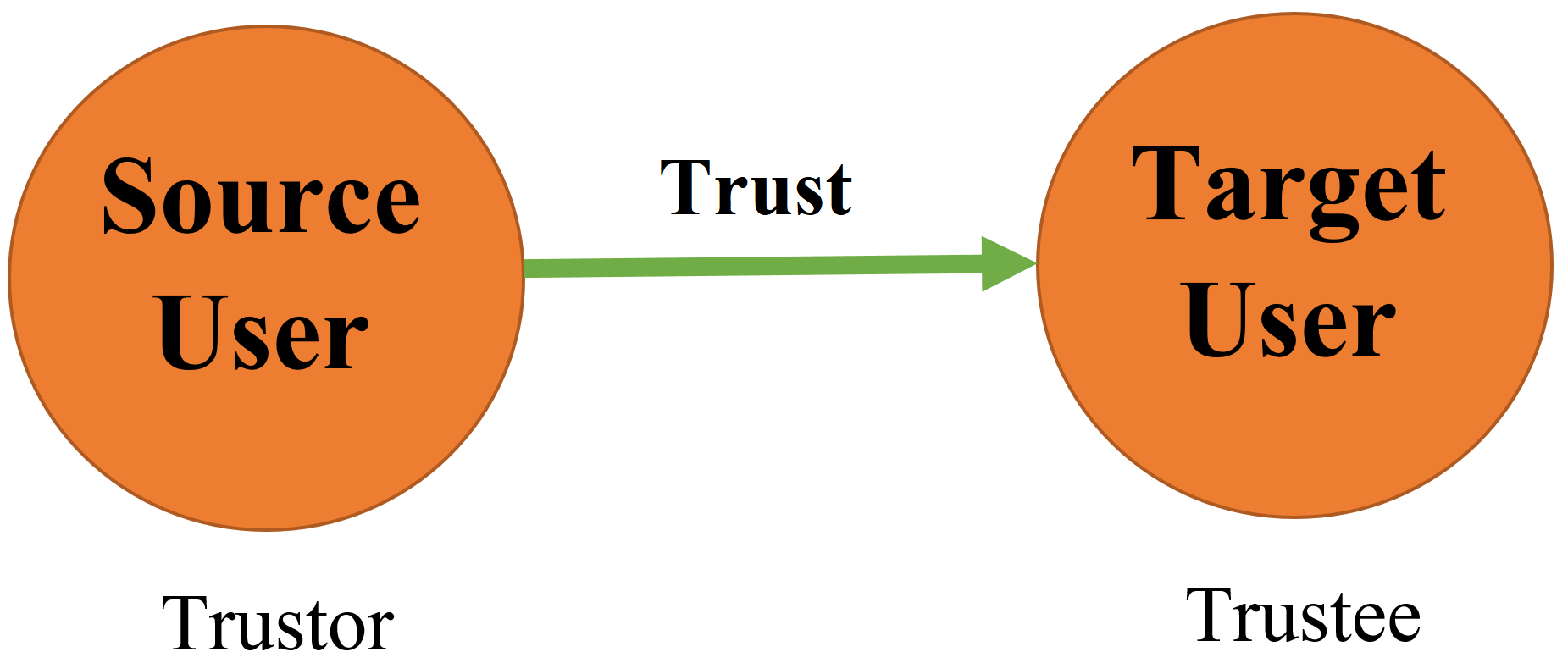}
  \caption{Representation of a pair-wise trust relation.}
  \label{fig:TrustPresentation}
\end{figure*}

A pair-wise trust relation (Figure~\ref{fig:TrustPresentation}) is a relationship between a source user (trustor) and a target user (trustee) that indicates that the trustor trusts the trustee. With the help of trust, the trustor may seek information from the trustee, to avoid being confused by the huge amount of available data (i.e., mitigated information overload) and to be confident about the credibility of the received information (i.e., increased information credibility)~\cite{TangBook}. In this chapter, we look at the properties of trust and how trust can be collected. We discuss how to represent trust relations and what the trust prediction process is. We also review existing trust prediction approaches and explain the relation between users' personality and trust, and applications of trust in OSNs.    

\section{Properties of Trust}
The properties of trust have been listed as context specific, dynamic, propagative, subjective, asymmetric and event sensitive~\cite{Sherchan/2013}.

\textbf{Context Specific:} Trust is a context-dependent notion. A trust relation in one context does not guarantee its existence in another context. 

\textbf{Dynamic:} Trust is a time-dependent concept. Trusting someone at one point in time does not mean the trust relation will exist at another point in time. Trust relations can change because of new experiences, new behaviours on the part of target users or a shift in interests of either or both users. 

\textbf{Propagative:}
`Because of its propagative nature, trust information can be passed from one member to another in a social network, creating trust chains'~\cite{Sherchan/2013}. As an example, if David trusts Sarah, and Sarah trusts Mathew, there is a trust relation between David and Mathew, whereby David may derive some amount of trust towards Mathew from the strength of the trust relations between David and Sarah, and Sarah and Mathew~\cite{Sherchan/2013}. 

\textbf{Subjective:} Trust is a subjective concept. Being trustworthy in one's mind does not imply a person is considered trustworthy by all others. For instance, suppose David and Sarah are two PhD students in the computer science department, and Mathew is a PhD supervisor and a lecturer in this department. David may believe that Mathew is trustworthy, while Sarah does not. Such differences in opinion arise from people's diverse expectations, biases and interests.   

\textbf{Asymmetric:} `Trust is typically asymmetric'~\cite{Sherchan/2013}. In other words, if David trusts Sarah, he may not necessary be trusted by her.  

\textbf{Event Sensitive:} Establishing a trust relation may take a great deal of effort and time, but a high-impact event can destroy it~\cite{Sherchan/2013}~\cite{Eventsensitive}.

\section{Collecting Trust Information}
There are three different sources from which to collect trust information~\cite{Sherchan/2013}; that is, attitude, experience and behaviour.   

\textbf{Attitude:} Our attitude is the way we think or feel (positively/negatively) about something. Information about a person's attitude can be captured by their online interactions using a measure such as a Likert scale~\cite{Sherchan/2013}. 

\textbf{Experience:}
Experience can refer to the `knowledge or skill that you get from doing, seeing, or feeling things, or the process of getting this'~\footnote{https://dictionary.cambridge.org/dictionary/english/experience}. In OSNs, users can gain experience information about other users by interacting with them. This experience can be captured by the feedback among users, and better feedback may result in more interactions in future~\cite{Sherchan/2013}. 

\textbf{Behaviour:}
Human behaviour refers to `the range of behaviours exhibited by humans ... [which are] typically influenced by culture, attitudes, emotions, values, ethics, authority,
persuasion, coercion and/or genetics'~\cite{behavior}~\cite{Sherchan/2013}. In OSNs, we may notice a sudden change in the frequency of interaction between two users, or the amount of activity of a user. The first case may indicate that the trust level between those users has decreased. While the second may represent a decline in the user's trust towards the community in which he or she participate~\cite{Sherchan/2013}. 

\section{Trust Representation}
\begin{figure*}
\centering
  \includegraphics[width=\textwidth]{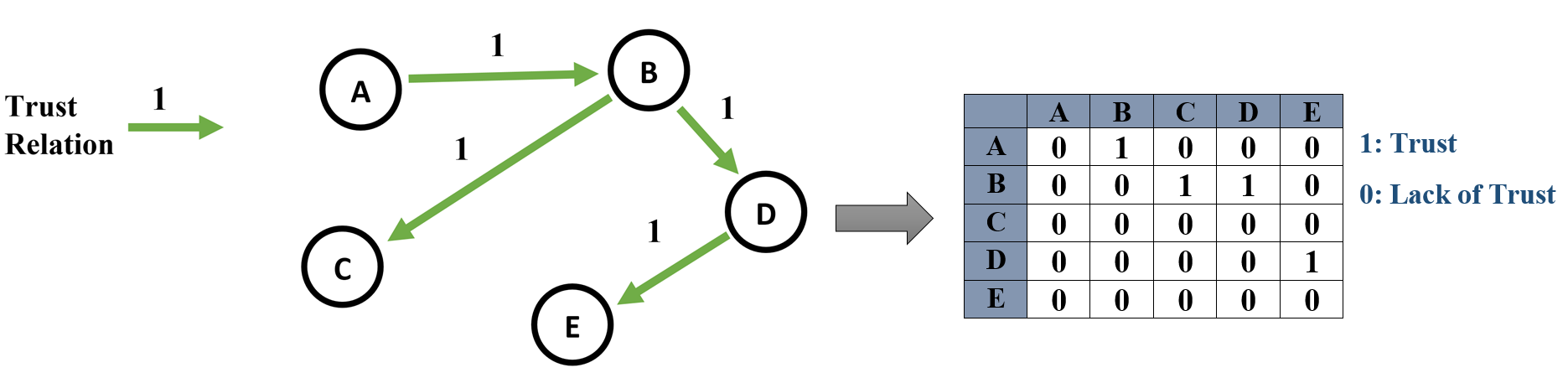}
  \caption{Representation of a trust network and its corresponding adjacency matrix. There are four trust relations in this figure: A trusts B, B trusts C and D, and D trusts E.}
  \label{fig:TrustPresentationgraph1}
\end{figure*}

To denote the naivest notion of trust (e.g., single-dimensional trust), one can use a representation similar to Figure~\ref{fig:TrustPresentationgraph1}. In this figure, there are five users (A, B, C, D and E). On the left side, there is a trust network representation among these users, where the green arrow with the label `1', indicates the existence of a trust relation between two users, and its absence means that there is not any trust relation between them. To the right of this figure, there is a corresponding adjacency matrix, showing the trust network between any two users. In this matrix, `0' represents the lack of trust and `1' illustrates the existence of trust between two users.

However, trust may have multiple dimensions. For instance, trust is a context-dependent concept. Context is the information about the condition of an entity~\cite{Zheng/ShortPaper/2014}. 
As
an illustration of a single context (focusing on the domain of the trust), consider Sarah, a football player, who trusts her coach in football. This does not necessarily mean that she also trusts her coach regarding music. 
Hence, to represent trust relations among users in different contexts, we need a representation with more dimensions. As another example, if Mathew trusts Jack (as two users in an OSN) at time $T_1$, this does not necessary mean that Mathew will also trust Jack at time $T_2$ (where $T_2=T_1 + h$, and $h$ is a fraction of time). Hence, matrices cannot appropriately represent a multi dimensional trust network. 
Instead, tensors are one of the most favoured representations for trust relations as they can store data in several dimensions. 

\begin{figure*}
\centering
  \includegraphics[width=\textwidth]{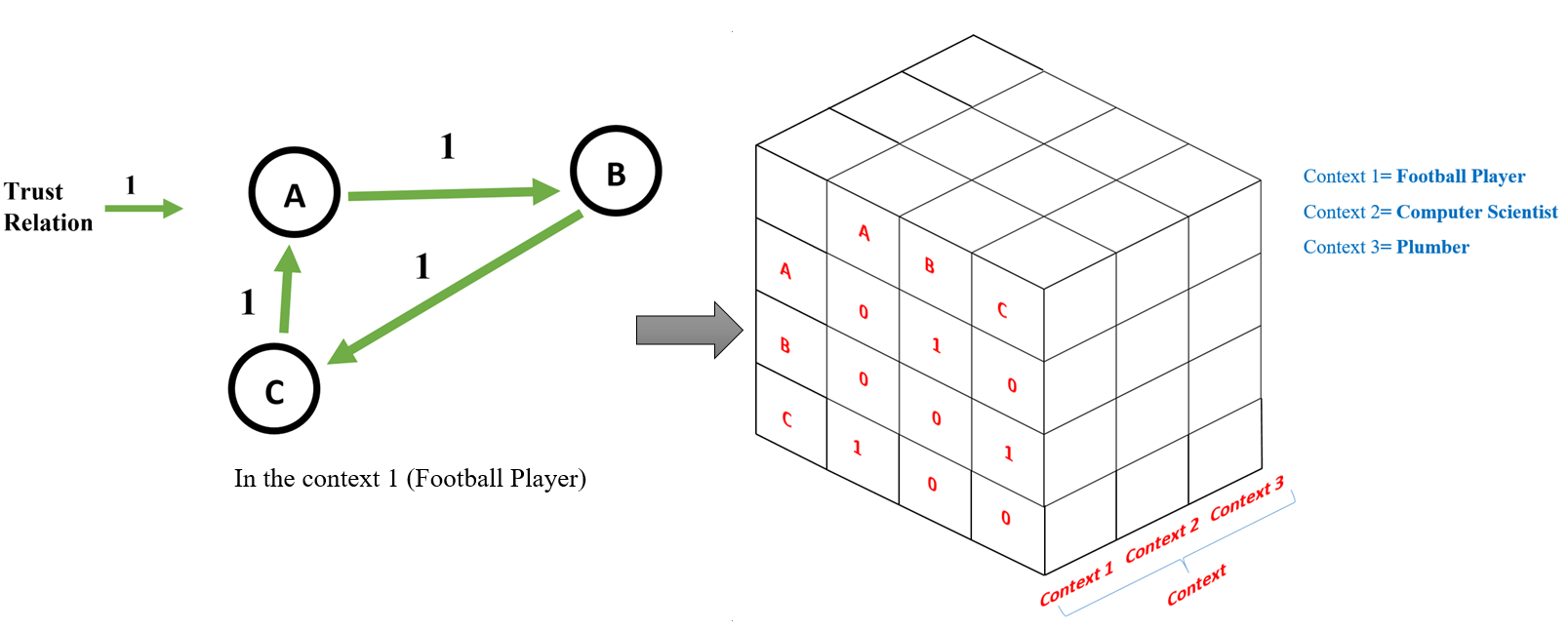}
  \caption{Representation of a trust network and its corresponding adjacency matrix. There are three trust relations and three contexts of trust in this figure: in the first context, A trusts B, B trusts C and C trusts A.}
  \label{fig:TrustPresentationgraph}
\end{figure*}

Figure~\ref{fig:TrustPresentationgraph} illustrates an example of representing trust relations in different contexts of trust. In this figure, which demonstrates a single dimension of context (e.g., domain of expertise), there are three contexts of trust (football player, computer scientist and plumber). There are also three users (A, B and C). As shown, there are three trust relations between users in the first context (football player). According to these trust relations, A trusts B, B trusts C and C trusts A as a football player. For representing these trusts relations and other trusts relations between these users in other contexts, we can use a tensor. For instance, Figure~\ref{fig:TrustPresentationgraph} shows a three dimensional tensor with two dimensions for representing users' relations and a third dimension denoting the contexts of trust. Since all the mentioned trust relations are related to the football player (context 1), they are stored in the matrix of context 1 of this tensor.

\section{Trust Prediction Process}
\begin{figure*}
\centering
  \includegraphics[width=\textwidth]{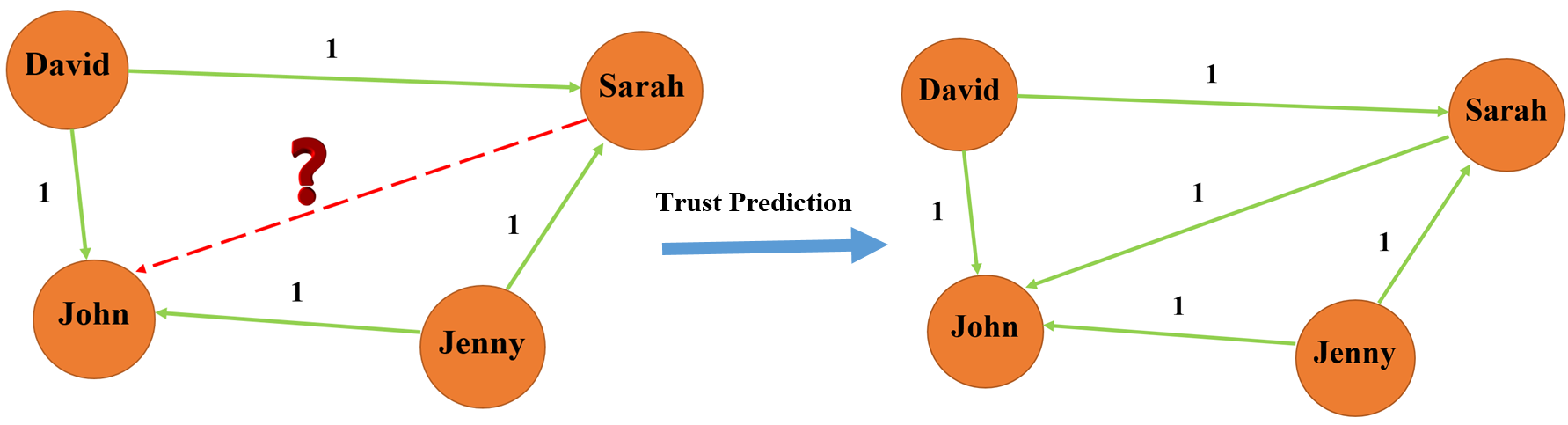}
  \caption{Trust prediction in OSNs: David, Sarah, John and Jenny are four users in an OSN.  Explicit trust relations are shown by the green arrow and the label `1'. On the left, we want to know if there is any trust relation from Sarah to John. On the right, using a trust prediction model, we can give a positive answer to that question. Sarah trusts John. }
  \label{fig:trustPrediction}
\end{figure*}

Trust networks in OSNs are usually sparse~\cite{TangGHL13}. They follow the power law distribution whereby a small number of users account for the majority of the trust relations~\cite{TangGHL13}. As a result, the explicit trust relations among many users in OSNs are unknown~\cite{TangBook}. 
Therefore, to employ trust information in different applications in OSNs (e.g., recommender systems and retweet behaviour prediction), we need to predict unknown trust relations among users. Figure~\ref{fig:trustPrediction} illustrates a simple example of the trust prediction procedure; on the left side, we have some users and their explicit trust relations, as shown by the green arrow and the label `1'. We want to know if there is a trust relation between Sarah and John. A trust prediction approach can be used to predict that the existence of this trust relation.

\section{Trust Prediction Approaches}
In this section, we describe related work in four areas: representation of the network, type of prediction algorithms, context-awareness and time-awareness. Finally, we classify the existing pair-wise trust prediction approaches (Table~\ref{long}).

\subsection{Representation of Trust Networks}
We broadly categorise trust prediction approaches into three categories: graph-based trust models, interaction-based trust models and hybrid trust models~\cite{Sherchan/2013}.  

\subsubsection{Graph-Based Trust Prediction Models}
Approaches in the category of graph-based trust models are mostly based on the concept of web-of-trust or Friend-of-a-Friend (FOAF). Each user is assumed to have a trust network that contains
friends (i.e., social network actors/users) as nodes, with the relationships (i.e., value of their trust relations) among them as the edges~\cite{Sherchan/2013}. This assumption can be invalid or too strong because, in many online communities, there is either no way to identify a web-of-trust or the connectivity is sparse~\cite{LiuLLLSSK08}. Moreover, in some cases, this kind of approach may fail to capture the actual interactions among members~\cite{Sherchan/2013}. Trust propagation-based~\cite{GOLBECK/2003} and inference-based~\cite{Yan/WWWJ/2015} methods belong to this category.

Golbeck et al.~\cite{GOLBECK/2003} proposed another trust inference approach based on the FOAF concept that can determine which pairs of users trust each other and on which topic. Similarly, Zhang et al.~\cite{ZHANG/network-based/2006} presented an approach by which the source user accepts the recommendation from similar neighbour nodes (i.e., other users directly connected to the target user). Kim et al.~\cite{KimLeLauw2008} proposed an approach to build a web-of-trust based on the implicit feedback of users in a certain context. Golbeck~\cite{GOLBECKb} proposed another trust prediction approach, TidalTrust, also based on the FOAF concept. In TidalTrust, if two neighbours have a high trust rating, it is more likely they would agree on other users' trustworthy levels~\cite{Sherchan/2013}. Ziegler and Lausen~\cite{ZIEGLER2004} developed another network-based trust prediction model, Appleseed, for use in the semantic Web. They focused on local group trust metrics to improve the computational complexity of the trust prediction procedure. Hang and Singh~\cite{HANGANDSINGH} introduced a new trust prediction approach based on the similarity of users' trust networks; they treated the recommendation problem as a graph
similarity problem~\cite{HANGANDSINGH}. Zuo et al.~\cite{ZUOandHU} proposed a trust prediction framework based on trust chains and a trust graph. This framework can
`calculate trust along a trust chain and evaluate a trust
based on a trust certificate graph'~\cite{ZUOandHU}. Caverlee et al.~\cite{CaverleeandLiu} developed another trust prediction framework, SocialTrust, focusing on social relationships and users' feedback. SocialTrust also allocates a weight value to feedback according to the PageRank algorithm. 

Zhang and Yu~\cite{ZhangandYu2012} designed a semantic-based trust reasoning mechanism for trust prediction in OSNs. They noted that trust is a category-dependent concept and traditional trust prediction approaches required much human effort to predict pair-wise trust relations. They also built `a domain ontology for data communication and knowledge sharing and exploit[ed]  role-based and behaviour-based reasoning functions to infer implicit trust relationships and category-specific trust relationships'~\cite{ZhangandYu2012}.
Liu et al.~\cite{Liu2012} proposed a heuristic approach, called the Heuristic
Social Context-Aware Trust Network Discovery algorithm, adopting the K-best-first search for addressing the trust network extraction problem by developing a contextual social network structure and proposing the concept of Quality of Trust Network~\cite{Liu2012}.

\subsubsection{Interaction-Based Trust Prediction Models}
Approaches in the previous category may fail to `capture actual interactions among
members. The volume, frequency and even the nature of interaction are important indicators
of trust in social networks'~\cite{Sherchan/2013}. By contrast, interaction-based trust prediction models mainly focus on the interactions among users. 
Liu et al.~\cite{LiuLLLSSK08} proposed a classification approach for trust prediction in OSNs based on the action and interactions of users. A similar approach presented by Nepal et al.~\cite{Eventsensitive} proposed a trust prediction model that considers two types of trust: the trust of other users towards a target user and the trust value that a user has towards a community. Adali et al.~\cite{ADALI} developed a trust prediction approach focusing on users' communication behaviour and more specifically on conversational trust (i.e., duration and frequency of communication between two users) and propagation trust.
Sacco and Breslin~\cite{SaccoBreslin} proposed a trust prediction approach centering on the subjective trust values of connected users, based on their social interactions~\cite{SaccoBreslin}. They stated that most of the existing trust prediction approaches are `propagating known trust values among peers in a trusted network and do not provide measures for asserting a trust value from user interactions between peers' ~\cite{SaccoBreslin}.
These approaches only focus on users' interactions and do not consider the social network structure, which may contain important information about users and the type of relations among them. 

\subsubsection{Hybrid Trust Prediction Models}
Hybrid trust models combine the network-based and interaction-based models. In particular, they simultaneously consider users' previous interactions and the social network's structure~\cite{TRIFUNOVIC/2010}.

\subsection{Type of Prediction Algorithms}
We now discuss past work from the perspective of the type of algorithms used. We can roughly categorise trust prediction approaches into supervised and unsupervised approaches. 

\subsubsection{Supervised Approaches}
Liu et al.~\cite{LiuLLLSSK08} developed a supervised trust prediction model and a classifier that works with a set of users' features and interactions. Ma et al.~\cite{MaLNSL09} proposed a personalised and cluster-based classification trust prediction model that creates user clusters and then trains a classifier for them.
Matsuo et al.~\cite{MatsuoY09} focused on a Japanese e-commerce website called @cosme, and became the first to explain the concept of community gravity: a two-way effect of trust and rating. They followed this with a model to formulate the trust prediction and rating prediction problems. Grana~\cite{supervised1} introduced a supervised trust prediction approach: a binary classification that focuses on users' reputation.
Wang et al.~\cite{supervised2} proposed a trust-distrust prediction approach that simultaneously employed Dempster-Shafer theory and neural networks. They also analysed the effects of homophily theory, emotion tendency and status theory in trust relations~\cite{supervised2}.
Zhao and Pan~\cite{supervised3} developed another supervised trust prediction approach: a classifier with a feature set that included several trust-related factors. They used the existing trust labels for training their classifier.  
Bachi et al.~\cite{Bachi2012} developed a new trust inference framework to infer trust-distrust relationships. Their approach was based on frequent subgraph mining, signed networks, social balance
theory, edge classification and rule-based link prediction~\cite{Bachi2012}. It decomposed a `trust network into its ego~\footnote{`A portion of a social network formed of a given individual,term edego, and the other persons with whom she has a social relationship, termed alters'~\cite{ARNABOLDI201626}} network components and mining on this
ego network set the trust relationships'~\cite{Bachi2012}.

Korovaiko and Thomo~\cite{supervised4} designed a classifier that works with users' provided ratings on product review websites. They analysed the effects of similarities in users' ratings on their trust relations. 
Borzymek and Sydow~\cite{supervised5} focused on analysing graph-based and users' rating-based attributes and employed a C4.5 decision tree-based algorithm to predict users' trust-distrust relations in OSNs.     
Lopez and Maag~\cite{supervised6} proposed a generic trust prediction framework as a multi-class classifier, employing the RESTful web-service architecture and support
vector machines technique~\cite{CortesVapnik}. 

Zolfaghar and Aghaie~\cite{ZolfagharAghaie2011} developed a supervised time-aware trust prediction approach. They considered the trust prediction problem as a temporal link prediction problem. Their main focus was analysing historical information on the trust relations (or links).
Raj and Babu~\cite{RAJBabu} presented a probabilistic reputation feature model as a supervised trust prediction approach. They proposed a  framework using reputation features
to solve the cold start problem in trust prediction. They also employed
the SMOTE-Boost algorithm to establish balanced classes in their datasets~\cite{RAJBabu}.
Zhao et al.~\cite{Zhao2016ATT} introduced a trust prediction approach to evaluate the trustworthiness of users and tweets on Twitter, focusing on Twitter data from Latin America. Their approach `jointly consider users' social and contextual relationships in a Twitter social graph'~\cite{Zhao2016ATT}. Their approach used a novel topic-focused trustworthiness estimator model based on a similarity metric. For instance, if a tweet is similar to trustworthy tweets, it can also be considered trustworthy. 

Zhang et al.~\cite{Zhang2014} with the aim of addressing the `all good reputation' problem, proposed a multidimensional trust prediction approach called CommTrust, which evaluated trust by mining users' feedback comments~\cite{Zhang2014}. Chakraverty et al.~\cite{Chakraverty2016} introduced a logistic regression-based model that focused on the ratings similarity of users to predict their pair-wise trust relations. Their experimental results is somewhat contradict those of Tang et al.~\cite{TangGHL13}. Chakraverty et al.'s study focused on the implicit similarity and co-rated item-count thresholds, finding low precision, recall and coverage for the similarity threshold and better precision, recall and coverage for the co-rated item-count threshold~\cite{Chakraverty2016}. Nunez-Gonzalez et al.~\cite{Nunez-Gonzalez2015} considered the trust prediction problem as a classification problem. They focused on the reputation features of users, because they believed that their reputation information could be used
to evaluate the trustworthiness of a user~\cite{Nunez-Gonzalez2015}.

\subsubsection{Unsupervised Approaches}
Tang et al.~\cite{TangGHL13} proposed an unsupervised trust prediction model called $hTrust$. It exploits the homophily effect on the trust prediction procedure by focusing on similar users. In this way, Tang et al. identified similar users based on the users' ratings similarity. They considered three factors for rating similarities: users who rated similar items, users who gave similar ratings for similar items and users who had similar ratings patterns.
Wang et al.~\cite{WangWTZC15} developed an unsupervised model, sTrust, using social status theory and the PageRank algorithm~\cite{pagerank}, based on $MF$. In this approach, if a user has a higher social status in an OSN, he or she is more likely to be trusted by other users. 

Guha and Kumar~\cite{Guha2004} developed a trust prediction model that propagate trust based on users' trust or distrust relations with others. Golbeck~\cite{Golbeck06} put forward a website called FilmTrust which used trust to produce movie recommendations. Wang et al.~\cite{IjCAI2018} proposed a trust prediction approach that, in addition to learning low-rank representations of users, also learned these sparse components of the trust network~\cite{IjCAI2018}. 
Zheng et al.~\cite{Zheng/ShortPaper/2014} suggested an unsupervised trust prediction model based on the concept of trust transference, to transfer trust between different contexts~\cite{Zheng/ShortPaper/2014}. 
Wang et al.~\cite{Yan/WWWJ/2015} introduced an unsupervised trust prediction model to infer trust among users with an indirect connection. Liu et al.~\cite{Liuwwwj} proposed a trust inference model, incorporating factors such as residential location and outdegree.
Liu et al.~\cite{WangCCSLLT18} proposed a novel trust prediction model, CATrust, for auction websites, using Bayesian inference based on  Markov Chain Monte Carlo. More importantly, their model considered the contexts of trust.

Moradi and Ahmadian~\cite{MORADIAhmadian20157386} proposed a trust-aware recommender system, Reliability-based Trust-aware Collaborative Filtering, to address the problem of the accuracy of ratings predictions in recommender systems. This system dynamically extracts trust networks among users based on similarity values and trust statements.
Sanadhya and Singh~\cite{SanadhyaSingh} designed a trust prediction approach based on ant colony optimization (ACO), called Trust-ACO, to calculate trust
path and trust cycle and identify the most trustworthy path to find trustworthy services~\cite{SanadhyaSingh}. Their approach is based on probabilistic trust rule, social intimacy pheromone.
Fazeli et al.~\cite{Fazeli2014} proposed a trust prediction approach based on social trust, using $MF$. They first studied the effect of existing trust metrics in predicting pair-wise trust relations, employing those they deemed most effective in their prediction approach. 

Massa and Avesani~\cite{Massa2005} stated that `predicting a distrust statement is harder than predicting a trust statement'; however,
Tang et al.~\cite{Tang2014} have proposed an approach to predict distrust in OSNs. Specifically, their approach facilitates computational understanding of distrust. Zhang et al.~\cite{ZhangICWS2011} proposed a context-aware trust prediction approach focusing on `the ratings of past transactions, the nature of both past transactions and the new transaction'~\cite{ZhangICWS2011}. This approach used transaction context similarities to `identify and prevent potentially malicious transactions with
the value imbalance problem'~\cite{ZhangICWS2011}. Matsutani et al.~\cite{Matsutani2015}  assumed that the trust prediction problem could be solved in the same way as a link prediction problem. They proposed an approach based on non-negative $MF$ (NMF) methods. This approach `incorporates people's evaluation of users' activities as well as trust-links and users' activities themselves'~\cite{Matsutani2015}. 

Tang et al.~\cite{Tang2012} delved into the evolution of trust as a result of interpersonal interactions. They proposed a dynamic MF-based trust prediction approach, called eTrust, which focused on the dynamic preferences of users on product review websites~\cite{Tang2012}.
Huang et al.~\cite{Huang2012} believed that `people who are in the same social circle often exhibit similar behaviour and tastes'. They treated the trust prediction problem as a link prediction problem and proposed a joint manifold factorisation method that aggregated heterogeneous social networks to explore `the user group level similarity between correlated graphs and simultaneously [learn] the individual graph structure'~\cite{Huang2012}. Moturu et al.~\cite{Moturu2011} proposed an unsupervised approach for evaluating the trustworthiness of shared content, particularly shared health content. They proposed an approach based on feature identification, for determining the features most relevant to trust and quantification.
Yao et al.~\cite{Yao2013} proposed a trust inference approach based on $MF$. They addressed the trust prediction problem as a recommendation problem. Their model `characterizes multiple latent factors
for each trustor and trustee from the locally-generated trust
relationships'. To improve the accuracy of their approach, they also employed prior knowledge (e.g., trust bias and trust propagation).
Huang et al.~\cite{Huang2013} stated that, since trust matrices are of low-rank, they could consider the trust prediction problem as a recommendation problem. Specifically, they proposed a rank-k matrix completion approach that was robust to noise.

\subsection{Context-awareness of Trust}
Existing trust prediction approaches can be classified into two groups based on their consideration of the context of trust: approaches that consider context and those that do not. Before discussing the approaches that fall into these categories, we first discuss the notion of the context of trust as it relates to OSNs.

\subsubsection{Definition of Context}
Context, which influences the building of a trust relationship between the trustor and the trustee~\cite{Uddin/2008}, is multi-faceted~\cite{Zheng/ShortPaper/2014}.
In a society, the interactions between two participants can form a \emph{context} that can provide information such as the time or location of that interaction. Uddin et al.~\cite{Uddin/2008} provided a definition for context of trust in OSNs: `a context is a situation, which influences in the building of a trust relationship
between the trustor and the trustee'.

\subsubsection{Context-less Approaches}
The context-less approaches do not consider context to predict a trust relation in OSNs. The majority of existing trust prediction approaches can be considered context-less (see Tang et al.~\cite{TangGHL13}, Y. Wang et al.~\cite{WangWTZC15}, Golbeck~\cite{Golbeck06} and X. Wang et al.~\cite{IjCAI2018}). These approaches assume that if John trusts Jack, this means John trusts Jack in all fields of expertise  (e.g., electronics, sports, music, movies and science), for a lifetime and in any location. This assumption is too simplistic for real-word scenarios, because people only trust each other in certain contexts~\cite{trustcontext}~\cite{TangBook}~\cite{GhafariWise}.

\subsubsection{Context-Aware Approaches}
Liu et al.~\cite{Liu-Yan/2012} and Zhang and Wang~\cite{ZhangandWang} highlighted the importance of the context of trust as an essential factor for trust prediction approaches.
However, little effort has been made to consider the context of trust for a first class citizen.
One exception is Zheng et al.~\cite{Zheng/ShortPaper/2014}, who proposed a context-aware approach that considers both user's properties and the features of contexts. Social trust proposed as a novel probabilistic social context-aware trust inference approach, exploits textual information to deliver better results~\cite{Yan/WWWJ/2015}. In Zheng et al.'s approach, trust is inferred along the paths connecting two users. Thus, if two users are not connected by any path, no trust among them can be predicted.
Similarly, Liu et al.~\cite{Liuwwwj} developed a context-aware trust prediction approach based on the web-of-trust concept, which considered social context factors, such as users' location, previous interactions, social intimacy degree with other users, existing trust relations and so on. Zolfaghar and Aghaie~\cite{Zolfaghar2012} proposed a supervised context-aware trust prediction approach. They investigated the effects on trust relations of certain social trust factors, such as contextual similarity, users' reputation and relationship-based trust factors. 

Zhang et al.~\cite{Zhang2012} proposed a novel context-aware trust prediction approach based on contextual transaction factors, categorised into those relating to service and those relating to transaction~\cite{Zhang2012}. This approach considered the context of past transactions and forthcoming transactions to evaluate the reputation profile of the seller~\cite{Zhang2012}. In another study, Zhang et al.~\cite{ZhangContextual2012} aimed to develop a context-aware trust prediction approach.  They designed a data structure to support the Contextual Transaction Trust (CTT) computation in e-commerce environments~\cite{ZhangContextual2012}. They also proposed `an approach for promptly responding to a buyer's CTT query'~\cite{ZhangContextual2012}. 
Liu et al. in \cite{LiuStrong2016}, \cite{Liu2013} and \cite{Liu2012-2} noted that `predicting the trust between two unknown participants based on the whole large-scale social network can lead to very high computation costs'~\cite{LiuStrong2016}. Hence, they proposed an approach to extract a sub-network of the trust network that contained the most important nodes and trust relations. Since this sub-network extraction problem is an NP-complete problem, they proposed a strong social component-aware trust
sub-network extraction model, So-BiNet, to address this~\cite{LiuStrong2016}. Zheng et al.~\cite{Zheng2015} proposed another solution to `extract a small-scale contextual network that contains most of the important participants as well as trust and contextual information'~\cite{Zheng2015}. They developed a context-aware trust sub-network extraction model. They also used ant colony algorithm sub-network extraction.


Liu and Datta~\cite{TimeAware1} introduced a new context-aware trust prediction approach based on the Hidden Markov Model (HMM). This approach can dynamically model a user's interactions in OSNs. Rettinger et al.~\cite{Rettinger2011} proposed a context-aware trust prediction approach, called the Infinite Hidden Relational Trust Model. They expressed that `from the truster's point of
view trust is best expressed as one of several relations that exist between the agent to be
trusted (trustee) and the state of the environment'.
Xiong and Liu~\cite{Xiong2003} developed a novel context-aware trust prediction model, PeerTrust, for e-commerce platforms, based on a transaction-based feedback
system. They also introduced the factors of transaction context and community context for capturing the contexts of trust relations.
Rehak et al.~\cite{Rehak2006} designed a situational (context-dependent) trust prediction approach. They proposed a mechanism that `describes the similarity
among the situations using their distance in a metric
space and defines a set of reference contexts in this space
to which it associates the trustfulness data'.

Uddin et al.~\cite{Uddin/2008} proposed an interaction-based context-aware trust prediction approach, called CAT. They also suggested the concept of context similarity, which can be used for decision making in similar situations~\cite{Uddin/2008}. Kim et al.~\cite{KimLeLauw2008} believed that existing trust prediction approaches mostly relied on the web-of-trust concept, which may fail to accurately predict trust relations among users because of the data sparsity problem. They developed a context-aware trust prediction approach focusing on users' expertise and affinity in a particular context (topic).
Li and Wang~\cite{LiandWang} developed a fuzzy comprehensive evaluation based method to evaluate the trustworthiness of a service provider in an upcoming transaction based on the trust ratings in its transaction history. This approach is grounded in context-based trust normalisation, which focuses on `the familiarity between each rater and the service client of the upcoming transaction'~\cite{LiandWang}.

\subsection{Time-Dependency of Trust}
Although time can be considered one of the elements of context, because of its importance we investigate it more deeply. The literature on time-aware trust prediction in OSNs can be divided into two categories based on the approach taken: static approaches and dynamic approaches.

\subsubsection{Static Trust Prediction Approaches}
Static trust prediction approaches assume that trust relations among users do not change over time. However, in real-world scenarios, trust relations among people may be terminated at any time for various reasons (e.g., changes in interests, expectations or opinions).
The majority of existing trust prediction approaches belong to this category
(see Liu et al.~\cite{LiuLLLSSK08}, Ma et al.~\cite{MaLNSL09}, 
Matsuo et al.~\cite{MatsuoY09}, Tang et al.~\cite{TangGHL13}, Wang et al.~\cite{WangWTZC15} 
Ghafari et al.~\cite{GhafariWise}~\cite{GhafariQUAT} and Wang et al.~\cite{IjCAI2018}).

\subsubsection{Dynamic Trust Prediction Approaches}

Dynamic trust-prediction approaches can be classified into three main categories: Beta models, HMM-based models and others.

In the Beta models, Beta probability density
functions consider reputation and feedback simultaneously (see Ismail~\cite{BetaModel1}). In another work ~\cite{BetaModel3}, a decay factor was used to give more weight to recent events based on Recency bias (i.e., a person will remember the most recent events more easily compared to older events). Zhang et al.~\cite{BetaModel2} introduced an approach that monitors the dynamic behaviour of an agent based on the concept of time windows. In each time window, the number of successful and unsuccessful transactions is considered.

HMM-based models use HMM to propose dynamic trust prediction models. These approaches are of two main types. The first type focuses on the outcomes of past transactions and observations of HMM~\cite{PayanName}~\cite{HMM2}~\cite{HMM1}. Although these may have better performance compared to the Beta models, they fail to consider contextual information about each transaction~\cite{PayanName}. In the second type, researchers seek to consider contextual information about the transactions (see Liu and Datta~\cite{TimeAware1}). Zheng et al.~\cite{Zheng2013} developed a dynamic trust prediction approach based on HMM, which focused on the
hidden characteristics of the HMM model as well as the outcomes. They used a service provider's historical transactions to predict its trust level. They considered `static features, such as the
provider's reputation and item price and the dynamic features,
such as the latest profile changes of a service provider and price
changes'~\cite{Zheng2013}. Malik et al.~\cite{MalikAkbar} presented a means of assessing reputation in a service oriented approach for service oriented environments based on HMM. This approach can predict trust-based interactions among Web services.

Falling under the third category of dynamic trust prediction approaches, Cai et al.~\cite{Cai2014} proposed a MF-based trust prediction model. They incorporated temporal dynamics to model the dynamics of users' preferences. Laifa et al.~\cite{Laifa2014} tested a research model using structural equation modeling and delivered the outputs to an artificial neural network and fuzzy logic model developing their dynamic prediction approach. Liu and Datta~\cite{Liu2011ATP} designed another dynamic trust prediction approach. They believed that modelling the behaviour of people is challenging as people may change their behaviour strategically to increase their profits~\cite{Liu2011ATP}. By measuring similarity among the
contexts of transactions, they estimated the trustworthiness of a transaction based on previous cases of similar
transactions.
Although these approaches give outstanding performance in some situations, they may fail when a user's `behaviour is highly dynamic or is changing strategically'~\cite{PayanName}.

To gain a better understanding of the existing trust prediction approaches, in Table~\ref{long} we classify the existing approaches according to whether they are supervised ($S$) or unsupervised ($U$), where $S$ denotes supervised and $U$ represents unsupervised; whether the context of trust is considered, where $Y$ represents that the property is satisfied and $N$ denotes that the method cannot satisfy the property; and whether the dynamic, time-dependent nature of trust is considered, where $Y$ likewise denotes that the property is satisfied, while $N$ means it is not.  Based on this analysis, we find that around 54\% of existing trust-prediction approaches do not consider the context of trust. This means, they assume all trust relations are the same and that if a user trusts another user in one context, he or she will trust that user across all contexts. Surprisingly, only 27\% of existing trust-prediction approaches are time-aware. Accordingly, the majority of existing approaches assume that trust relations last a lifetime.

\begin{longtable}[c]{| c |c |c |c |}
 \caption{Classification of existing trust prediction approaches classification}
 \label{long}\\
 
 \hline
 Approach & \underline{S}upervised/\underline{U}nsupervised & Context-Aware & Dynamic  \\
 \hline
 \endfirsthead
 \hline
 \multicolumn{4}{|c|}{Continuation of Table~\ref{long}}\\
 \hline
 Approach & \underline{S}upervised/\underline{U}nsupervised & Context-Aware & Dynamic  \\
 \hline
 \endhead
 \hline
 \endfoot
 \endlastfoot
Moradi and Ahmadian~\cite{MORADIAhmadian20157386}  & U & N & Y  \\
 \hline
Zolfaghar and Aghaie~\cite{ZolfagharAghaie2011} & S & Y & Y\\
 \hline
Sanadhy and Singh~\cite{SanadhyaSingh} & U  & N  & Y  \\
 \hline
Raj and Babu~\cite{RAJBabu} &  U &  N &  N  \\
 \hline
Zhao et al.~\cite{Zhao2016ATT} &  S &  Y &  N  \\
 \hline
Zhang et al.~\cite{ZhangICWS2011} & U  &  Y & Y  \\
 \hline
Bachi et al.~\cite{Bachi2012} &  S &  N &  N  \\
 \hline
Zhang et al.~\cite{Zhang2012} & S  & Y  & N   \\
 \hline
Zhang et al.~\cite{Zhang2014} & S & N  & N   \\
 \hline
Zhang et al.~\cite{ZhangContextual2012} & S  & Y  &  N  \\
 \hline
Liu et al.~\cite{LiuStrong2016} & U  & Y  &  N \\
 \hline
Zheng et al.~\cite{Zheng2015} & U  & Y  & N   \\
 \hline
Matsutani et al.~\cite{Matsutani2015} &  U & N  & N   \\
 \hline
Tang et al.~\cite{Tang2012} & U  & N  &  Y  \\
 \hline
Zhang and Yu~\cite{ZhangandYu2012} & U  & N  & N   \\
 \hline
Chakraverty et al.~\cite{Chakraverty2016} & S  & N  & N  \\
 \hline
Sacco and Breslin~\cite{SaccoBreslin} & S  & N  & N   \\ 
\hline
Huang et al.~\cite{Huang2012} &  U & N  & N   \\
 \hline
Li and Wang~\cite{LiandWang} & U  & Y  & N   \\
 \hline
Fazeli et al.~\cite{Fazeli2014} &  U &  N &  N  \\
 \hline
Tang et al.~\cite{Tang2014} &   U &  N &  N   \\
 \hline
Moturu and Liu~\cite{Moturu2011} &   U &  N &  N  \\
 \hline
Nunez-Gonzalez et al.~\cite{Nunez-Gonzalez2015} & S  & N  & N   \\
 \hline
Yao et al.~\cite{Yao2013} & U &  N &  N  \\
 \hline
Huang et al.~\cite{Huang2013} &  U &  N &  N  \\
 \hline
Liu et al.~\cite{LiuLLLSSK08} & S  &  N   &  Y  \\
 \hline
Ma et al.L~\cite{MaLNSL09} & S &  N  &  N   \\
 \hline
Matsuo and Yamamoto~\cite{MatsuoY09} & S & N   &  Y   \\
 \hline
Grana et al.~\cite{supervised1} & S &  N   &N     \\
 \hline
Wang et al.~\cite{supervised2} & S &  N   &    N \\
 \hline
Bachi et al.~\cite{Bachi2012} & S &     Y&  N  \\
 \hline
Korovaiko and Thomo~\cite{supervised4} &S &  N   &    N  \\
 \hline
Borzymek and Sydow~\cite{supervised5} & S & N    &    N  \\
 \hline
Laspez and Maag~\cite{supervised6} & S & Y    &N    \\
 \hline
Zolfaghar and Aghaie~\cite{ZolfagharAghaie2011} & S &  Y   &    Y \\
 \hline
Tang et al.~\cite{TangGHL13} & U  &    N &N   \\
 \hline
Wang et al.~\cite{WangWTZC15} & U  &    N &N     \\
 \hline
Guha et al.~\cite{Guha2004} & U &  N   & N   \\
 \hline
Golbeck~\cite{Golbeck06} &  U &  N   & N     \\
 \hline
Wang et al.~\cite{IjCAI2018} &  U &  N   & N     \\
 \hline
Zheng et al.~\cite{Zheng/ShortPaper/2014} &  U &  Y   & N     \\
 \hline
Wang et al.~\cite{Yan/WWWJ/2015} &  U &  Y   & N    \\
 \hline
Liu et al.~\cite{Liuwwwj} & U &  Y   & N    \\ 
\hline
Wang et al.~\cite{WangCCSLLT18} &U &  Y   & Y   \\
 \hline
Liu et al.~\cite{Liu-Yan/2012} & U &  Y   &  N  \\
 \hline
Zhang and Wang~\cite{ZhangandWang} & U & Y   &  N    \\
 \hline
Zolfaghar and Aghaie ~\cite{Zolfaghar2012} & S  &  Y   & N  \\
 \hline
Liu and Datta.~\cite{TimeAware1} & S &  Y   &    Y  \\
 \hline
Rettinger et al.~\cite{Rettinger2011} & S & Y &  N  \\
 \hline
 Xiong and Liu~\cite{Xiong2003} & U  &  Y   &  N   \\
 \hline
 Rehak et al.~\cite{Rehak2006} & S &    Y & Y  \\
 \hline
 Uddin et al.~\cite{Uddin/2008} &  U &  Y   & Y    \\
 \hline
 Kim et al.~\cite{KimLeLauw2008} &  U&    Y & N   \\
 \hline
 Ismail and Josang~\cite{BetaModel1} & U   &     N&  N  \\
 \hline
  Teacy et al.~\cite{BetaModel2} & U &  N   &  Y  \\
 \hline
 Moe et al.~\cite{HMM1} & S  &  N   &Y     \\
 \hline
 Elsalamouny et al.~\cite{HMM2} &  S  &  N  &Y    \\
 \hline
 Zheng et al.~\cite{Zheng2013} & S & Y   &  Y  \\
 \hline
 Malik et al. ~\cite{MalikAkbar} & U  &  N   &Y    \\
 \hline
 Liu and Datta~\cite{Liu2011ATP} & U  &  Y   &Y  \\
 \hline
 Laifa et al.~\cite{Laifa2014} & S &   Y  & N  \\
 \hline
 Golbeck~\cite{GOLBECK/2003} &  U&  N   & N    \\
 \hline
  Trifunovic et al.~\cite{TRIFUNOVIC/2010} & U  &  N   &  N    \\
 \hline
  Zhang et al.~\cite{ZHANG/network-based/2006} &  U  &  N   &  N  \\
 \hline
  Kim et al.~\cite{KimLeLauw2008} &  U  & Y   &  N    \\
 \hline
  Ziegler and Lausen~\cite{ZIEGLER2004} & S &      Y& N  \\
 \hline
  Hang and Singh ~\cite{HANGANDSINGH} &  U &      N& N     \\
 \hline
  Zuo et al.~\cite{ZUOandHU} &U &      N& N    \\
 \hline
  Caverlee and Liu ~\cite{CaverleeandLiu} & U  &     Y &    Y \\
 \hline
 Liu et al.~\cite{Liu2012} & U  &     Y &    Y \\
 \hline
 \end{longtable}


\section{Personality and Trust}
Alarcon et al.~\cite{ALARCON201869}, in investigating the relation between personality and trust, focused on the relations between propensity to trust, the five-factor model~\cite{BigFive}, trust beliefs and behaviours. 
Thielmann et al~\cite{Thielmann} researched the impact of HEXACON, another trait-based personality mechanism, on trustworthiness by designing three trust games. Their work demonstrated the relation between honesty/humility and trustworthiness, independent of the prior level of trust. 
Another study by Evans and Revelle~\cite{EVANS20081585} considered the trust inventory and personality traits and validated this inventory through an economic task. They discovered that trust can be related to the Extraversion personality trait. 
Sicora~\cite{Sicora} focused on trust among co-workers and workplace leaders and its relationship with two personality models. Their aim was to create greater trusting relationships in organisations~\cite{Sicora}.
Gerris et al.~\cite{Gerris} studied the influence of the Big Five personality traits of couples on their marriages.  
Solomon et al.~\cite{PersonalityLinkPrediction} studied Twitter users based on the Big Five personality model~\cite{BigFive} and the Schwartz sociological behaviour model~\cite{Schwartz} to understand the psycho-sociological homophilic nature of personal networks. 
We proposed a pattern-based word embedding technique,
personality2vec~\cite{WSDMPersonality2vec} as a novel data analytics pipeline that enables analysis of users' personality patterns and behavioural disorders, based on their activities in OSNs. We also proposed to use domain knowledge to
design cognitive services to automatically contextualise raw
social data and prepare them for behavioural analytics.

\section{Summary}

In this chapter, we presented an overview of the work related to predicting pair-wise trust relations, categorising them based on different factors. First, we considered the approaches that only focused on the trust network structure (i.e., graph based approaches), which were mostly based on the concept of web-of-trust or FOAF. Since trust networks in OSNs suffer from the data sparsity problem, these approaches are limited in their effectiveness for handling real-world scenarios. We also reviewed those approaches that mostly focus on the previous interactions of users. However, as most users in OSNs either do not know each other or do not have any interactions with each other, these approaches are also ineffective for use in OSNs. 

Next, we analysed the existing trust prediction approaches in terms of whether they considered the context-dependency of trust. We stated that many existing trust prediction approaches do not consider the context of trust, before then reporting on those approaches that are context-aware. Finally, we considered the time-dependent nature of trust and discussed the static and dynamic trust prediction approaches. Finally, we analysed the existing trust prediction approaches in terms of three properties (context-awareness, time-awareness and whether they were supervised or unsupervised), as shown in Table~\ref{long}. In this table, Y represents that the property is satisfied and N represents the method cannot satisfy the property (in context-aware and dynamic columns). 
In the following chapters, we propose a series of trust prediction models, including an approach capable of predicting trust relations in the presence of the data sparsity problem, context-aware trust prediction approaches, and a dynamic trust prediction model that accounts for the time-dependency of trust relations.


\chapter{Experimental Setup}
\label{chap:DataandMetrics}

This chapter discuses the experiment setup, including the datasets used in our experiments, the evaluation metrics and the baseline methods, for the sets of approaches proposed in this dissertation. 

\section{Datasets}

\subsection{Epinions and Ciao Datasets}
To evaluate the trust prediction approaches that we propose in Chapters~\ref{chap:chapter4} and~\ref{chap:context}, we use two benchmark datasets from real-world websites: Epinions and Ciao~\cite{Tang2012}~\cite{trustcontext}. These are review websites, frequently used by trust prediction studies~\cite{WangWTZC15}~\cite{TangGHL13}, which contain users' ratings, reviews and trust relations. 
In these datasets, each pair of users has a Boolean label associated with its trust relation (which acts as the ground truth for our experiments). The trust labels were generated by explicitly asking users to give a `0' or `1' value as the trust value to other users. These datasets also contain the attributes of the users and their reviews, and use a 5-star (one-to-five) rating system to rate reviews. The characteristics of these datasets are presented in Table~\ref{tab:datasetEpinionCiao}. The reviews in these datasets are categorised by topic, such as travel, books, food and drink, house and garden and family.

\begin{table}[t!]
\begin{center}
 \caption{Characteristics of the Epinions and Ciao datasets.}
 \begin{tabular}{||c | c | c ||} 
 \hline
Dataset  & Epinions & Ciao\\
\hline
\hline
Number of users  &  1050 & 1000  \\
\hline
Trust network density  & 0.0093 & 0.0087 \\
\hline
Number of trust relations  & 10264 & 8726 \\
\hline
Minimum number of reviews per users & 3 & 3 \\
\hline
\end{tabular}
   \label{tab:datasetEpinionCiao}
\end{center}
\end{table}

Consider an example from the Epinions dataset: `Jo.com' is a user in this dataset who has provided some reviews. One of their reviews reads,
\begin{center}
\textit{`San Simeon, California's Oceanfront Resort can be Cavalier about being the Best in the West'}.
\end{center}

This review is in the category of `Hotels and Travel' and received an overall rating from other users of four out of five. Along with this information, the dataset provides additional information about each user, such as their number of followers and location. 
The trust network density given in Table~\ref{tab:datasetEpinionCiao} refers to the proportion of known trust relations compared to the potential trust relations among users. For example, the Epinions dataset has the potential for around one million trust relations among all its users (the number of users $\times$ the number of users). However, it has only 10264 known trust relations, making the trust network density 0.0093. Thus, in this dataset, the users' specified trust relations network is very sparse. This is also the case in the Ciao dataset, where the trust network density is 0.0087.

Unlike in the Ciao dataset, the Epinions dataset does not include information about which users rated the reviews of which user. This means we cannot use this dataset for evaluating the previous interactions among users (a context factor) in one of our proposed trust prediction approaches in Chapter~\ref{chap:context} ($TDTrust$). Hence, we ignore this context factor when we test $TDTrust$ on the Epinions dataset.


\subsection{Twitter Dataset}
The Ciao dataset does not provide any timestamp for individual trust relations. The Epinions dataset only provides one timestamp for the date of establishing the trust relation. Hence, we cannot evaluate our time-aware trust prediction approach with these datasets. Instead, we collected data from Twitter to create a time-aware dataset.      
Retweet behaviour in the Twitter platform~\footnote{https://twitter.com} can be a proxy for trust~\cite{retweet5}~\cite{retweet4}~\cite{retweet3}~\cite{retweet2}~\cite{ADALI}: if user $A$ retweets user $B$'s tweet, it may indicate that $A$ trusts $B$. We assume that the person who wrote a tweet is the target user and the person who retweeted it is the source user in a pair-wise trust relation (the source user trusts the target user). To analyse trust relations in Twitter, with the following steps, we obtained a set of tweets from Twitter.

We collected tweets using Twitter search API~\footnote{https://developer.twitter.com/en/docs/tweets/search/api-reference/get-search-tweets} during August and September 2019, using the queries `RT' (representing retweets) and `@9NewsAUS' (@9NewsAUS is the Twitter account of a popular Australian news service). We collected 2,096,000 tweets in this step. 
We then removed all tweets where either the source or the target users were other news services, including `9News Australia'. This left 3,620 retweets for which the source and target users were 2,334 ordinary users. We then crawled the timelines of these users (more than 4,000,000 tweets), looking for other retweet behaviours of our 2,334 with other ordinary users (not news services). Our Twitter dataset and statistics are reported in Table~\ref{tab:capter3dataset}. This dataset contains 66,572 retweets, with each user retweeting around 19 tweets on average. The average number of tweets per day is about 71 tweets.

\begin{table}[t!]
\begin{center}
 \caption{Characteristics of the Twitter dataset}
 \begin{tabular}{||c | c ||} 
 \hline
Number of individual users  &\quad \quad \quad \quad 2334 \\
\hline
Number of retweets  &\quad \quad \quad \quad 66,572  \\
\hline
Maximum number of retweets between a pair of users &\quad \quad \quad \quad 1,438 \\
\hline
Minimum number of retweets between a pair of users  &\quad \quad \quad \quad 1  \\
\hline
Average number of retweets per user pair  &\quad \quad \quad \quad 19.179  \\
\hline
Average number of retweets per day  &\quad \quad \quad \quad 71.5059  \\
\hline
Maximum number of retweets on a day  &\quad \quad \quad \quad 7,831  \\
\hline
Minimum number of retweets on a day  &\quad \quad \quad \quad 1  \\
\hline
Average length (in words) of a retweet  &\quad \quad \quad \quad 15.642  \\
\hline
Average length (in words) of users' Twitter bios  &\quad \quad \quad \quad 14.172  \\
\hline
Number of unique pair of users  &\quad \quad \quad \quad 3,475  \\ 
 \hline
\end{tabular}
   \label{tab:capter3dataset}
\end{center}
\end{table}


\section{Evaluation Metric}
In this section, we introduce the evaluation metrics that we employed for our experiments.

\subsection{Ranking-Based Evaluation}
One of the most well-known trust prediction evaluation metrics is ranking-based evaluation ~\cite{TangBook}~\cite{TangGHL13}~\cite{WangWTZC15}.
For this evaluation metric, we divide each of our datasets into two parts. The first part includes users who do not have any trust relations
(N). The second part includes users who have trust relations with other users (T). We sort these trust relations based on their time of establishment. At that point, we
select the first $A\%$ trust relations as old trust relations and denote $1-A\%$ of them as the $New$ trust relations to predict. We consider four percentage values for A=\{60,70,80,90\}. Further, we employ a trust prediction metric
from Liben-Nowell and Kleinberg ~\cite{Liben-NowellandKleinberg} to evaluate the performance of our approaches. Based on this, we first merge all $New$ (new trust relations) and $N$ (non-trust relations) such that $N \cup New$ and call them $M$. Then, we predict the trust relations in $M$ and extract the $|New|$ number of trust relations and call this $Predict$. Based on these sets, the performance of any trust prediction approach is determined by the following formula:
\begin{equation}
TPA= \dfrac{|New \cap Predict|}{|New|}
\end{equation}

where $TPA$ is the trust prediction quality. The value of $TPA$ is usually small and `to demonstrate the significance of performance, [a] randomly guessing predictor is usually used as a baseline method'~\cite{TangBook}. As we increase the size of $A$, the size of $New$ decreases. This makes it difficult to accurately predict trust relations in $M$; thus, the $TPA$ is expected to decrease. We employ this metric in Chapter~\ref{chap:context}.

\subsection{The Mean Absolute Error and Root Mean Squared Error Metrics}
Two widely used prediction accuracy metrics for trust prediction approaches are mean absolute error (MAE) and root mean squared error (RMSE)~\cite{TangBook}~\cite{DCAT}~\cite{GhafariWise}. Similar to the settings for the ranking based metric, we create $M$, $New$ and $N$. Then, the trust values for the pairs of users in $N$ are computed. MAE and RMSE can be defined as follows:
\begin{equation}
MAE= \frac{1}{|New|} \sum_{{i,j}\in New}| T_{AC_{ij}}-T_{{Pre}_{ij}} |,
\end{equation}
\begin{equation}
  RMSE= \sqrt{\dfrac{1}{|New|} \sum_{{i,j}\in New} (T_{AC_{ij}}-T_{{Pre}_{ij}})^2}
\end{equation}

where $ T_{AC_{ij}}$ is the actual trust relations between $u_i$ and $u_j$, and $T_{{Pre}_{ij}}$ is the predicted trust relations. A lower MAE and RMSE indicate a better performance. A small improvement in terms of RMSE or MAE has a significant effect on the quality of the top-few recommendations ~\cite{TangBook}.






\section{Baselines}
We compare our proposed trust prediction approaches with the state-of-the-art trust prediction methods and other approaches. In this section, we introduce these approaches.

\textbf{A. hTrust}~\cite{TangGHL13}: \textit{hTrust} is a state-of-the-art trust prediction approach. It exploits the effects of homophily in trust relations. Specifically, \textit{hTrust} investigates the effect of similarities among users (particularly their ratings similarities) on their trust relations. It proposes an MF-based model that incorporates the homophily effects into its mathematical model. The initial $MF$ model to seek a low-rank representation  of $U$ is defined as~\cite{TangGHL13}:
\begin{equation}
\ min_{U,H} \Vert G -   UHU^T \Vert ^2 _F +\ \alpha \times({\Vert U \Vert ^2 _F + \Vert H \Vert ^2 _F})
\  , U > 0 , H > 0  
\label{eq:chapter3-1}
\end{equation}

where $U$ represents users' interests, the $H$ matrix contains compact correlations among $U$~\cite{WangWTZC15}, $G$ is a trust matrix with the trust relations between users, and $\alpha$ is a controlling parameter. \textit{hTrust} then adds a homophily regularisation:
\begin{equation}
\ min_{U,H} \Vert G -   UHU^T \Vert ^2 _F +\ \alpha \times({\Vert U \Vert ^2 _F + \Vert H \Vert ^2 _F}) +\lambda Tr(U^T \zeta U) 
\  , U > 0 , H > 0
\end{equation}

where $Tr(U^T \zeta U)$ represents the homophily regularisation, and $\lambda$ is a controlling  parameter. $U^T$ is the transpose of matrix U, and $|.|_F$ denotes the Frobenius norm. To capture similarities among users and calculating the homophily effects in OSNs, Tang et al, investigated three 
similarity measures for $\zeta$ as follows~\cite{TangGHL13}: 
\begin{equation}
 \zeta (i,j)= JC(u_i, u_j)=  {\dfrac{|I(i)\cap I(j)|}{|I(i)\cup I(j)|}},
\end{equation}

where JC is the Jacard's Coefficient and represents the number of common rated items divided by the total number of unique rated items~\cite{TangGHL13}. $I(i)$ and $I(j)$ are the sets of items that $u_i$ and $u_j$ rated.  
\begin{equation}
 \zeta (i,j)= RS(u_i, u_j)=  {\dfrac{\sum_k R_{ik}\times R_{jk}}{\sqrt{\sum_k R_{ik}^2}\sqrt{\sum_k R_{jk}^2}}},
 \label{eq:chapter3-2}
\end{equation}

where $RS$ is the cosine similarity among rating vectors of users. $R_{ik}$ is the rating that $u_i$ gave to the $k^{th}$ item.
\begin{equation}
 \zeta (i,j)= PCC(u_i, u_j)=  {\dfrac{\sum_{k \in I(i) \cap I(j)} (R_{ik} - R'_{i}) \times (R_{jk} - R'_{j})}{\sqrt{\sum_{k} (R_{ik} - R'_{i})^2} \sqrt{\sum_{k}(R_{jk} - R'_{j})^2}}},
\end{equation}

where PCC is the Pearson correlation coefficient, $R'_{i}$ is the average rate of $u_i$ and $k$ is the subset of items rated by both $u_i$ and $u_j$.

\textbf{B. Rating Similarity}~\cite{TangGHL13}:
Although rating similarity is employed as part of the $hTrust$ model, it has also been used as a baseline in other previous trust prediction studies~\cite{TangGHL13}~\cite{ICDM}.

\textbf{C. sTrust}~\cite{WangWTZC15}: This approach investigates the relation between the statuses of users in OSNs and their trust relations. This approach suggests that a user with a higher status is more likely to be trusted by other users.
This approach identifies the users' status with the help of the PageRank algorithm~\cite{pagerank}. 
$sTrust$ employs an $MF$ model based on the Formula~\ref{eq:chapter3-1}. It also models status theory as follows~\cite{WangWTZC15}:
\begin{equation}
\sum_i^n \sum_{j!=i}^n (max\{0,f(r_i-r_j)(U_iHU_j^T -U_jHU_i^T)\})^2,
\end{equation}

where $U$ and $H$ are similar to their definitions in $hTrust$. $f(x)$ is a function with the same sign as x~\cite{WangWTZC15}; and $r_i$ and $r_j$ are the statuses of $u_i$ and $u_j$, respectively, computed based on the PageRank algorithm~\cite{pagerank}. Finally, $sTrust$ uses the following equation to model the status theory in trust prediction~\cite{WangWTZC15}: 
\begin{equation}
\begin{aligned}
\begin{split}
& \ min_{U,H} \Vert G -   UHU^T \Vert ^2 _F +\ \lambda \times\sum_i^n \sum_{j=i+1}^n (max\{0,f(r_i-r_j)(U_iHU_j^T -U_jHU_i^T)\})^2 + \\
& \alpha (||U||_F^2 + ||H||_F^2) , \\
& \ U > 0 , H > 0
\end{split}
\end{aligned}
\end{equation}

where $\lambda$ is a controlling parameter and G is the trust matrix.

\textbf{D. Zheng}~\cite{Zheng/ShortPaper/2014}:
This is a context-aware trust prediction approach that investigates the impact of considering context in the trust prediction process and proposes a method to find a low-rank representation of users as follows~\cite{Zheng/ShortPaper/2014}:
\begin{equation}
\ min_{U,H} \dfrac{1}{2} \sum_{i=1}^n \sum_{j=1}^n (I_{ij}+ \eta I'_{ij})(r_{ij}-u_i^T h_j)^2
+\ \dfrac{\lambda_1}{2}\Vert U \Vert ^2 _F +  \dfrac{\lambda_2}{2} \Vert H \Vert ^2 _F),
\end{equation}

where $\eta $, $\lambda_1$ and $\lambda_2$ are controlling parameters; $I_{ij}$ and $I'_{ij}$ indicate whether $u_i$ and $u_j$ trust each other in any context; $r_{ij}$ is the inner product of the trustor and trustee vectors. 

\textbf{E. MF}~\cite{Zhu2007}: This approach is based on a link $MF$ model as follows:
\begin{equation}
\ min_{U,V,Z} \{ \Vert A -   ZUZ^T \Vert ^2 _F + \alpha \times \Vert C-ZV^T \Vert ^2 _F+
\gamma \times({\Vert U \Vert ^2 _F + \Vert V \Vert ^2 _F}),
\label{eq:chapter3-1}
\end{equation}

where $A$ is a link matrix and $U$ is a low-rank matrix; $\gamma$ is a small positive number (controlling parameter); $\alpha$ is a controlling parameter; $Z$ is a matrix that contains the feature vector of a web page; $C$ is a matrix consisting of the content of the web pages; and $V$ is a matrix of latent space for words.

\textbf{F. Liu and Datta [2011]}~\cite{Liu2011ATP}: This is a state-of-the-art dynamic trust prediction approach. Liu and Datta [2011] aims to estimate the trustworthiness of a potential transaction related to an agent in a platform (e.g., e-market or peer-to-peer networks). The authors specifically focus on the previous transactions in similar contexts to evaluate the potential transaction. The following formula can calculate the similarity between two transactions:
\begin{equation}
\sigma_{\Theta_i,\Theta_i'}= \dfrac{1}{\sqrt{\sum_{j=1}^{|F_{\Theta}|} (f_{\Theta_i}^j -f_{\Theta_i'}^j)^2}}
\label{eq:chapter3-similaritytr}
\end{equation}

where $\sigma_{\Theta_i,\Theta_i'}$ denotes the similarity between two transaction windows at $i^{th}$ transaction; $\Theta_i$ and $\Theta_i'$ are two transaction windows; and $F_{\Theta}$ is the feature vector of the transaction. Next, based on the outcome of previous similar transactions, Liu and Datta estimate the trustworthiness of the potential transaction. Finally, they proposed multiple transaction window sizes and apply Dirichlet distribution~\cite{Dirichletdistribution}
to model multiple trust indicators~\cite{Liu2011ATP}. 

\textbf{G. Liu and Datta [2012]}~\cite{TimeAware1}: This is a state-of-the-art dynamic trust prediction approach based on HMM, capable of considering the contextual information of trust relations. Liu et al.~\cite{TimeAware1} employed information theory and multiple discriminant analysis to create their required feature vector. They assumed that an interaction has $n$ levels (e.g., good, medium and bad) and that the outcome for each level affects the trustworthiness of the target agent on an online auction site. In this approach, 
the probability distribution of the outcome of the next transaction with HMM $\lambda_{x,y}^l$ is calculated by:
\begin{equation}
P(s_m | F_{m+1}, \lambda_{x,y}^l)= \dfrac{P(s_m =L_j, F_{m+1},\lambda_{x,y}^l)}{P(F_{m+1},\lambda_{x,y}^l)}
\label{eq:chapter3-HMM}
\end{equation}

where the l-state HMM constructed by $a_{x}$ (trustor) for modelling the dynamic behaviour of $a_y$ (trustee) is denoted as $\lambda_{x,y}^l$. $F$ represents the features associated with the transaction, $m$ is the number of transactions, and $s_m$ denotes the next transaction. $L_j$ is the rating that $a_{x}$ gave to a transaction. Liu et al.~\cite{TimeAware1} proposed the following formula for obtaining the next most likely outcome:
\begin{equation}
s_m = arg \quad max_{L_j \in 	\zeta } [P(s_m = L_j | F_{m+1}, \lambda_{x,y}^l)]
\label{eq:chapter3-HMM}
\end{equation}

where $\zeta$ is a set of ratings that $a_{x}$ gave to other transactions.

\textbf{H. Random}~\cite{TangGHL13}:
We also employed a baseline that randomly assigns a trust value to a pair of users. This baseline gives the prediction performance if we randomly predict pair-wise trust relations.
This baseline has been used in ~\cite{ICDM} and ~\cite{TangGHL13}.

\section{Summary}
In this chapter, we discussed the datasets that we employed to evaluate our proposed trust prediction approaches. The Ciao and Epinions datasets are used in the experiments of proposed methods in Chapters~\ref{chap:chapter4} and ~\ref{chap:context}. Since these two datasets do not provide dynamic timestamps of trust relations, for evaluating our dynamic trust prediction approach in Chapter~\ref{chap:time}, we created our own dataset using Twitter search API. We employed MAE and RMSE metrics and a ranking-based evaluation metric (TPA) to evaluate our proposed trust prediction approaches. We also presented the evaluation metrics required for our experiments. Finally, we introduced the baselines with which we compare the performance of our proposed trust prediction approaches.


\chapter{Modelling Personality Traits in Trust Prediction}
\label{chap:chapter4}

The association between trust and
the personality traits of the people involved in establishing relations has
been proven by social science theories. In this chapter, we attempt to alleviate
the effect of the sparsity of trust relations by extracting implicit
information about users. We achieve this by focusing on users' personality
traits and seeking a low-rank representation of users. We
investigate the potential impact of incorporating users' personality traits based on the Big Five
personality model for the prediction of trust relations. We evaluate the impact of similarities of
users' personality traits and the effect of each personality trait on
pair-wise trust relations. Next, we formulate a new unsupervised
trust prediction model based on TD. Finally, we
empirically evaluate this model on the datasets presented in Chapter~\ref{chap:DataandMetrics}. Our model's
superior performance compared to the state-of-the-art
approaches highlights the value of personality traits for predicting trust. The proposed approach in this chapter has been published in the $13^{th}$ ACM International WSDM Conference 2020 (Core Rank A*), Pacific Asia Conference on Information Systems (PACIS) 2019 (Core Rank A) and Cyber Defence Next Generation Technology and science Conference (CDNG) 2020.

\section{Introduction}

As discussed in Chapter~\ref{chap:introduction}, trust networks in OSNs suffer from the sparsity of user-specified trust relations and trust relations in OSNs follow the power law distribution~\cite{TangGHL13}. In other words, many trust relations can be attributed to a small number of users, while a large number of users participate in only a few trust relations. 
As a result, for any solution for the trust prediction problem, to be capable of predicting unknown trust relations~\cite{TangGHL13}, it needs to deal with the problem of sparsity. 

\begin{figure*}
\centering
  \includegraphics[width=0.90 \textwidth]{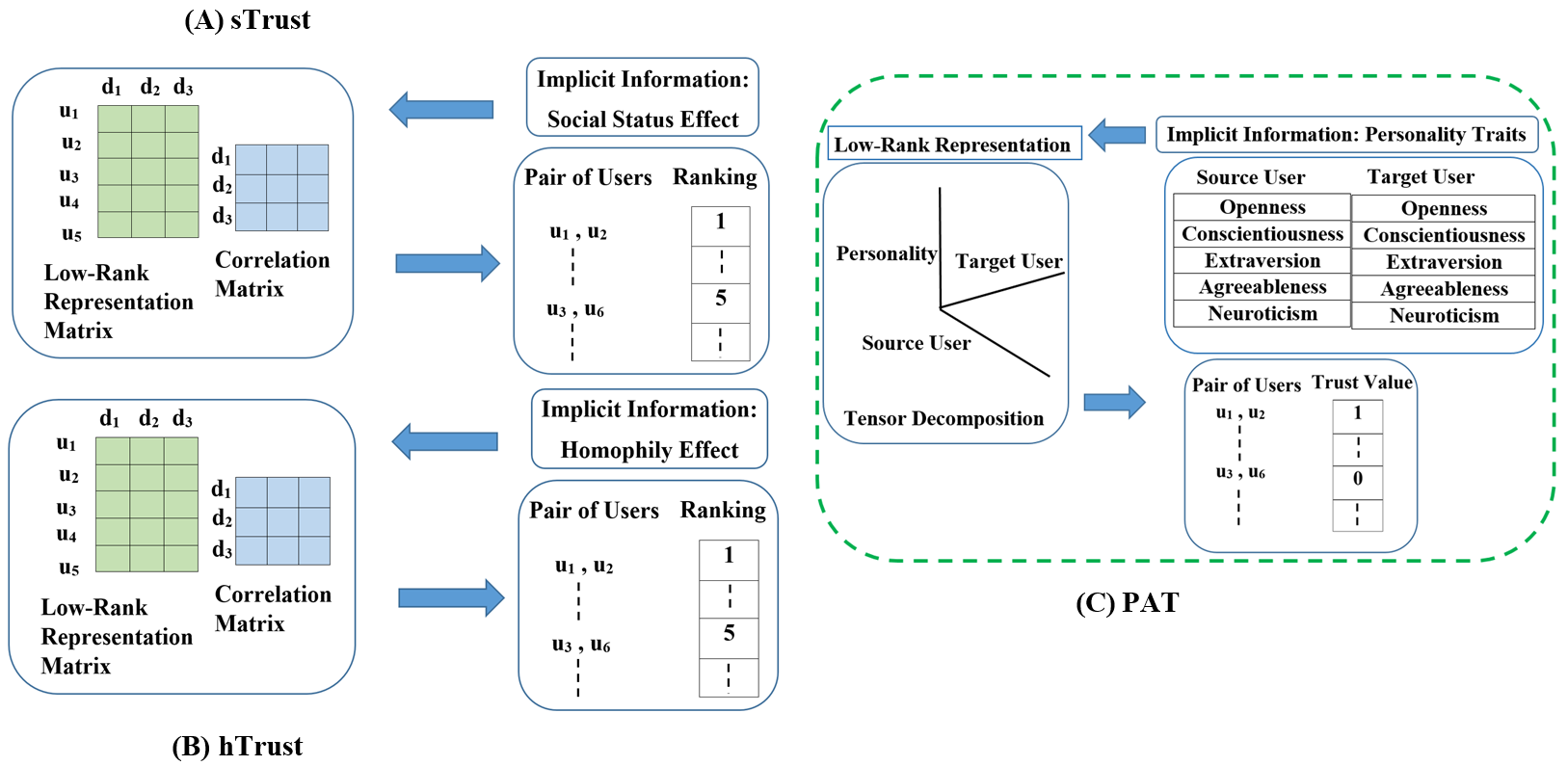}
  \caption{(A) $sTrust$~\cite{WangWTZC15} and (B) $hTrust$~\cite{TangGHL13}, incorporate implicit information, such as users' status (considering social status theory~\cite{WangWTZC15}) and users' ratings similarity (considering homophily theory~\cite{TangGHL13}), to improve their trust prediction performance. Our proposed method (C), Personality-Aware Trust prediction, is an unsupervised trust prediction approach that incorporates users' personality traits as implicit information, and uses TD to find a low-rank representation of trust relations.}
  \label{fig:archaa}
\end{figure*}

As discussed in Chapter~\ref{chap:chapterRelatedWorks}, we can classify the existing trust prediction approaches into two categories~\cite{TangGHL13}: supervised approaches~\cite{LiuLLLSSK08}~\cite{Nguyen2009}~\cite{TangandLiu} and unsupervised approaches~\cite{TangGHL13}~\cite{WangWTZC15}~\cite{Borzymek2009}~\cite{IjCAI2018}~\cite{Guha2004}. The shortcoming of supervised approaches is that they may face an imbalance classification problem~\cite{TangGHL13} due to the data sparsity problem. Likewise, the performance of unsupervised approaches may be limited due to the lack of sufficient trust relations~\cite{WangWTZC15}. To overcome the problem of the sparsity of trust relations, some of the unsupervised trust prediction approaches incorporate \textit{implicit or additional information} (e.g., users' ratings similarity~\cite{TangGHL13} or users' social status~\cite{WangWTZC15}) (Figure~\ref{fig:archaa}-A,B).

Many social studies have attempted to explore the reasons behind establishing trust relations among people. Although these studies consider trust a situational construct, some of them investigate individual characteristics in their trusting behaviour predictions~\cite{EVANS20081585}. Personality has been considered an important characteristic. For example, Alarcon et al.~\cite{ALARCON201869} stated that `personality can assist researchers in understanding the processes underlying trust interactions'. Studies in social science consider people's personality as part of developing trust relations in their face-to-face interactions. 
However, this important attribute remains unexplored for pair-wise trust relations prediction in OSNs.
Based on one of the well-known psychology theories, the Big Five personality model~\cite{BigFive}, people's personality can be characterised by five personality traits: Openness, Conscientiousness, Extraversion, Agreeableness and Neuroticism. 
Incorporating these personality traits (as \textit{implicit or additional information}) in a low-rank representation of users in our unsupervised trust prediction approach can help us to alleviate the data sparsity problem.  


In this chapter, we investigate: (i) the relation between users' personality traits and their trust relations; and (ii) how to make use of personality traits in our trust prediction approach. Our answers to these questions are used to generate a novel trust prediction model, which we call the Personality-Aware Trust prediction approach ($PAT$).
$PAT$ is an unsupervised model that seeks a low-rank tensor representation of users while investigating the effects of users' personality traits. 

Our main contributions are as follows:

\begin{itemize}
  \item Demonstrating the effects of users' personality traits on pair-wise trust relations: users who have similar personality traits are likely to establish trust relations.
    \item Demonstrating the effects of specific user personality traits on pair-wise trust relations:  the impact of the Extraversion and Conscientiousness personality traits is significant. 
  \item Proposing PAT as an unsupervised approach based on TD, for addressing the trust prediction problem by incorporating users' personality traits.
\end{itemize}

The rest of the chapter is organised as follows. Section~\ref{PATsec:section2} provides a data analysis on the two datasets described in Chapter~\ref{chap:DataandMetrics}. Section~\ref{sec:section3} discusses the proposed method. We then present the experimental results in Section~\ref{sec:section4}, before concluding the chapter in Section~\ref{sec:section5}.

\begin{table*}[t!]
\centering
\caption{The relation between the Big Five personality traits and different LIWC categories~\cite{Mairesse} (in no particular order),  based on the analysis from Pennebaker and King~\cite{LIWC}. The Linguistic Inquiry and Word Count (LIWC) categorises words into more than 88 categories. Here, the categories of LIWC related to the Big Five personality traits, `Open.', `Consc.', `Extra.', `Agree.', `Neuro.' represent Openness Conscientiousness, Extraversion, Agreeableness and Neuroticism, respectively.
  }
\begin{tabular}{lllll}
\toprule
\textbf{Open.} & \textbf{Consc.} & \textbf{Extra.} & \textbf{Agree.} & \textbf{Neuro.}\\  
\midrule
\small
Punctuation & \small Affect Words &\small Total pronouns     &\small Exclamation marks &\small Affect Words \\
\small Affect Words&\small Death        &\small Exclamation marks   &\small Dictionary words  &\small Anger \\
\small Apostrophes &\small Future       &\small Article             &\small Feel             &\small Anxiety \\
\small Achievement  &\small Home        &\small Friends             &\small Home             &\small Article \\
\small Anger      &\small Prepositions  &\small Periods             &\small Singular Pronoun &\small Feel \\
\small Home       &\small Anger         &\small Pronoun            &\small Anger            &\small Leisure \\
\small Article    & \small Body          &\small Question marks      &\small Negative emotion &\small Music \\
\small Positive Feeling &\small Hear    &\small Positive Emotion    &\small Positive emotion &\small Number \\
\small Assent     &\small Apostrophes    &\small Punctuation         &\small Body             &\small Apostrophes \\
\small Causation  &\small Certainty     &\small Apostrophes         &\small Family           &\small Exclamation marks \\
\small Death      &\small Hear          &\small Parentheses         &\small Motion           &\small Family \\
\small Family     &\small Job           &\small Body                &\small Negations        &\small Friends \\
\small Feel       &\small Music         &\small Certainty           &\small Parentheses      &\small Singular Pronoun \\
\small Friends    &\small Negations     &\small Family              &\small Pronoun          &\small Negations \\
\small Singular Pronoun&\small Negative emotion&\small Fillers      &\small Future           &\small Negative emotion \\
\small Job        &\small Prepositions  &\small Other punctuation   &\small Periods          &\small Total pronouns \\
\small Motion     &\small Question marks& \small Singular Pronoun    &\small Achievement      &\small Prepositions \\
\small Music      &\small Nonfluencies  &\small Music              &\small Anxiety          &\small Present focus  \\
\bottomrule
\end{tabular}
   \label{tab:categoriessss}
\end{table*}


\section{Background and Data Analysis}
\label{PATsec:section2}
We first aim to verify the correlation between trust relations and users' personality traits. In particular, we seek answers to the following research questions:
\\\\
\textbf{RQ1.} Do users who bear a trust relation between them have similar personality traits?

\noindent \textbf{RQ2.} What is the relationship between personality traits and trust relations?
\\\\

Assume a set of users $U=\{u_1, u_2,..., u_n\}$, where $n$ is the number of users. A trust relation between $u_i$ and $u_j$, as the $i^{th}$ and $j^{th}$ users, implies that $u_i$ (as the source user) trusts $u_j$ (as the target user).


\subsection{Acquiring Personality Traits}
\label{Acquiring}
Personality refers to `the characteristic set of behaviour, cognition and emotional patterns that evolve from biological and environmental factors'~\cite{Mairesse}. Users' personality traits can be either identified `explicitly by filling a questionnaire or implicitly through observing users' behavioural patterns'~\cite{Yakhchi}. Among several personality trait detection models, the Big Five model is one of the most studied in psychology. It characterises five personality traits~\cite{BigFive}:
\begin{itemize}
  \item Openness: This trait includes characteristics such as active imagination and aesthetic sensitivity.
\item Conscientiousness: This trait relates to purpose, strong will and determination.
 \item Extraversion: This trait includes characteristics such as sociability,
assertiveness, activity and talkativeness,
inner feelings, a preference for variety, intellectual curiosity
and independence of judgement.
\item Agreeableness: This trait is related to altruism, sympathy towards others and eagerness to help them and the belief that others are equally helpful.
\item Neuroticism: People with this trait tend to experience fear, sadness, embarrassment,
anger, guilt and disgust.
\end{itemize}



To identify the personality traits of users, we first gather all the users' reviews/tweets/posts. We then analyse the textual content of these reviews/tweets/posts using the Linguistic Inquiry and Word Count (LIWC)~\cite{LIWC}~\footnote{LIWC is a standard text analysis tool to identify personality traits from text}. This tool categorises words into more than 88 categories (such as `word count', `negative emotion', `anxiety' and `anger'). Next, inspired by Roshchina et al.~\cite{Twin} and our own work~\cite{Yakhchi}, we use a linear regression model to calculate the users' personality trait values as follows: 
\begin{equation}
\small
\begin{aligned}
Openness= w_1\times X_1+w_2\times X_2+w_3 \times X_3+...
\end{aligned}
\end{equation}

where $X_t$, $t=\{1, 2, 3, ..., b\}$ denotes the categories of LIWC and $b$ is the number of these categories. Out of more than 88 linguistic categories of LIWC, we only consider those related to the Big Five personality traits (e.g., Affect Words, Anger and Anxiety, according to Table~\ref{tab:categoriessss}). 
Moreover, $w_c$, $c=\{1, 2, 3, ..., d\}$, where $d$ is the number of categories of LIWC related to a particular personality trait, represents the $d^{th}$ weight. This is based on the extracted weights by Mairesse et al.~\cite{Mairesse}. Table~\ref{tab:categoriessss} shows the relationships between the LIWC categories and the personality traits. The procedure for calculating the values of the other four personality traits is the same. 
\begin{table*}[t!]
\centering
\caption{The similarities of users' personality traits in the Ciao and Epinions datasets, measured by the cosine similarities of the personality vectors for source and target users. `Open.', `Consc.', `Extra.', `Agree.' and `Neuro.' represent Openness Conscientiousness, Extraversion, Agreeableness and Neuroticism, respectively.
  }
\begin{tabular}{llllll}
\toprule
\textbf{Dataset}&\textbf{Open.} & \textbf{Consc.} & \textbf{Extra.} & \textbf{Agree.} & \textbf{Neuro.}\\  
\midrule
Ciao & \quad \textbf{0.9} &  \quad \textbf{0.99}  & \quad  \textbf{0.99}  & \quad  0.25 & \quad  0.24 \\
Epinions &  \quad  \textbf{0.91} & \quad  \textbf{0.98}  & \quad  \textbf{0.97} & \quad  0.26 & \quad  0.22 \\
\bottomrule
\end{tabular}
   \label{tab:categories14}
\end{table*}

\begin{table*}
\centering
\caption{The average personality traits values of source
  users and target users in the existing pair-wise trust relations in the Cioa and Epinions datasets. `Open.', `Consc.', `Extra.', `Agree.' and `Neuro.' represent Openness Conscientiousness, Extraversion, Agreeableness and Neuroticism, respectively.
  }
\begin{tabular}{lllllll} 
\toprule
\textbf{Personality of}&\textbf{Dataset}&\textbf{Open.} & \textbf{Consc.} & \textbf{Extra.} & \textbf{Agree.} & \textbf{Neuro.}\\  
\midrule
Source Users & Ciao & \quad 0.32&\quad     \textbf{0.69}   &\quad  \textbf{0.76}   &    \quad0.57  & \quad  0.21  \\
 \quad & Epinions& \quad 0.35  & \quad    \textbf{0.73}   &\quad   \textbf{0.79}  &    \quad0.49  & \quad  0.38   \\
 \hline
 Target Users & Ciao &\quad 0.39  & \quad    \textbf{0.68}   &\quad   \textbf{0.75}  &    \quad0.54 & \quad 0.34  \\
 \quad & Epinions&  \quad  0.32 & \quad    \textbf{0.72}  &\quad   \textbf{0.73}  &  \quad 0.59  & \quad 0.36  \\
\bottomrule
\end{tabular}
   \label{tab:categories2}
\end{table*}





\subsection{Similar Personality Traits and Trust Relations} 

In this section, we address RQ1: Do users who bear a trust relation between them have similar personality traits? Homophily theory~\cite{HomophilyTheory}, a well-known social psychology theory, can explain the reasons for establishing trust relations. Based on this theory, two similar users are more likely to trust each other~\cite{TangGHL13}. Tang et al.~\cite{TangGHL13} have previously investigated the relation between this theory and trust in OSNs, based on the similarities in users' ratings on product review websites. Here, we aim to investigate the relationship between trust relations and the similarity of users' personality traits. We first collect all the trust relations from the Ciao and Epinions datasets, presented in Chapter~\ref{chap:DataandMetrics}. Next, we identify the personality traits of the users who were involved in a trust relation based on the methods discussed in Section~\ref{Acquiring}. At that point, we use the cosine similarity metric to calculate the similarities between the personality traits of the source and target users, where the outcome is bounded in $[0, 1]$. The results are reported in Table~\ref{tab:categories14} and demonstrate that people who established a pair-wise trust relation (the users in these datasets explicitly mentioned the trust value they gave to other users, and we have access to these values as the ground truth of our experiments) have a cosine similarity of more than 0.9 for their Openness, Conscientiousness and Extraversion personality traits. We can conclude that the users involved in the trust relations in these two datasets are similar in  three out of five of their personality traits. We further investigate the impact of personality trait similarities on our trust prediction model.



\subsection{Impact of Personality Traits on Trust Relations} 

To answer RQ2---What is the relationship between personality traits and trust relations?---we calculate the average value of each of the personality trait of the source and target users. The result, reported in Table~\ref{tab:categories2}, demonstrates that the users who established a pair-wise trust relation in these datasets have high Extraversion and Conscientiousness personality trait values. This result can be an indicator of the relationship between these personality traits and trust relations. We further investigate this relationship in our trust prediction model.

\textbf{In summary,}
the dataset analysis demonstrates that users in pair-wise
trust relations may have similar personality traits and some personality
traits may have a higher impact on trust relations. Next, we present a method that incorporates these personality traits into trust prediction.


\section{The Personality-Aware Trust Prediction Approach}
\label{sec:section3}
In this section, we propose a TD-based pair-wise unsupervised trust prediction model, $PAT$, that incorporates users' personality traits. First, we discuss the problem statement. Next, we formulate the personality traits of users. Finally, we exploit them in our trust prediction model. 

 
\subsection{Problem Statement}
With $n$ users $U=\{u_1, u_2,...,u_n\}$ and $H_e$ as their personality traits $p=\{1, 2,..., 5\}$, 
G is a three-way trust tensor that represents trust relations among users together with their personality traits, where $G \in R^{n\times n \times p}$. Considering the personality traits of $u_i$ and $u_j$, if $u_i$ trusts $u_j$, this can be shown as $G(i,j,p)=1$. Conversely, $G(i,j,p)=0$ indicates the lack of a trust relation between them. Since G is very sparse~\cite{GhafariWise}, we are looking for a low-rank representation. Hence, we model our trust prediction approach on TD, following the approach proposed by Wang et al.~\cite{WangZLGSH13} and the CPD/Parafac model~\cite{SidiropoulosLFH17}, to learn three f-dimensional matrices: $U \in R^{n\times f}$, $U' \in R^{n\times f}$ and $H \in R^{p\times f}$, where $U$ and $U'$ indicate the source and target users, respectively. 
Finally, the sum of the inner product of these matrices creates the trust prediction tensor:
\begin{equation}
 \tilde{G}= \sum_{r=1}^{f} U_r U'_r H_{p_{r}} = \textless U, U', H_{p} \textgreater
 \label{For:PATFormula2}
\end{equation}

\subsection{Personality-Awareness}
This section focuses on how we incorporate users' personality traits into our trust prediction model. Based on the technique discussed in Section 2.1, we identify user's personality traits. Each user has five personality trait values, $V_{ip}$, with $i$ indicating the $i^{th}$ user and $p$ indicates $p^{th}$ personality trait, where $p=\{1,...,5\}$. 
We consider the order of the personality traits as Openness, Conscientiousness, Extraversion, Agreeableness and Neuroticism. Hence, $V_{i2}$ refers to the Conscientiousness personality trait value of the $i^{th}$ user. Thus,
$V_{ijp}=\sum_{p=1}^{z} V_{ip}+V_{jp}$, z=\{1, 2, ..., 5\}
captures the effects of the five personality traits of the $i^{th}$ and $j^{th}$ users.  
In addition, $SV_{i,j}$, which captures the similarity of the personality trait values of the source and target users, can be calculated by the cosine similarity metric as follows: 
\begin{equation}
SV_{i,j}= \frac{ \sum_p^z V_{ip} \times V_{jp}}{\sqrt{\sum_p^z V_{ip}^2} \times \sqrt{\sum_p^z V_{jp}^2}},
 \label{For:PATFormula3}
\end{equation}

where $SV_{i,j}$, $SV_{i,j} \in R^{n\times n}$, captures the similarity of the personality traits of $u_i$ and $u_j$.  

\subsection{Proposed Model}
We use the following regularisation to incorporate the personality traits and the impact of their similarities:
\begin{equation}
\begin{aligned}
\begin{split}
&\beta \times( \sum_{i}^{n} \sum_{j !=i}^{n} \sum_{p=1}^{z} ( min \{0,f((V_{ijP})(SV_{i,j})((H \odot \\ & U')U^T )\}))^2)
\end{split}
\end{aligned}
 \label{For:PATFormula4}
\end{equation}

where f(y) is a function that has the same sign as y. $U$ dimension is fixed to overcome the non-convex problem and to turn this problem into a linear one. $U^T$ indicates the transpose of $U$, $z$ is the number of users' personality traits (five), and $\beta$ is a controlling parameter for the effect of this regularisation. In addition, $\odot$ is the Hadamard
product.
We follow the same procedure for fixing $H$ and $U'$. 

With the definition of the above regularisation, $PAT$ is based on TD while exploiting the effect of users' personality traits: 
\begin{equation}
\begin{aligned}
\begin{split}
&\ min_{U,H,U'} \Vert G -  (H \odot U') U^T \Vert ^2 _F + \beta \times(\sum_{i}^{n} \sum_{j !=i}^{n} \sum_{p=1}^{z}  ( min \{0,f(\\
& (V_{ijP})(SV_{i,j})((H \odot U'_{i} )U^T )\}))^2)+ \alpha \times( \Vert U \Vert ^2 _F + \Vert H \Vert ^2 _F + \Vert U' \Vert ^2 _F) \\
& U \geq 0, U' \geq 0, H \geq0
\end{split}
\end{aligned}
 \label{For:PATFormula5}
\end{equation}

where $\alpha$ controls the $U$, $H$ and $U'$, and `$\Vert  _F$' is the Frobenius norm. Further, after applying the Lagrangian function, we have:
\begin{equation}
\begin{aligned}
\begin{split}
&L(G;U,H,U')=Tr((G - (H\odot U')U^T)(G- (H\odot U')U^T)^T)+\\
&\beta \times(Tr((V SV)(H\odot U')U^T)(V SV(H\odot U')U^T)^T)+ \alpha \times Tr(  \\
& UU^T))+ \alpha \times Tr(HH^T) + \alpha \times Tr(U'U'^T)
\end{split}
\end{aligned}
\label{For: PATformula6}
\end{equation}

where $Tr$ indicates the trace of a matrix in linear algebra. The procedure is the same when we fix $H$ and $U'$. Now, we use the alternating least squares algorithm and the updating rule presented by Krompaas et al.~\cite{Nonnegativetensor} to update $U$, $U'$ and $H$ as follows:
\begin{equation}
 \Theta _i=\Theta _i \Bigg({\dfrac{\dfrac{\partial C(\Theta)^-}{\partial \Theta _i}}{\dfrac{\partial C(\Theta)^+}{\partial \Theta _i}}}\Bigg)^a
  \label{For:PATFormula7}
\end{equation}

where $\Theta$ is a non-negative variable, and $C(\Theta)$ is the negative part of the derivation. We use two element-wise operations for multiplication and division as $\bullet$ and /, respectively. 
Then, we calculate the partial derivative of Formula~\ref{For: PATformula6} with respect to $U$, $H$ and $U'$ and make them equal to zero. Next, based on Karush Kuhn Tucker complementary condition ~\cite{TRIFUNOVIC/2010}~\cite{Lenhart/2010} and the approach presented by Tang et al.~\cite{TangGHL13}, we have the updating rule as follows:
\begin{equation}
\begin{aligned}
\begin{split}
\small
&  U\longleftarrow U\bullet \Bigg((2G^T(H \odot U'))\Bigg/((H \odot U')U(H\odot U')+ (H \odot U')^T \\
& U(H \odot U')+ \beta \times VSV(H \odot U')U VSV(H \odot U')+\beta \times V^T SV^T  (H  \\  
& \odot U')^T U VSV(H\odot U')+2\alpha U )\Bigg)
\end{split}
\end{aligned}
 \label{For:PATFormula8}
\end{equation}

We can also follow the same updating approach for $H$ and $U'$. The exact algorithm of $PAT$ can be found in Algorithm 1.

\begin{algorithm}[h]
\caption{Trust prediction with $PAT$}
\begin{algorithmic}[1]
 \State Input: G, $\beta$, $\alpha$, n, z  
 \State Output:  $\tilde{G}$ which is a low-rank representation of G
 \State Calculate the personality trait values
  \State Calculate the similarities in the source and target users' personality traits
 \State Randomly initialise $U$, $U'$, and $H$
 \State \textbf{While} {It is not the convergent state}
 \State $A=-2G^T(H \odot U')$
 \State $B=(H \odot U')U(H\odot U')+ (H \odot U')^T U(H \odot U')+ \beta \times VSV(H \odot U')U VSV(H \odot U')+\beta \times V^T SV^T  (H \odot U')^T U VSV(H\odot U')+2\alpha U$
 \State    for j = 1 to n do
\State       \quad for j = 1 to n do
 \State      \quad\quad for r = 1 to z do
 \State     \quad \quad \quad   $  U\longleftarrow U\bullet(\dfrac{A_{ijr}}{B_{ijr}})$
 \State        \quad \quad end for
 \State   \quad end for
  \State    end for
  \State Repeat the same procedure for updating $H$ and $U'$
 \State \textbf{end While}
 \State Calculate $\tilde{G}$ in Formula~\ref{For:PATFormula2}
 \State return $\tilde{G},U, H, U'$
 \end{algorithmic}
  
\end{algorithm}

\section{Experiments}
\label{sec:section4}
For our experiments, we use datasets (Ciao and Epinions), evaluation metrics (MAE and RMSE) and baseline approaches, as discussed in Chapter~\ref{chap:DataandMetrics}.

\subsection{Experimental Setup}
We use four-fold cross-validation method and consider the average performance of all folds as the final performance value. The trust labels already provided in these datasets are used as the ground truth. The controlling parameters of our model were defined by applying cross-validation, and it reached its best performance when: $\beta = 0.5$, $\alpha = 0.1$ and $f=100$. 

\subsection{Impact of Similarity of Personality Traits}
\begin{figure*}
\centering
  \includegraphics[width=0.80 \textwidth]{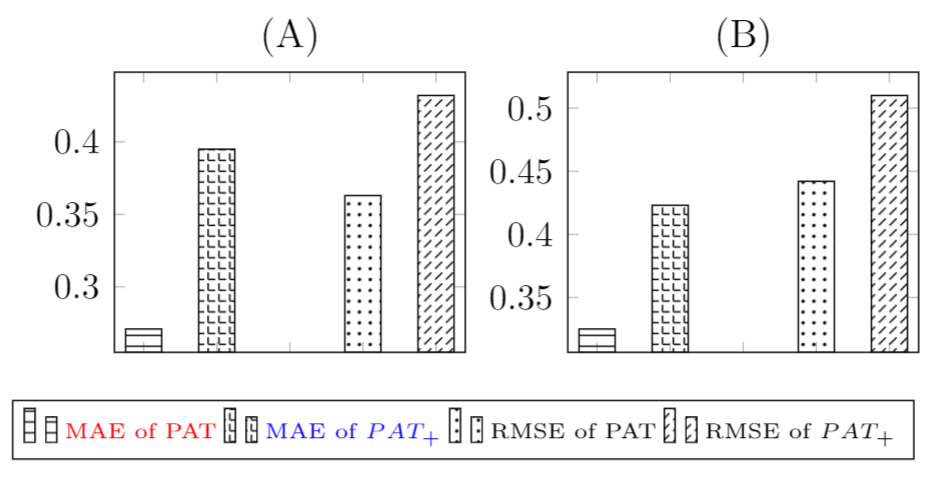}
  \caption{Comparison of the performance of $PAT$ (when $PAT$ considers the similarity of users' personality traits) and $PAT_{+}$ (when $PAT$ does not consider the similarity of users' personality traits), with respect to the MAE and RMSE metrics on (A) the Ciao dataset and (B) the Epinions dataset.}
  \label{PlusPlusCombinedBar}
\end{figure*}

To further investigate \textbf{RQ1}---Do users who bear a trust relation between them have similar personality traits?---in this section, we explore the impact on $PAT$ of considering the similarity of source and target users' personality traits. We remove the similarity metric $SV_{i,j}$, proposed in Formulas ~\ref{For:PATFormula3} and \ref{For:PATFormula4}, from our model, to observe the performance of $PAT$ when it does not consider the similarity of the personality traits of users. 
This new version of $PAT$ is given the name $PAT_{+}$. Figure~\ref{PlusPlusCombinedBar} compares the performance of $PAT$ and $PAT_{+}$ on the Ciao and Epinions datasets with respect to the MAE and RMSE metrics.
It demonstrates that adding $SV_{i,j}$ to $PAT$ (when $PAT$ considers the similarity of users' personality traits) significantly improves the model's performance. $PAT$ has around $32\%$ and around $24\%$ lower MAE and RMSE compared to $PAT_{+}$. Hence, considering the similarity of the source and target users' personality traits can significantly improve the performance of our trust prediction approach ($PAT$).

\begin{table}
\centering
\caption{Comparison of the trust relation prediction performance of $PAT$ on the Ciao and Epinions datasets, with respect to the MAE metric. 
  }
\begin{tabular}{lll}
\toprule
\textbf{The Personality Trait of \quad } & \quad \textbf{Ciao} & \textbf{\quad  Epinions} \\  \midrule
Source Users (Scenario 1)  &\quad 0.271 & \quad \quad 0.325\\
Target Users (Scenario 2)& \quad 0.321 & \quad \quad  0.363 \\
Both Users (Scenario 3)& \quad 0.348 & \quad \quad  0.395\\
\bottomrule
\end{tabular}
   \label{tab:ciaorescc}
\end{table}

\begin{table}
\centering
\caption{Comparison of the trust relation prediction performance of $PAT$, on the Ciao and Epinions datasets, with respect to the RMSE metric. 
  }
\begin{tabular}{lll}
\toprule
\textbf{The Personality Trait of \quad } & \quad \textbf{Ciao} & \textbf{\quad  Epinions} \\  \midrule
Source Users (Scenario 1) &\quad 0.363 & \quad \quad 0.442\\
Target Users (Scenario 2) & \quad 0.401 & \quad \quad  0.472 \\
Both Users (Scenario 3) & \quad 0.419 & \quad \quad  0.498\\
\bottomrule
\end{tabular}
   \label{tab:ciaorescc2}
\end{table}

\subsection{Best Performance of the Personality-Aware Trust Prediction}
Tables~\ref{tab:ciaorescc} and \ref{tab:ciaorescc2} show the trust relation prediction performance of $PAT$ in response to the following three research questions:
\\\\
\noindent \textbf{RQ3.} What is the performance of $PAT$ if only the personality traits of source users or $V_{ip}$ are considered (Scenario 1, Tables~\ref{tab:ciaorescc} and \ref{tab:ciaorescc2})?

\noindent \textbf{RQ4.} What is the performance of $PAT$ if only the personality traits of target users or $V_{jp}$ are considered (Scenario 2, Tables~\ref{tab:ciaorescc} and \ref{tab:ciaorescc2} )?

\noindent \textbf{RQ5.} What is the performance of $PAT$ if the personality traits of source users and target users or $V_{ijp}$ are considered simultaneously (Scenario 3, Tables~\ref{tab:ciaorescc} and \ref{tab:ciaorescc2})?
\\\\
To answer \textbf{RQ3} and \textbf{RQ4}, we replace $V_{ijp}$ in the Formula~\ref{For:PATFormula4} of our model with the personality traits of the source users, $V_{ip}$, and target users, $V_{jp}$, respectively. For answering \textbf{RQ5}, there is no need to modify Formula~\ref{For:PATFormula4}, as it already considers the personality traits of the source and target users. It should be noted that for answering these questions, we only modify $V_{ijp}$ in Formula~\ref{For:PATFormula4}, which contains the personality trait values of source or target users; the similarity metric of $SV_{ij}$ remains unchanged. This similarity metric is the similarity of the personality trait values of the source or the target users.  

As seen in Tables~\ref{tab:ciaorescc} and \ref{tab:ciaorescc2}, $PAT$ in Scenario 1 has the lowest MAE and RMSE compared to scenarios 2 and 3 for both datasets. In the Ciao dataset, it has about 15\% and 23\% lower MAE in Scenario 1 compared to scenarios 2 and 3, respectively. In the Epinions dataset, it has approximately 11\% and 18\% lower MAE in Scenario 1 compared to scenarios 2 and 3, respectively. This superior performance can also be seen in Table~\ref{tab:ciaorescc2}, with respect to the RMSE metric.  
  
\textbf{In summary,} the best performance of $PAT$ is achieved in both the CIao and Epinions datasets when it only considers the personality trait values of source users. We conclude that considering the personality trait values of source users and the similarity value of source and target users simultaneously (as two separate factors) can improve the performance of $PAT$. These results demonstrate that, in addition to considering the similarity of both users, focusing on the personality trait values of the source user may be an important indicator for trust prediction approaches.
In the following experiments, we only consider the personality traits of the source users or $V_{ip}$, in Formula~\ref{For:PATFormula4} of our model.

\begin{table*}
\centering
\begin{tabular}{llllllllll} 
\toprule
\textbf{Metric}&\textbf{Dataset}&\textbf{Random}& \textbf{RS}& \textbf{MF} & \textbf{sTrust} & \textbf{hTrust}  &\textbf{PAT}\\  
\midrule
MAE & Ciao &   5.63& 3.1 & 1.95 & 1.8   &  1.17    &\textbf{0.27}\\
 \quad & Epinions&   5.98& 3.19   &    2.05   & 1.47   &  1.36    &\textbf{0.32}\\
 \hline
 RMSE & Ciao & 6.65& 3.11   &   2.09     &  1.93  & 1.32   & \textbf{0.36}\\
 \quad & Epinions&   6.85& 3.29  &    2.16     &  2.15 & 1.56  & \textbf{0.442}\\
\bottomrule
\end{tabular}
\caption{Comparison of the performance of our trust prediction model (PAT) with other baseline approaches with respect to the MAE and RMSE on the Ciao and Epinions datasets.} 
   \label{tab:categories}
\end{table*}

\subsection{Comparison of Different Trust Prediction Approaches}
In this section, we compare $PAT$ with various baseline approaches.
Table~\ref{tab:categories} compares the performance of $PAT$ and other pair-wise trust prediction approaches with respect to the MAE and RMSE metrics in the Ciao and Epinions datasets. We see that $PAT$ has the lowest MAE and RMSE in both datasets. In the Ciao dataset, the MAE of $PAT$ is about 4, 7, 7, 11 and 21 times less than for $hTrust$, $sTrust$, $MF$, $RS$ and $Random$, respectively.
Likewise, in the Ciao dataset, the RMSE of $PAT$ is approximately  4, 5, 6, 9 and 18 times less than for $hTrust$, $sTrust$, $MF$, $RS$ and $Random$, respectively. Similar superior performance for $PAT$ can be observed on the Epinions dataset (Table~\ref{tab:categories}). 

\textbf{In summary,} $PAT$ outperforms the baseline trust prediction approaches with respect to the MAE and RMSE metrics on the Ciao and Epinions datasets.

\begin{table}
\centering
\caption{The impacts of each personality trait on $PAT$ with respect to the MAE and RMSE metrics on the Ciao dataset. $\Delta_{MAE}$ and $\Delta_{RMSE}$ represent the difference between the performance of $PAT$, when it includes all the personality traits, and $PAT_{new}$, when it excludes a particular personality trait value, with respect to the MAE and RMSE, respectively. 
}
\begin{tabular}{lllll} 
\toprule
\textbf{Approach}&\textbf{MAE-Ciao}&\textbf{$\Delta_{MAE}$} & \textbf{RMSE-Ciao} & \textbf{$\Delta_{RMSE}$} \\  
\midrule
PAT & \quad 0.271 & \quad - &\quad \quad 0.363 &  \quad - \\
\hline
PAT-Openness & \quad 0.346 &  0.075 &\quad \quad 0.426 &    0.063\\
PAT-Conscientiousness & \quad 0.393 &  \textbf{0.122} &\quad \quad 0.465 &    \textbf{0.102}\\
PAT-Extraversion &  \quad 0.395 &  \textbf{0.124}  &\quad \quad 0.482 &    \textbf{0.119} \\  
PAT-Agreeableness & \quad 0.308 &   0.037 &\quad \quad 0.398 &    0.035\\
PAT-Neuroticism & \quad 0.309 &  0.038 &\quad \quad 0.391 &   0.028 \\
\bottomrule
\end{tabular}
   \label{tab:categories21}
\end{table}

\begin{table}
\centering
\caption{The impacts of each personality trait on $PAT$ with respect to the MAE and RMSE metrics on the Epinions dataset. $\Delta_{MAE}$ and $\Delta_{RMSE}$ represent the difference between the performance of $PAT$, when it includes all the personality traits and $PAT_{new}$, when it excludes a particular personality trait value with respect to the MAE and RMSE, respectively. For example, PAT-Extraversion refers to the model of \textbf{PAT} that ignores Extraversion. The MAE of $PAT-Extraversion$ in the Epinions dataset is 0.412, whereas if $PAT$ considers all of the Big Five personality traits, its MAE is 0.325. Hence, $\Delta_{MAE} = 0.421-0.325 = 0.087$.}
\begin{tabular}{lllll} 
\toprule
\textbf{Approach}&\textbf{MAE-Epinions}&\textbf{$\Delta_{MAE}$} &\textbf{RMSE-Epinions} & \textbf{$\Delta_{RMSE}$} \\  
\midrule
PAT & \quad 0.325   &\quad  -   & \quad   0.442 &  -\\
\hline
PAT-Openness & \quad 0.362   &  0.037   &  \quad 0.481 &  0.039\\
PAT-Conscientiousness & \quad  0.408   &   \textbf{0.083}  & \quad 0.509 &  \textbf{0.067}\\
PAT-Extraversion &  \quad  0.412   &   \textbf{0.087}   & \quad 0.513 & \textbf{0.071}\\  
PAT-Agreeableness & \quad 0.328   &   0.003   &  \quad 0.472 &0.03\\
PAT-Neuroticism & \quad  0.331   &   0.006   & \quad 0.468 & 0.026\\
\bottomrule
\end{tabular}
   \label{tab:categories2121}
\end{table}

\subsection{Impact of Each Personality Trait}
To further investigate \textbf{RQ2}---What is the relationship between personality trait and trust relations?---in this section, we explore the effect of each personality trait on the performance of $PAT$. 
To do so, we remove the personality trait values one by one from the $V_{ijp}$ in Formula~\ref{For:PATFormula4} of our model and evaluate the performance. In other words, in each iteration, we only consider four of the users' personality trait values in our model, calling this the new version of $PAT$ (i.e., $PAT_{new}$). In this way, we can investigate the following question:

\textbf{RQ6.} Ignoring which personality trait can have a higher negative impact on the performance of PAT?

We define two metrics called {$\Delta_{MAE}$} and {$\Delta_{RMSE}$} as follows:
\begin{equation}
 \Delta_{MAE}= |MAE _{PAT_{new}} - MAE_{PAT}| 
\end{equation}
\begin{equation}
 \Delta_{RMSE}= |RMSE_{PAT_{new}} - RMSE_{PAT}| 
\end{equation}\\

where $\Delta_{MAE}$ and $\Delta_{RMSE}$ represent the differences between the performance of $PAT$, when it includes all of the Big Five personality traits, and $PAT_{new}$, when it excludes a particular personality trait value with respect to the MAE and RMSE metrics, respectively.
Tables~\ref{tab:categories21} and ~\ref{tab:categories2121} show the results on the Ciao and Epinions datasets, respectively. We see that when we remove the Extraversion trait from the personality vector in the Ciao dataset, the MAE increases to 0.395 from 0.271. Hence, $\Delta_{MAE} = 0.395-0.271 = 0.124$. 

To answer \textbf{RQ6}, we should consider the fact that higher $\Delta_{MAE}$ and $\Delta_{RMSE}$ indicate a greater negative impact on $PAT$'s performance of ignoring a particular personality trait. Accordingly, we can identify the most important personality traits for the trust relation prediction procedure.

As illustrated in Tables~\ref{tab:categories21} and ~\ref{tab:categories2121}, removing the Extraversion or Conscientiousness traits from our model increases the MAE and RMSE of $PAT$ significantly. The $\Delta_{MAE}$ and $\Delta_{RMSE}$ for these cases are more than 0.1 in the Ciao dataset.
Ignoring Agreeableness or Neuroticism does not lead to significant changes in the MAE or RMSE, indicating the low negative impact of ignoring Agreeableness or Neuroticism. 



\begin{figure*}[b!]
\centering
  \includegraphics[width=0.90\textwidth]{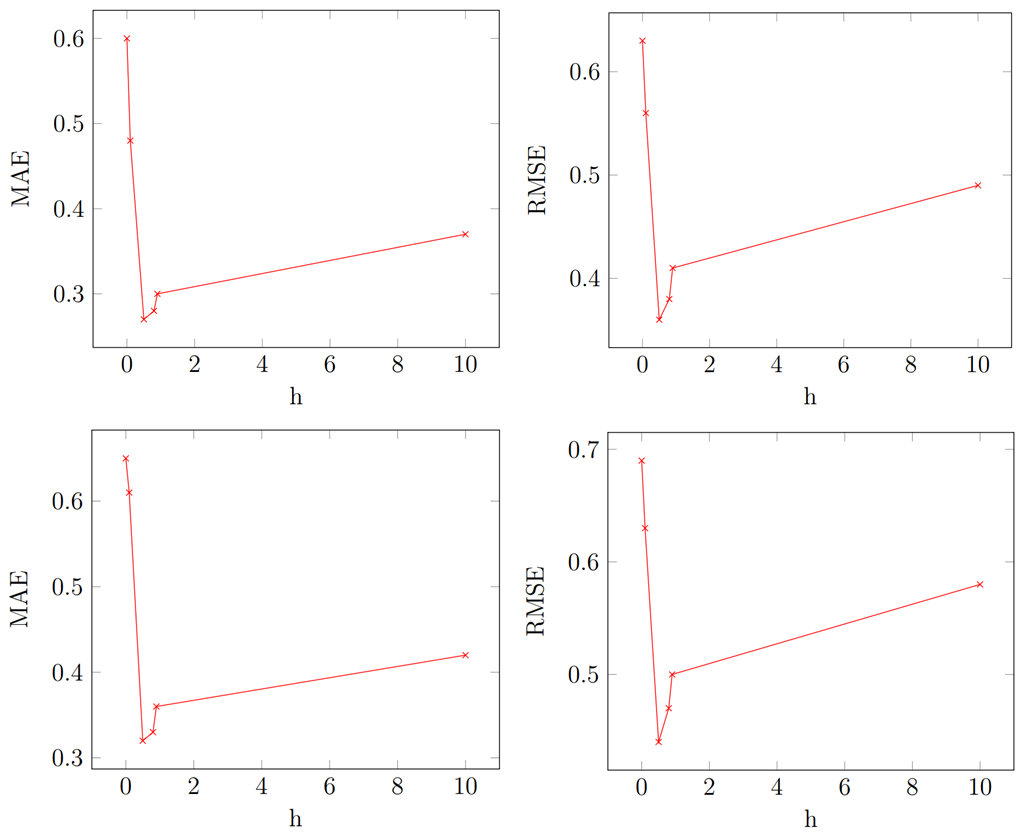}
  \caption{The PAT regularisation effects with respect to the MAE and RMSE on the Ciao (the upper images) and Epinions (the lower images) datasets.}
  \label{fig:regularizationPAT}
\end{figure*}

\subsection{The PAT Regularisation Effects}
We applied different values for $\beta$ as $\beta=\{0, 0.1, 0.5, 0.8, 0.9, 10\}$, finding that the best performance of \textbf{PAT} was achieved by $\beta = 0.5$ (Figure~\ref{fig:regularizationPAT}). The performance of \textbf{PAT} is increased by increasing $\beta$ from `0' to `0.5'. However, when $\beta$>0.5, the performance decreases.






\begin{figure*}[b!]
\centering
  \includegraphics[width=0.90 \textwidth]{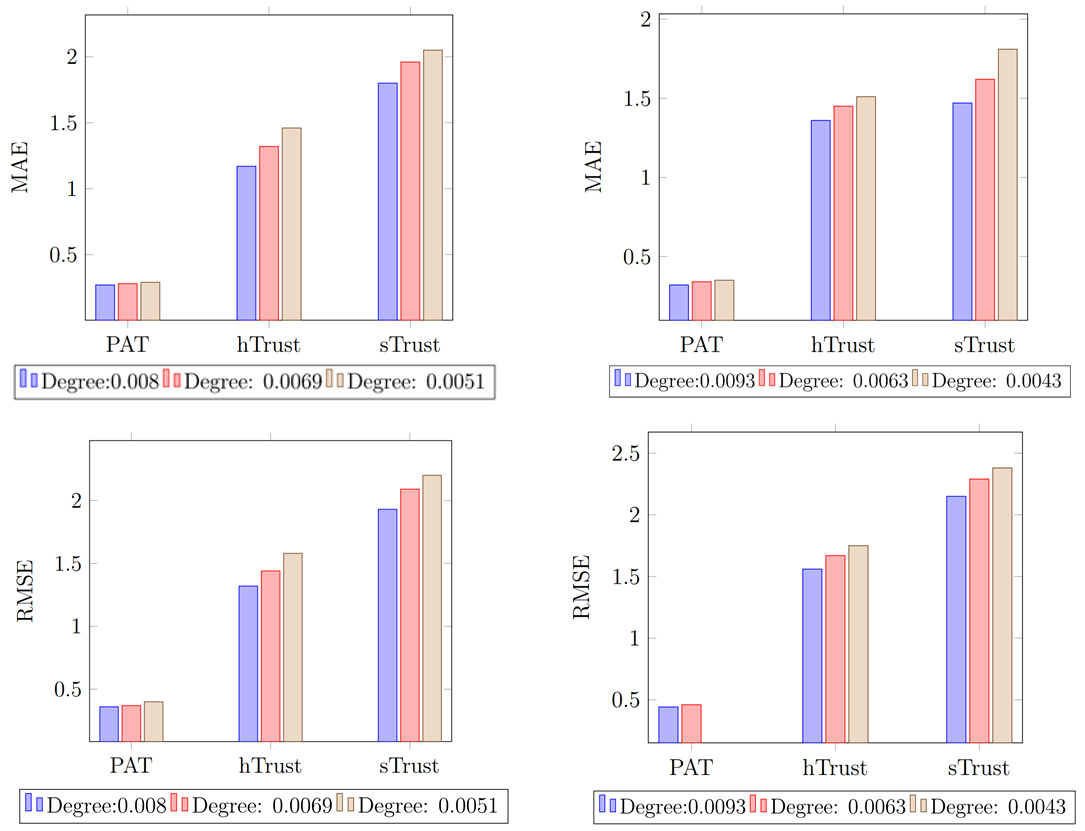}
  \caption{The impact of degree of sparsity on the performance of trust prediction approaches. The left images illustrate the performance of trust prediction approaches with respect to the MAE and RMSE in the Ciao dataset. The right images demonstrate the performance of these approaches with respect to the MAE and RMSE in the Epinions dataset.}
  \label{fig:sparsity}
\end{figure*}
\subsection{Impact of Data Sparsity Degree}

Finally, we investigate the impact of the degree of data sparsity on $PAT$. According to Wang et al.~\cite{ICDM}, the data sparsity degree (how sparse a dataset is~\cite{ICDM}) can be calculated by: 
\begin{equation}
Degree = \dfrac{NT}{n \times n} 
\end{equation}

where $Degree$ is the data sparsity, NT is the number of existing trust relations, and $n$ is the number of users. A smaller $Degree$ indicates a sparser dataset. We follow the same approach proposed by Wang et al.~\cite{ICDM} and evaluate $PAT$ on the Epinions and Ciao datasets with different $Degrees$ of 0.0093, 0.0063, and 0.0043, and 0.008, 0.0069 and 0.0051, respectively. Figure~\ref{fig:sparsity} demonstrates that unlike the two state-of-the-art trust prediction approaches to which it is compared in the figure, $PAT$ is insensitive to the sparsity degree of the trust relations. $PAT$ has a close to stable prediction performance in the presence of different degrees of sparsity, whereas $hTrust$ and $sTrust$ are negatively affected by increasing the degree of sparsity of the datasets.


 \section{Summary}
 \label{sec:section5}
In this section, we proposed $PAT$ as a novel unsupervised trust prediction model capable of incorporating users' personality traits. We first analysed the relation between trust and the similarity of source and target users' personality traits, and examined the relationship between each personality trait and trust relations. Then, we proposed a new TD-based trust prediction model, incorporating users' personality traits, for predicting pair-wise trust relations in OSNs.
The experimental results demonstrated the effectiveness of our approach compared to other state-of-the-art approaches. However, $PAT$ is not context- nor time-aware. In Chapter~\ref{chap:context}, we propose trust prediction approaches that consider the contexts of trust relations. We also propose a deep learning-based model that predicts trust relations in different time windows.



\chapter{Proposing Context-Aware Trust Prediction Approaches}
\label{chap:context}

To address our second research problem, in this chapter, we present context-aware trust
prediction approaches that consider the notion of context (which conceptually refers to any knowledge to specify the condition of an entity) and the social actor's behaviour (supported by theories from social psychology) as first class citizens. We present novel algorithms that employ
social context factors inspired by social psychology theories and mathematically model our approach based on MF and TD. We also propose a deep context-aware trust prediction approach. We perform extensive empirical studies, and present evaluations of the
effectiveness and quality of the results using real-world datasets. The proposed approaches in this chapter have been accepted and published by the Web Information Systems Engineering Society (WISE) 2018 (Core Rank A), Data Quality and Trust in Big Data (QUAT) 2018 and International Conference on Advances in Mobile Computing and Multimedia (MOMM) 2019 (Core Rank B).

\section{Introduction}

In this chapter, we propose three novel context-aware trust prediction approaches. In the first approach, we employ SET, a well-known social psychology theory~\cite{Thibaut1959}~\cite{Eisenstadt1965}~\cite{Homans1958}. Based on this theory, two people may establish
a relationship (in the context of this dissertation, it is to establish a trust relation), if and only if the cost of that relationship is less than its benefit. We also propose some social context factors to capture the contexts of trust relations. Finally, we incorporate the context factors and SET effects into an MF-based model to propose our first context-aware trust prediction approach, $SETTrust$ (a SET-based trust prediction approach)~\cite{GhafariQUAT}.

Next, we propose a trust prediction model that directly considers the context of trust. We employ a three dimensional tensor to present users' trust relations in different contexts. We also consider several social context factors to capture the context of trust relations and incorporate them into a TD model, called $TDTrust$ (a TD-based trust prediction approach)~\cite{GhafariWise} to predict pair-wise trust relations among users. The experimental results indicate a substantial improvement in $TDTrust$'s trust prediction performance compared to $SETTrust$.  

Finally, in our third context-aware trust prediction approach, we aim to address the pair-wise trust prediction problem by proposing a supervised approach, called $DCAT$~\cite{DCAT} (a deep context-aware trust prediction approach). This approach is based on a deep learning-based structure, making it one of the first deep trust predictors for use in OSNs, and like the previous models, it employs some social context factors to capture the contexts of the trust relations. In addition, we analyse the textual content provided by users in OSNs to improve the performance of $DCAT$. The experimental results demonstrate the superior performance of $DCAT$ compared to the state-of-the-art approaches, $TDTrust$ and $SETTrust$.

Our contributions in this section are summarised as follows:

\begin{itemize}
\item Proposing $SETTrust$, a trust prediction model based on SET and MF that, in contrast to many of the existing approaches, considers the context of trust relations.
\item Proposing a TD-based trust prediction model, TDTrust, and mathematically modelling our approach. This model can directly consider the context of trust.
\item Proposing algorithms to capture the context of trust in OSNs and employing social context factors inspired by psychological theories such as social penetration theory~\cite{Altman/1973}.
 \item Proposing $DCAT$, one of the first deep learning-based and context-aware graph analytics models for trust prediction in OSNs.
\end{itemize}

The rest of this chapter is organised as follows. In Section~\ref{ch:chapt5sec1}
we discuss the social exchange theory. We present our first trust prediction model, $SETTrust$, in Section~\ref{ch:chapt5sec2}. Next, we present the experimental results for evaluating $SETTrust$ in Section~\ref{ch:chapt5sec3}. We introduce our second context-aware trust prediction approach, $TDTrust$, in Section~\ref{ch:chapt5sec4}. We evaluate $TDTrust$ in Section~\ref{ch:chapt5sec5}. Our third context-aware trust prediction approach, $DCAT$, and its required background knowledge are discussed in Sections~\ref{ch:chapt5sec6}, ~\ref{ch:chapt5sec7}, ~\ref{ch:chapt5sec8}. Section~\ref{ch:chapt5sec9} presents the procedure for evaluating $DCAT$. Finally, the chapter is concluded in Section~\ref{ch:chapt5sec10}.

\section{Social Exchange Theory and its Applications in OSNs}
\label{ch:chapt5sec1}
SET~\cite{Thibaut1959,Eisenstadt1965,Homans1958}, a well-known theory from social psychology, explains the nature of social interaction with the help of a cost--benefit structure. Based on this theory,
people prefer to participate in relationships that have low cost while bringing them maximum benefit~\cite{Thibaut1959,Eisenstadt1965,Homans1958}; that is,
\begin{equation}
SET = Benefit - Cost > 0
\end{equation}

Hence, if the SET of a relationship is negative for
one or all of the participants, that relationship will probably break down in the future. It is worth mentioning that there are some other factors that may have an impact on breaking down or establishing
a relationship (e.g., the level of alternatives or level of expectations); however, we do not consider these in this dissertation, and instead leave them for our future work. Example of cost includes the time and money already
spent or that will be spent on the relationship, the information to be shared in the relationship, the negative points of the other party in the relation, and so on. Benefit includes the services that a person receives as a result of a relationship, the knowledge or money that he or she acquire and the happiness that may be obtained. 

There have been some studies on the effects of SET on OSNs. For instance, Surma et al.~\cite{Surma16} investigated the existence of SET on Facebook, noting that any relationship requires a starting point and that one of the participants must make the first move. The authors considered the \textit{Likes} on their posts that users had received from friends during the past week. They also studied how many users \textit{Liked back} the posts of other participants, and considered some other factors, such as gender, age, posts sent and comments received. Their experimental results demonstrated the existence of SET effects in OSNs. Another study, conducted by Barak~\cite{Shtatfeld}, looked at the factors influencing the selection of potential partners on online dating websites. Their SET-based approach focused on users' features, such as marital status, the level of education, income and appearance. Their study showed that users look for people with the same marital status and a higher level of income and education than they have themselves.

\begin{figure}
\centering
  \includegraphics[width=140mm]{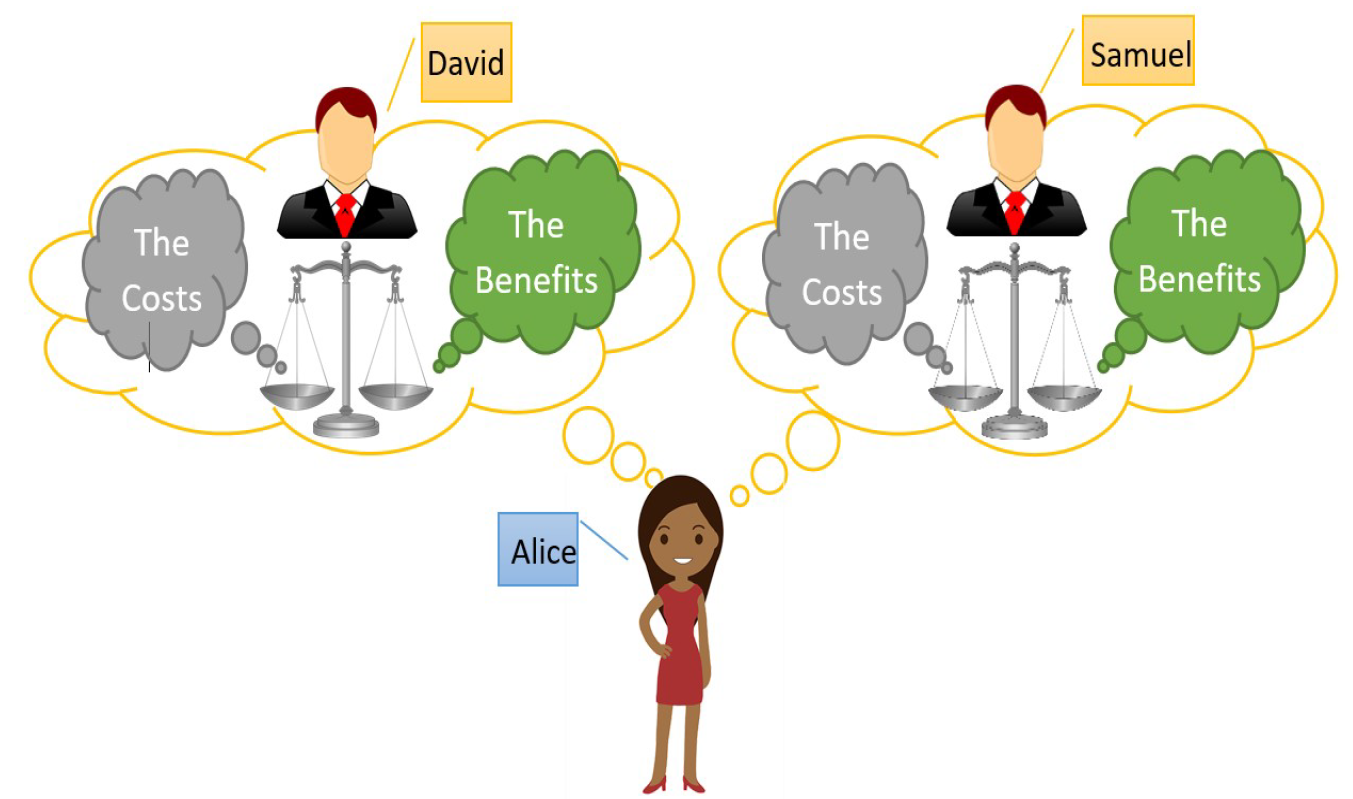}
  \caption{A motivating example.}
  \label{fig: SETTRustMotivation}
\end{figure}
 
\textbf{Motivating Example.} Alice is a postgraduate student in the computer science field and David and Samuel are university professors (Figure~\ref{fig: SETTRustMotivation}). Alice wants to continue her studies
at the PhD level, so she needs to find a supervisor. In making her decision, Alice will compare the benefit and cost of supervision by David and Samuel. First, consider that Alice is interested in working in the area of social media analysis. As for her potential supervisors, David's expertise relates to cloud computing, he is a kind man, and he has been published several times in top venues (benefit); however, he has not supervised any PhD students before (cost). Conversely, Samuel works on social media analysis, has supervised many PhD students before and has also been extensively published (benefit); however, he is known as a harsh supervisor (cost). From Alice's perspective, the benefit of having Samuel as her supervisor outweighs the cost. Moreover, she considers the cost of Samuel's supervision (his harshness) to be less than the cost of the David's supervision (having a different area of interest and no supervisory experience). Hence, Alice chooses Samuel as her PhD supervisor.

\section{SETTrust: A Trust Prediction Approach Based on Social Exchange Theory}
\label{ch:chapt5sec2}

\subsection{Social Context Factors}
\label{sec:levelofexpertise}

In this section, we describe the social contextual factors that we employ in our trust prediction approach; that is, level of expertise, interest, number of followers and bad language detection.

\textbf{A. \textit{Level of Expertise}}. `A recommendation from an expert person in a certain domain is more acceptable compared to the [recommendation of a] less knowledgeable person'~\cite{Yan/WWWJ/2015}. The level of expertise of a target user will be calculated by
(i)~evaluating his or her activeness in a certain context, where we assume that the target user is active in a certain context if the number of his or her posts/reviews is equal to or more than the average number of other users' posts/reviews; and
(ii)~considering other users' opinion about the target users' posts/reviews, as measured by whether they liked or highly rated those posts/reviews. For instance, consider that David (a user) has written many reviews/posts on the topic of sports that are highly rated by other users.
In contrast, Sarah only posted two reviews/posts on the same topic, neither of which have been rated.
Therefore, it can be argued that, from the perspective of other users, David is more trustworthy in the context of sports.

Let $U = \{u_1, u_2,..., u_m\}$ denote the set of users and $C = \{c_1, c_2,...,c_k\}$ the set of contexts of trust relations. Let $n_i$ denote the total number of posts/reviews by user $u_i$, for $i = \{1, ..., m\}$ and $n_{i,{c_k}}$ the total number of posts/reviews by user $u_i$ in context $c_k$ for $i = \{1, ..., m\}$  and $k = \{1, ..., z\}$. Let $a_{i,{c_k}}$ denote the status of user $u_i$, where $a_{i,{c_k}}= 0$ means $u_i$ is inactive and $a_{i,{c_k}}= 1$ means $u_i$ is active. Each review/post is evaluated by other users in the form of a score $s$ (on a Likert scale of 1 to 3):
\begin{equation}
\small
  a_{i,{c_k}}=\begin{cases}
    1, & \text{if $n_{i,{c_k}} >= {\dfrac{\sum_{r=1}^{m} n_{r,{c_k}}}{m}} $}\\
    0, & \text{otherwise}
  \end{cases}
\end{equation}

where we calculate the average score of posts/reviews that $u_i$ received from other users in $c_k$ by:
\begin{equation}
\small
 S_{i,{c_k}} = {\dfrac{1}{n_{i,{c_k}}}} \sum_{r=1}^{n_{i,{c_k}}} s_r ^{i,{c_k}}
\end{equation}

where $s_r ^{i,{c_k}}$ is the score value of the $r^{th}$ review of $u_i$ that was achieved in $c_k$. Finally, the level of expertise $u_i$ in $c_k$ can be calculated by:
\begin{equation}
 v_{i,{c_k}} = S_{i,{c_k}} \times n_{i,{c_k}}
\end{equation} 

\textbf{B. \textit{Interest}.} `Interest could be conceived of as an individual's attitude towards a set of objects'~\cite{Yan/WWWJ/2015}.
We consider $p_{ic_k}$ as the interest of $u_i$ in the context $c_k$, which represents the topics/categories of items to which a user's posts/reviews belong. $p_{ic_k}=1$ is within the scope of $c_k$ and is not when  $p_{ic_k}=0$. Further, if both the source user $u_i$ and the target user $u_j$ have the same interest in the same context $c_k$, $P_{ijc_k}=1$; otherwise, $P_{ijc_k}=-1$.

\textbf{C. \textit{Number of Followers.}}
$NoF_j$ denotes the number of followers of user $u_j$. It is equal to the `number of followers' divided by the `number of people who read and rated the user's reviews'. Assume that David has a higher number of followers than does Sarah.
It has been validated in social science theories~\cite{Liuwwwj}, that a recommendation from David (who has a larger number of followers) is more credible than one from Sarah.

\textbf{D. \textit{Bad Language Detection.}} Machiavellianism, one of the dark triad traits, was significantly positively correlated with swear words~\cite{SumnerBBP12}. Using swear words and bad language in OSNs places a user at risk of losing his or her intimacy with other users~\cite{Dussault2013}. Traits reflecting the dark triad (i.e., Machiavellianism, psychopathy and narcissism) seem to render someone as a candidate for avoidance.
Given the traits associated with psychopathy, it would likely be advantageous for others to avoid such individuals~\cite{Dussault2013}. 
We employ LIWC~\cite{LIWC} in our text analysis step to detect the bad language usage in posts/reviews, which we consider to negatively affect the trustworthiness of a user. LIWC provides the percentage of swear words that a user has in his or her published posts/reviews.
Thus, we can evaluate the level of bad language usage of $u_i$ in the context $c_k$  as follows:
\begin{equation}
Bl_{ic_k}= {\dfrac{B_{i,{c_k}}}{W_{i,{c_k}}}},
\end{equation}

where $B_{i,{c_k}}$ is the total number of swear words written by $u_i$ in the context $c_k$, and $W_{i,{c_k}}$ represents the total number of words $u_i$ wrote in his or her posts/reviews in the same context.

\subsection{Trust Prediction Mechanism}
We present algorithms for predicting trust values in OSNs, employing social context factors inspired by SET.
For trust prediction, we propose the following formula:
\begin{equation}
SET_{ijc_k}=w_1 \times P_{ijc_k}+ w_2 \times v_{jc_k}+ w_3 \times NoF_{j} - w_4 \times Bl_{jc_k},
\end{equation}

where $SET_{ijc_k}$ is the trust degree that we employ to predict the trust value between $u_i$ and $u_j$ in the context $c_k$, and $w_z$, $z=\{1,\dotsm,4\}$ is a controlling parameter to control for the impact of social contextual parameters.
The above formula, calculates the cost and benefit of a potential relationship between $u_i$ and $u_j$. 
For example, consider the relation between user $u_i$ and $u_j$ in context $c_k$. Our aim is to predict whether $u_i$ will form a trust relation with $u_j$ in that context. On the one hand, the cost of this relation will include the values for $Bl_{jc_k}$, $u_i$ and $u_j$ having different interests ($P_{ijc_k}=-1$), and $u_j$ having a low level of expertise and a small number of followers.

On the other hand, the benefit of this relation will include a lack of $Bl_{jc_k}$, $u_i$ and $u_j$ having the same interests ($P_{ijc_k}=1$), and $u_j$ having a high level of expertise and a large number of followers.
Since $NoF_j$ could be a large number, we normalise it to be in a range between 0 and 1 by feature scaling:
\begin{equation}
NoF_j^{'}={\dfrac{NoF_j - min(NoF)}{max(NoF)-min(NoF)}}
\end{equation}

\subsubsection{Problem Statement.}

Suppose we have $n$ users where $U = \{u_1,\dotsm, u_n\}$. Also assume that we have $k$ contexts of trust as $C = \{c_1, c_2,\dotsm, c_k\}$. In each context, $G \in R^{n\times n}$ is a square matrix that contains the trust relations among users. $u_i$ trusts $u_j$ if $G(i, j) = 1$, and if $G(i, j) = 0$, there is no trust relation between $u_i$ and $u_j$. The $G$ matrix is sparse in OSNs~\cite{TangGHL13}. Thus, to deal with this data sparsity problem, we must extract the $U$ low-ranked matrix, as $U \in R^{n\times d}$, $d << n$. Tang et al.~\cite{TangGHL13} proposed a trust prediction approach based on an MF model as follows:
\begin{equation}
\ min_{U,H} \Vert G -   UHU^T \Vert ^2 _F +\ \alpha \times({\Vert U \Vert ^2 _F + \Vert H \Vert ^2 _F}),
\  U > 0, H > 0
\end{equation} 

where $U$ represents users' interests, $d$ represents the facets of these interests and the $H$ matrix contains compact correlations among $U$~\cite{WangWTZC15}. Following Wang et al.~\cite{WangWTZC15} and Tang et al.~\cite{TangGHL13}, we propose two terms to capture the $SET$ effects in trust relations, as follows:
\begin{equation}
\begin{aligned}
\begin{split}
  SET_{ijc_k} \times(UHU^T) =0  \\
  SET_{ijc_k} \times(UHU^T) =1
\end{split}
\end{aligned}
\end{equation}

The aim of SET regularisation is to maximising this term:
\begin{equation}
\begin{aligned}
\sum_i^n\sum_{j!=i}^{n}(max\{0,f(SET_{ijc_k})(UHU^T)\}^2)
\end{aligned}
\end{equation}

Considering $\lambda$ a controlling parameter, we propose a trust prediction model as follows:
\begin{equation}
\begin{aligned}
\begin{split}
\ min_{U,H} \Vert G -   UHU^T \Vert ^2 _F + \lambda \times(\sum_i^n\sum_{j!=i}^{n}(max\{0,f(SET_{ijc_k})(UHU^T)\}^2)) +\\
 \alpha \times \Vert U \Vert ^2 _F + \alpha \times \Vert H \Vert ^2 _F)
\  , U > 0 , H > 0
\end{split}
\end{aligned}
\end{equation}

\subsubsection{Modeling Proposed Trust Prediction}
\label{sec: SetTrustModel}
Next, assuming $B = SET_{ijc_k}$, we propose an updating schema for both $U$ and $H$ based on Ding et al.~\cite{DingLJ08}:
\begin{equation}
\begin{aligned}
\begin{split}
U(i,j) \leftarrow U(i,j) \sqrt{\dfrac{D(i,j)}{E(i,j)}}
\end{split}
\end{aligned}
\end{equation}\\
\begin{equation}
\begin{aligned}
\begin{split}
H(i,j) \leftarrow H(i,j) \sqrt{\dfrac{P(i,j)}{Y(i,j)}}
\end{split}
\end{aligned}
\end{equation}

where:

\begin{equation}
D= 2GUH^T +2G^TUH
\label{formula:chap5-1}
\end{equation}\\
\begin{equation}
\begin{aligned}
\begin{split}
E=2H^TUU^THU + 2HUU^THU +\lambda \times UU^T HU \odot B \odot B^T + \\
\lambda  \times H^T U \odot B  \odot B^T  \odot UTH^TU +\\
\lambda \times HU \odot B \odot B^T \odot UTHU + \lambda \times \\
H^TUU^TH^TU \odot B \odot B^T + 2\alpha U
\end{split}
\end{aligned}
\label{formula:chap5-2}
\end{equation}\\
\begin{equation}
P= 2U^TGHU + 2G^TUHU^T
\label{formula:chap5-3}
\end{equation}\\
\begin{equation}
\begin{aligned}
\begin{split}
Y=U^TU^ThU^TU+UUHU^TU+\lambda \times (B\odot UHU^T \odot B \odot U+ \\
U\odot B^T \odot U^THU \odot B^T \odot U^T)+2\alpha H
\end{split}
\end{aligned}
\label{formula:chap5-4}
\end{equation}\\

\begin{algorithm}[h]
\caption{Trust prediction with SETTrust}
\begin{algorithmic}[1]
 \State Input: G, $\lambda$, $\alpha$, r
 \State Output: $U, H$
 \State Establish the SET in each context
 \State Randomly initialise $U, H, C$
 \State \textbf{While}  It is not the convergent state \textbf{do}
 \State \quad Initialise D, E, P, Y, based on Formulas~\ref{formula:chap5-1}, ~\ref{formula:chap5-2}, ~\ref{formula:chap5-3} and ~\ref{formula:chap5-4}
 \State \quad \textbf{for} i = 1 to m \textbf{do}
 \State \quad \quad \quad \textbf{for} j = 1 to r \textbf{do}
 \State \quad\quad\quad \quad $U(i,j) \leftarrow U(i,j) \sqrt{\dfrac{D(i,j)}{E(i,j)}}$
 \State \quad\quad\quad \quad $H(i,j) \leftarrow H(i,j) \sqrt{\dfrac{P(i,j)}{Y(i,j)}}$
 \State  \quad \quad \quad \textbf{end for}
 \State \quad \textbf{end for}
\State \textbf{end While}
 \State return $U, H$
 \end{algorithmic}
 \label{alg: SETTrustAlgo}
\end{algorithm}

\textbf{In summary,}
we presented a context-aware trust prediction approach based on SET. It considers the context of trust, proposing some social context factors. When we have a pair of users (a source user and a target user), $SETTrust$ checks whether the target user is an expert in a specific context, whether both types of users have the same preference in that particular context, how many people follow the target user and what sort of behaviour the target user has (based on his or her use of swear words) in OSNs.
Finally, $SETTrust$ sets a trust degree between pairs of users and with the help of Algorithm~\ref{alg: SETTrustAlgo}, predicts pair-wise trust relations. 

\section{Experiments}
\label{ch:chapt5sec3}
\label{experiment}
In this section, we employ datasets (Ciao and Epinions), an evaluation metric (TPA) and baseline methods, as presented in Chapter~\ref{chap:DataandMetrics}.

\begin{table}
\centering
\caption{Experimental results on the Ciao dataset}
\begin{tabular}{ || c | c| c | c| c|| }
\hline
\hline
\backslashbox{Approach}{A\%} & 60\% & 70\% & 80\% & 90\% \\
\hline
\hline
SETTrust & 0.6 & 0.42 & 0.36 & 0.31 \\
\hline
Zheng & 0.398  & 0.367 & 0.3& 0.29\\
\hline
hTrust  & 0.322  & 0.319& 0.28& 0.259\\
\hline
sTrust & 0.141 & 0.139 & 0.102 & 0.060\\
\hline
MF &  0.123 &  0.118 & 0.117 & 0.109 \\
\hline
RS  &  0.095 &  0.078  & 0.062 & 0.029 \\
\hline
Random & 0.001  & 0.00097 & 0.00089& 0.0008\\
\hline
\hline
\end{tabular}
\label{tab:Chapter5-results1}
\end{table}

\begin{table}[t!]
\centering
\caption{Experimental results on the Epinions dataset}
\centering
\begin{tabular}{ || c | c| c | c| c|| }
\hline
\hline
\backslashbox{Approach}{A\%} & 60\% & 70\% & 80\% & 90\% \\
\hline
\hline
SETTrust & 0.55& 0.53 & 0.51 & 0.49    \\
\hline
Zheng & 0.426  & 0.402  & 0.386 & 0.341  \\
\hline
hTrust & 0.297 & 0.290 & 0.289 & 0.288  \\
\hline
sTrust & 0.243 & 0.241  & 0.238 & 0.236 \\
\hline
MF &  0.195 & 0.191  & 0.184 & 0.181 \\
\hline
RS  & 0.05  &0.0456  & 0.044 & 0.043  \\
\hline
Random &0.00023  &0.00019 &0.00015 & 0.0001  \\
\hline
\hline
\end{tabular}
\label{tab:Chapter5-results2}
\end{table}

\subsection{Experimental Results}
The experimental results on the Ciao and Epinions datasets are illustrated in Tables~\ref{tab:Chapter5-results1} and ~\ref{tab:Chapter5-results2}, respectively.
$SETTrust$ can reach its highest performance with $\lambda = 0.5$ and $\alpha = 0.1$. We investigate the effects of the different $\lambda$ values on $SETTrust$ in Section~\ref{sec:regulSEtttttrust}. 
Based on our experimental results, it can be observed that $SETTrust$ outperforms other approaches in both datasets.
For example, when the size of $A$ is 90\% in the Ciao dataset, $SETTrust$ has about 7\%, 20\%, 5 times, 3 times, 15 times and 387 times higher prediction accuracy compared to $Zheng$, $hTrust$, $sTrust$, $MF$, $RS$ and $Random$, respectively. For the other values of $A$ (60\%, 70\% and 80\%) $SETTrust$ also outperforms other approaches. 


\begin{figure*}[t!]
\centering
  \includegraphics[width=110mm]{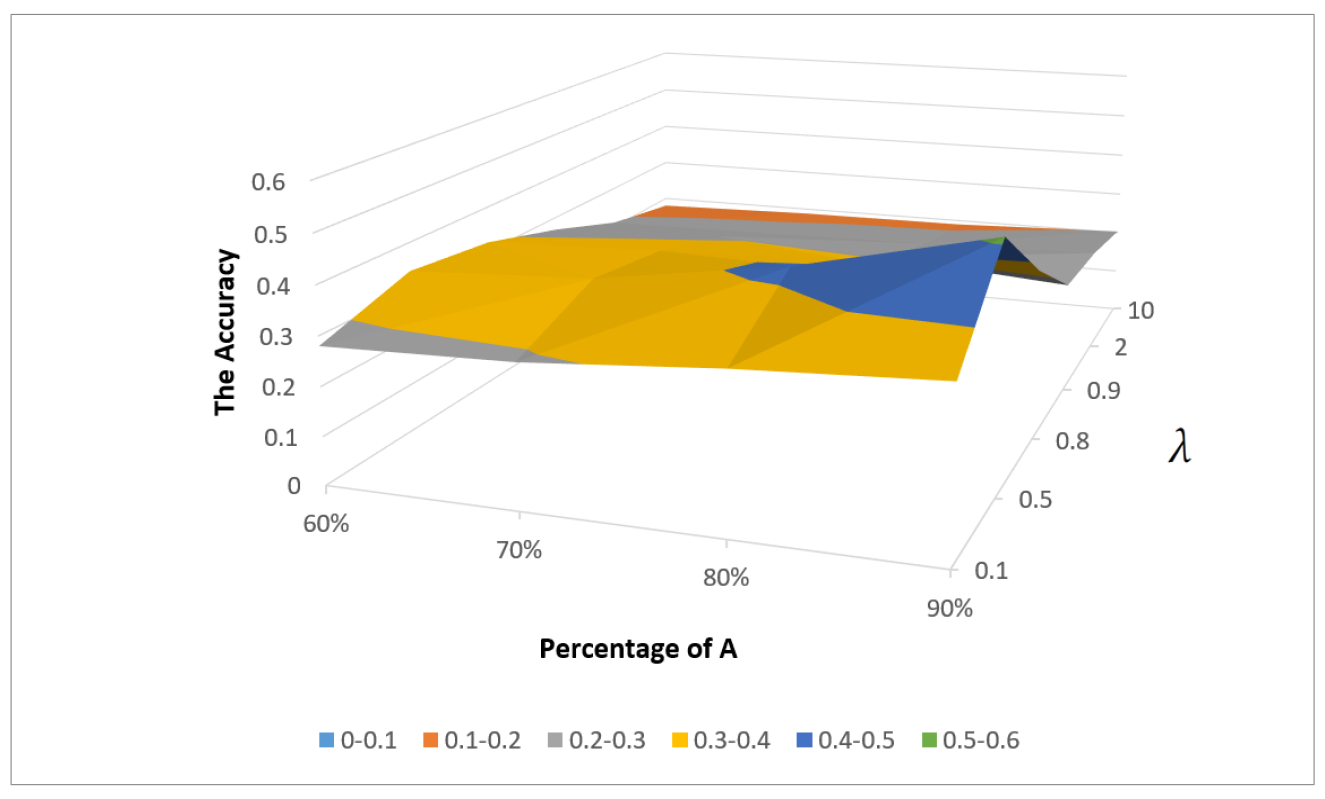}
  \caption{$SETrust$ regularisation effects in the Ciao dataset}
  \label{fig:Chapter5-regula1}
\end{figure*}
\begin{figure*}[t!]
\centering
  \includegraphics[width=110mm]{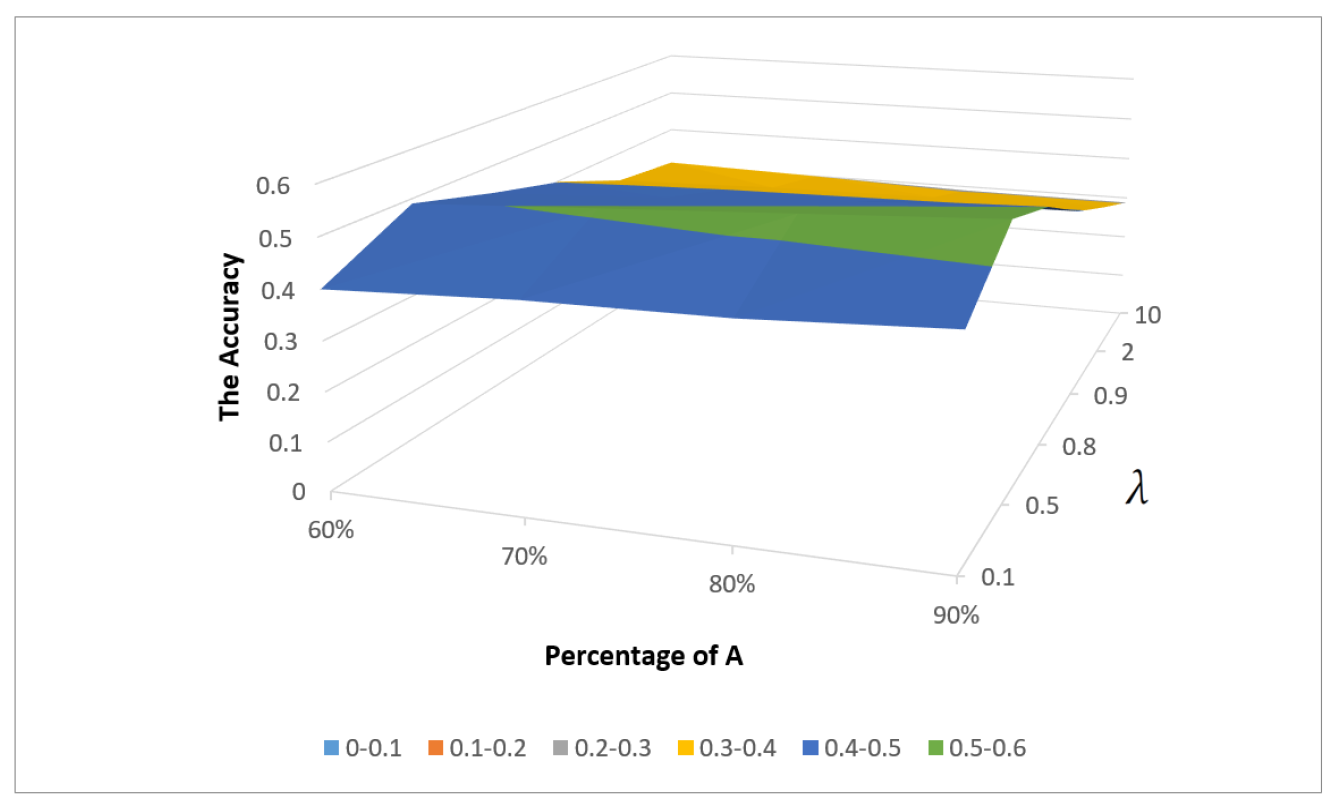}
  \caption{$SETrust$ regularisation effects in the Epinions dataset}
   \label{fig:Chapter5-regula2}
\end{figure*}

\subsection{The SETTrust Regularisation Effects}
\label{sec:regulSEtttttrust}
$\lambda$ is a controlling parameter for $SETTrust$ effects. Here, we investigate the effects of $\lambda$ on $SETTrust$'s prediction performance. Hence, we considered $\lambda =\{0.1, 0.5, 0.8,$
$0.9, 10\}$ and reported the prediction performance of $SETTrust$ in Figures~\ref{fig:Chapter5-regula1} and ~\ref{fig:Chapter5-regula2}, accordingly. The results illustrate that:
(i)~the highest accuracy of $SETrust$ is when $\lambda = 0.5$;
(ii)~whenever $\lambda$ increases from 0 to 0.5, we see an increase in the accuracy of $SETTrust$; and
(iii)~when $\lambda > 0.5$, the accuracy of $SETrust$ decreases, especially for $\lambda > 1$.

\textbf{In summary,}
the experimental results show that $SETrust$ has the highest prediction performance compared to the other approaches for all sizes of A ($A= \{60,70,80,90\}$).
$SETrust$'s high level of accuracy can be attributed to its consideration of the context of trust. If a source user trusts a target user in one context, $SETrust$ does not simply assume that trust carries over into contexts as well. 
However, the model still does not directly consider the context of trust, meaning that to predict the pair-wise trust relations in OSNs, we have to run $SETrust$ for each of the contexts of trust, separately.



\section{TDTrust: A Social Context-Aware Trust Prediction Approach}
~\label{ch:chapt5sec4}
Although our previous proposed approach, $SETrust$, has achieved a great prediction performance improvement over the state-of-the-art approaches, it could be further improved by making it capable of directly considering the context of trust and by employing some other contextual factors. In this section,
we propose a TD-based context-aware approach, $TDTrust$, for predicting pair-wise trust relations in OSNs.
\begin{figure}
\centering
  \includegraphics[width=\textwidth]{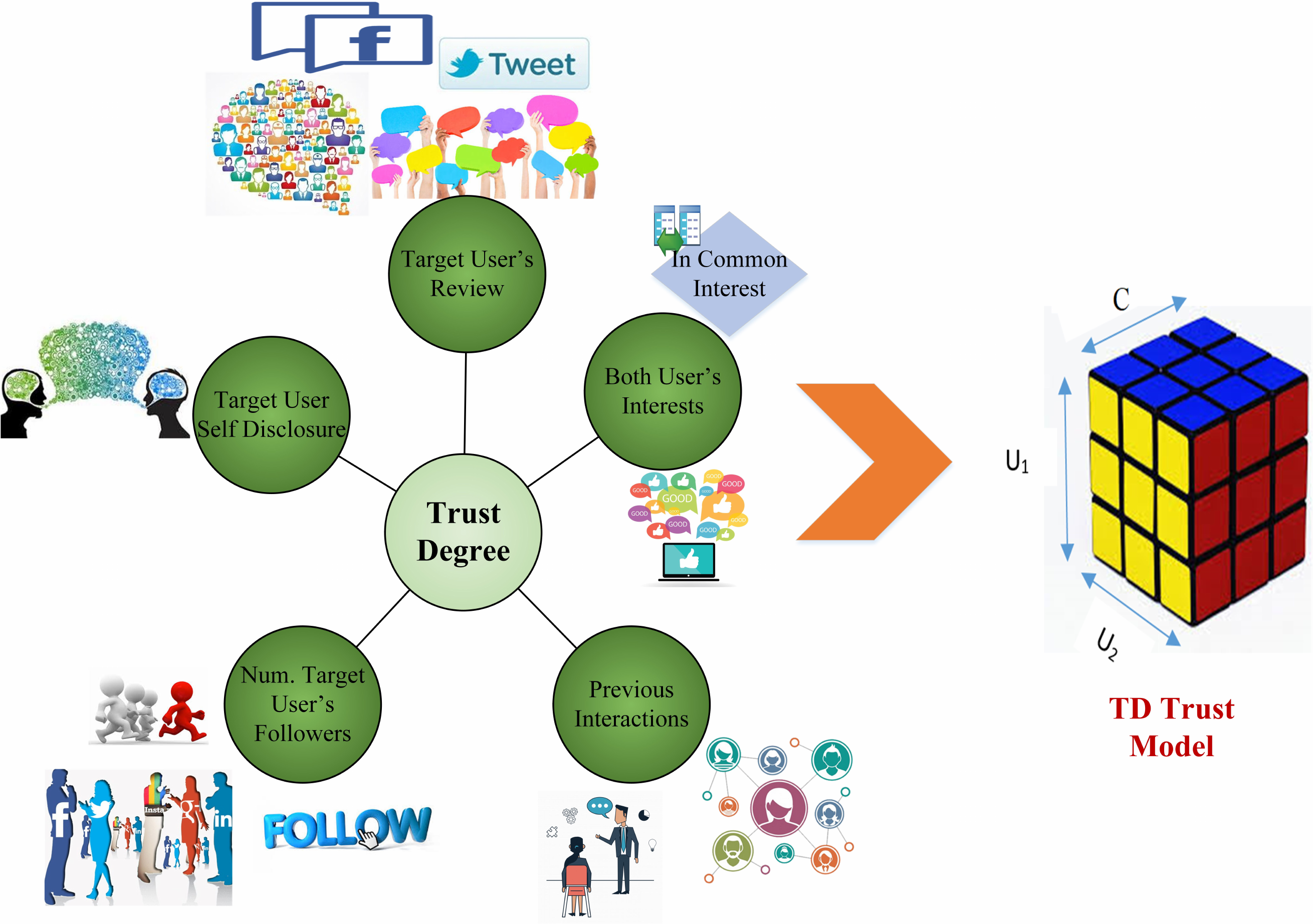}
  \caption{The proposed TDTrust framework}
  \label{fig: proposedTDTRUST}
\end{figure}

\subsection{Social Context}
In this section, we describe the social context factors that we employ in our trust prediction approach.

\subsubsection{Interaction-Based Social Context Factor}

 \textbf{\textit{Frequency and Quality of Previous Interactions (FQPI).}}
We consider the interactions that a source user has with a target user. If a source user highly rates/likes a target user's review/post, there is a high potential that he or she may trust that user. We calculate FQPI as follows:
\begin{equation}
\small
 FQPI_{{i},{j},{c_k}} = {\dfrac{1}{n_{e,{c_k}}}} \sum_{b=1}^{n_{e,{c_k}}} R_{b,{c_k}} ^e
\end{equation}

where $FQPI_{{i},{j},{c_k}}$ is the average value of the ratings ($R_{b,{c_k}}$) that $u_i$ gave to $u_j$ in the context $c_k$ and $n_{e,{c_k}}$ is the total number of ratings that $u_j$ received from $u_i$ in $c_k$. 

\subsubsection{Social Context Factors Related to User Behaviour and Demographic Features}
\label{sec: SelfdisclouserTDTrust}

 \textbf{\textit{Self-disclosure.}}
Social penetration theory proposes that `as relationships develop, interpersonal communication moves from relatively shallow, non-intimate levels to deeper, more intimate ones'~\cite{Altman/1973}. Based on this theory, self-disclosure---sometimes known as self-presentation, and defined as
revealing personal information such as personal motives or desires, feelings, thoughts and experiences to others~\cite{Altman/1973}---can be the reason that a relationship develops or intimacy level increases. Moreover, since there is a relationship between level of intimacy and level of trust~\cite{Yan/WWWJ/2015}, we can assume that self-disclosure may influence pair-wise trust relations.

We calculate the self-disclosure of a target user through analysing his or her textual contents in OSNs. We employ LIWC~\cite{LIWC} to identify words related to personal feelings, emotions and personal thoughts, which are related to self-disclosure in social presentation theory. We can evaluate the level of self-disclosure of user $u_i$ in the context $c_k$  with the following formula:
\begin{equation}
Sd_{jc_k}= {\dfrac{self_{jc_k}}{W_{j,{c_k}}}}
\end{equation}

where $self_{jc_k }$ is the number of self-disclosure related words that $u_j$ wrote in the context $c_k$ and $W_{j,{c_k}}$ is the total number of words that he or she wrote in his posts/reviews in the context $c_k$.
In addition to the self-disclosure context factor described in this section, we also consider the context factors of level of expertise, interest and number of followers, which are defined as in Section~\ref{sec:levelofexpertise}.

\subsection{Trust prediction Mechanism}

We propose the following formula to set a trust degree between pairs of users in OSNs:
\begin{equation}
B_{ijc_k}=w_1 (P_{ic_k} \times P_{jc_k})+ w_2 \times v_{jc_k}+w_3 \times NoF_j^{'}+ w_4 \times Sd_{jc_k} + w_5 \times FQPI_{ijc_k}
\end{equation}

where $B_{ijc_k}$ is the trust degree that we employ to predict the trust value between $u_i$ and $u_j$ in the context $c_k$ and $w_z$, $z=\{1,\dotsm,5\}$, is a controlling parameter to control the impact of social context parameters.
We use $(p_{ic_k} \times p_{jc_k})$ to ensure both users have the same interest.
Since $NoF_j$ could be a large number, we normalise it (as we did for $FQPI$) to the range of 0 and 1 by feature scaling:
\begin{equation}
\small
NoF_j^{'}={\dfrac{NoF_j - min(NoF)}{max(NoF)-min(NoF)}}
\end{equation} 


\subsubsection{Problem Statement}

In this section, we assume that we have $m$ users $U = \{u_1,\dotsm, u_m\}$ and $k$ context of trust as $C = \{c_1, c_2,\dotsm, c_k\}$. We also assume that $G \in R^{n\times n \times k}$ is a three-way tensor that contains the trust relations among users. $G(i, j, k) = 1$ means $u_i$ trusts $u_j$ in the context of $c_k$, while $G(i; j; k) = 0$ indicates there is no trust relation between $u_i$ and $u_j$ in $c_k$.
We can have the following assumptions:
\begin{equation}
\begin{aligned}
\begin{split}
  B_{ijc_k} \times(C \odot U_{2j})U_{1i}^T =0  \\
  B_{ijc_k} \times(C \odot U_{2j})U_{1i}^T =1
\end{split}
\end{aligned}
\end{equation}

where $U_1$ and $U_2$ are the first and second users' dimensions of $G$, and $U_{1i}$ and $U_{2j}$ represent the $i^{th}$ user in the first dimension of $G$ and the $j^{th}$ user in the second dimension of $G$, respectively.
Based on Wang et al.~\cite{WangZLGSH13} and also considering the CPD/Parafac model~\cite{SidiropoulosLFH17} for learning three f-dimensional matrices, $U_1 \in R^{n\times f}$, $U_2 \in R^{n\times f}$ and $C \in R^{k\times f}$, the low-ranked representation of users can be calculated by the sum of the inner products of dimensions:
\begin{equation}
\small
 \tilde{G}= \sum_{r=1}^{f} U_{1_{u_1r}} U_{2_{u_2r}} C_{cr} = \textless U_{1_{u_1}},U_{2_{u_2}},C_c \textgreater
\end{equation} 

\subsubsection{Modelling Proposed Trust Prediction}

To solve any non-convex problem in tensors, we can fix some of the dimensions to change the problem to a linear one.
Hence, we propose the following formula to fix $U_1$:
\begin{equation}
\begin{aligned}
\begin{split}
\small
\lambda \times( \sum_{i}^{n} \sum_{k=1}^{m} ( min \{0,f(B_{ijc_k})((C \odot U_2)U_1^T )\})^2)&
\end{split}
\end{aligned}
\label{form: TDTrusttttt1}
\end{equation}

where $\lambda$ is a controlling parameter. The same procedure can be used to fix $C$ and $U_2$. Then, based on CPD/Parafac~\cite{WangZLGSH13}~\cite{SidiropoulosLFH17}, a ranking decomposition approach for tensors~\cite{SidiropoulosLFH17}, we propose a TD model as follows:
\begin{equation}
\begin{aligned}
\begin{split}
&\ min_{U_1,C,U_2} \Vert G_1 -  (C \odot U_{2j} )U_{1i}^T \Vert ^2 _F + \lambda \times(\sum_{i}^{n} \sum_{j!=i}^{n} ( min \{0,f(B_{ijc_k })((C \odot\\
& U_{2j} )U_{1i}^T )\})^2)+ \alpha \times( \Vert U_1 \Vert ^2 _F + \Vert C \Vert ^2 _F + \Vert U_2 \Vert ^2 _F) \\
& U_1 \geq 0, U_2 \geq 0, C \geq0
\end{split}
\end{aligned}
\label{form: TDTrusttttt2}
\end{equation}

where $\alpha$ is a controlling parameter to control $U_1$, $C$ and $U_2$. Moreover, the Lagrangian function of the above formula would be:
\begin{equation}
\begin{aligned}
\begin{split}
\small
&L(G_1;U_1,C,U_2)=Tr((G_1 - (C\odot U_2)U_1^T)(G_1- (C\odot U_2)U_1^T)^T)+\\
&\lambda \times(Tr((B(C\odot U_2)U_1^T)(B(C\odot U_2)U_1^T)^T)+ \alpha \times Tr(U_1U_1^T))  \\
& + \alpha \times Tr(CC^T) + \alpha \times Tr(U_2U_2^T)
\end{split}
\end{aligned}
\label{form: TDTrusttttt3}
\end{equation} 

The same procedure with Formulas~\ref{form: TDTrusttttt1}, ~\ref{form: TDTrusttttt1} and ~\ref{form: TDTrusttttt3} should be repeated after fixing $C$ and $U_2$. Next, with the help of the alternating least squares algorithm~\cite{SidiropoulosLFH17}, we can update $U_1$, $C$ and $U_2$. We employ an updating rule as presented by Krompaas et al.~\cite{Nonnegativetensor}:
\begin{equation}
\small
 \Theta _i=\Theta _i \Bigg({\dfrac{\dfrac{\partial C(\Theta)^-}{\partial \Theta _i}}{\dfrac{\partial C(\Theta)^+}{\partial \Theta _i}}}\Bigg)^a
\end{equation}

where $C(\Theta)$ is the cost function and $\Theta$ is the variable that is not negative. The operators $X \bullet Y$ and $X/Y$, which are used in the following formula, are element-wise operations. Moreover, $(\partial C(\Theta)^-)/(\partial \Theta _i)$ and $(\partial C(\Theta)^+)/(\partial \Theta _i)$ are the negative and positive parts of the derivative, respectively. In addition, based on the partial derivative of $L$ with respect to $U_1$, $C$ and $U_2$ and the assumption that $L(G_1;U_1,C,U_2)/ \partial U_1 = 0$, $L(G_1;U_1,C,U_2)/\partial C = 0$, $L(G_1;U_1,C,U_2)/ \partial U_2= 0$, and since it is difficult to find the optimal solution for $U_1$, $U_2$ and $C$ simultaneously~\cite{TRIFUNOVIC/2010}~\cite{Lenhart/2010} with respect to the Karush Kuhn Tucker complementary condition, we leverage the approach presented by Tang et al.~\cite{TangBook} and propose an updating rule for $U_1$ (which can also be applied to $C$ and $U_2$ following the same procedure):
\begin{equation}
\resizebox{.9\hsize}{!}{$ U_1\longleftarrow U_1\bullet \Bigg(\dfrac{2G_1^T(C \odot U_2)}{(C \odot U_2)U_1(C\odot U_2)+ (C \odot U_2)^T U_1(C \odot U_2)+ \lambda \times B(C \odot U_2)U_1 B(C \odot U_2)+\lambda \times B^T (C \odot U_2)^T U_1 B(C \odot U_2) +2\alpha U_1 }\Bigg) $}
\end{equation}

The step-by-step process for $TDTrust$ can be found in Algorithm~\ref{alg: TDTrustAlgo}.

\begin{algorithm}[h]
\caption{Trust prediction with TDTrust}
\begin{algorithmic}[1]
 \State Input: G, $\lambda$, $\alpha$, k
 \State Output: $U_1, C, U_2$
 \State Establish the trust degree matrix in each context
 \State Randomly initialize $U_1 , U_2 , C$
 \State \textbf{While} {It is not the convergent state} \textbf{do}
 \State \quad  $D=-2G_1^T(C \odot U_2)$
 \State \quad  $E=(C \odot U_2)U_1(C\odot U_2)+ (C \odot U_2)^T U_1(C \odot U_2)+ B(C \odot U_2)U_1 B(C \odot U_2)+B^T (C \odot U_2)^T U_1 B(C \odot U_2) +2\alpha U_1$
 \State  \quad   \textbf{for} j = 1 to m \textbf{do}
 \State    \quad   \quad \textbf{for} r = 1 to k \textbf{do}
 \State   \quad   \quad \quad   $  U_1\longleftarrow U_1\bullet(\dfrac{D_{ijr}}{E_{ijr}})$
 \State    \quad     \quad \textbf{end for}
 \State  \quad   \textbf{end for}
  \State\quad  Do the same procedure for updating C and $U_2$
 \State \quad  \textbf{end for}
 \State \textbf{end While}
 \State return $U_1, C, U_2$
 \end{algorithmic}
 \label{alg: TDTrustAlgo}
\end{algorithm}

\section{Experiments}
~\label{ch:chapt5sec5}
In this section, we employ datasets (Ciao and Epinions), an evaluation metric (ranking based evaluation) and baseline methods, as presented in Chapter~\ref{chap:DataandMetrics}.

\subsection{Experimental Results}

The parameters of the approaches in this experiment were defined by applying cross-validation and are set as $\lambda =0.5$, $\alpha = 0.1$ and f=100. Our approach reaches its highest performance when $\alpha = 0.1$. We investigate the effects of the different $\lambda $ values on $TDTrust$ in the following subsections.

\begin {table}
\begin{center}
\caption {Experimental results on the Ciao dataset}
\label{tabel:ciaotd}
\begin{tabular}{ || c | c| c| c| c | c| c|c|c|c|| }
\hline
A\%   & Random & RS& MF &sTrust & hTrust & Zheng & SETTrust& TDTrusr \\
\hline
\hline
60\% &  0.001  & 0.095& 0.123  &0.141 & 0.322 & 0.398  & 0.601 & 0.685 \\
\hline
70\% & 0.00097 & 0.078 & 0.118 &0.139 & 0.319  & 0.367 & 0.421 & 0.591 \\
\hline
80\% &  0.00089  & 0.062 & 0.117  &0.102  & 0.288  & 0.306 & 0.362& 0.456 \\
\hline
90\% & 0.0008 & 0.029& 0.109 &0.060 & 0.259 & 0.29 & 0.31 & 0.395 \\
\hline
\end{tabular}
\end{center}
\end {table}

\begin {table}
\begin{center}
\caption {Experimental results on the Epinions dataset}
\label{tab:TDTrustEpinionsRes}
\begin{tabular}{ || c | c| c| c| c | c| c|c|c|c|| }
\hline
A\%   & Random & RS& MF &sTrust & hTrust & Zheng & SETTrust& TDTrusr \\
\hline
\hline
60\%  & 0.00023  & 0.051 & 0.195  &0.243  & 0.297 & 0.426  & 0.550 & 0.630 \\
\hline
70\%  & 0.00019 & 0.045 & 0.191  &0.241  & 0.290 & 0.402  & 0.532 & 0.594 \\
\hline
80\%  & 0.00015  & 0.044& 0.184 &0.238  & 0.289 & 0.386 & 0.517& 0.552 \\
\hline
90\%  & 0.0001 & 0.043& 0.181 &0.236 & 0.288 & 0.341 & 0.498& 0.523 \\
\hline
\hline
\end{tabular}
\end{center}
\end {table}


\subsubsection{Ciao Dataset}

As Table~\ref{tabel:ciaotd} demonstrates, as $A$ increases, so does the accuracy of the models. Moreover, in all experiments, $TDTrust$ outperformed other approaches in the Ciao dataset. For example, when $A$ is 60\%, $TDTrust$ has around 12\% and 43\% better prediction performance compared to $SETTrust$ and $Zheng$, respectively. $TDTrust$'s prediction performance is also twice that of $hTrust$, five times that of both $sTrust$ and $MF$ and significantly better than both $RS$ and $Random$. 
For other sizes of $A$ (70\%, 80\%, and 90\%), $TDTrust$ likewise outperforms the other approaches.

\begin{figure*}
\centering
  \includegraphics[width=0.7 \textwidth]{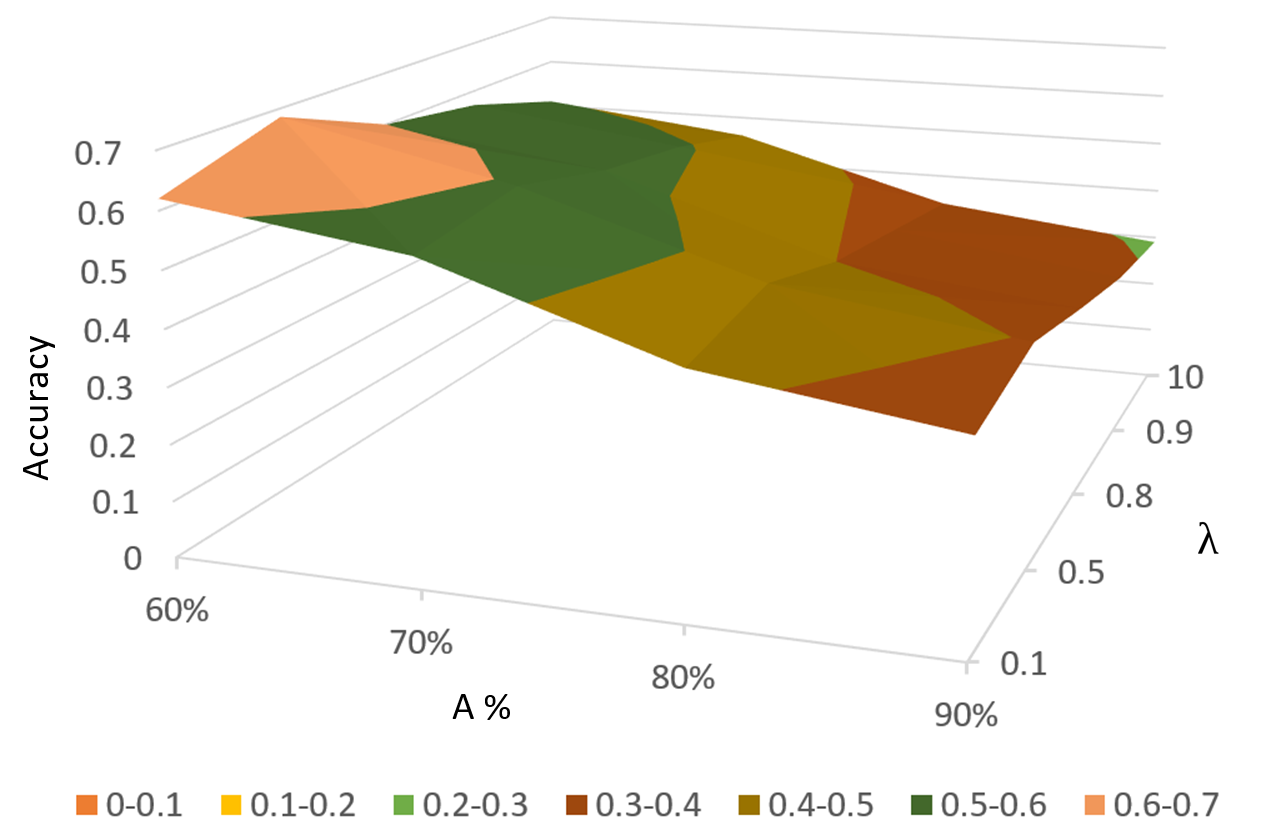}
  \caption{$TDTrust$ regularisation effects in the Ciao dataset}
   \label{fig:TDTrustregularization1}
\end{figure*}

\begin{figure*}
\centering
  \includegraphics[width=0.7 \textwidth]{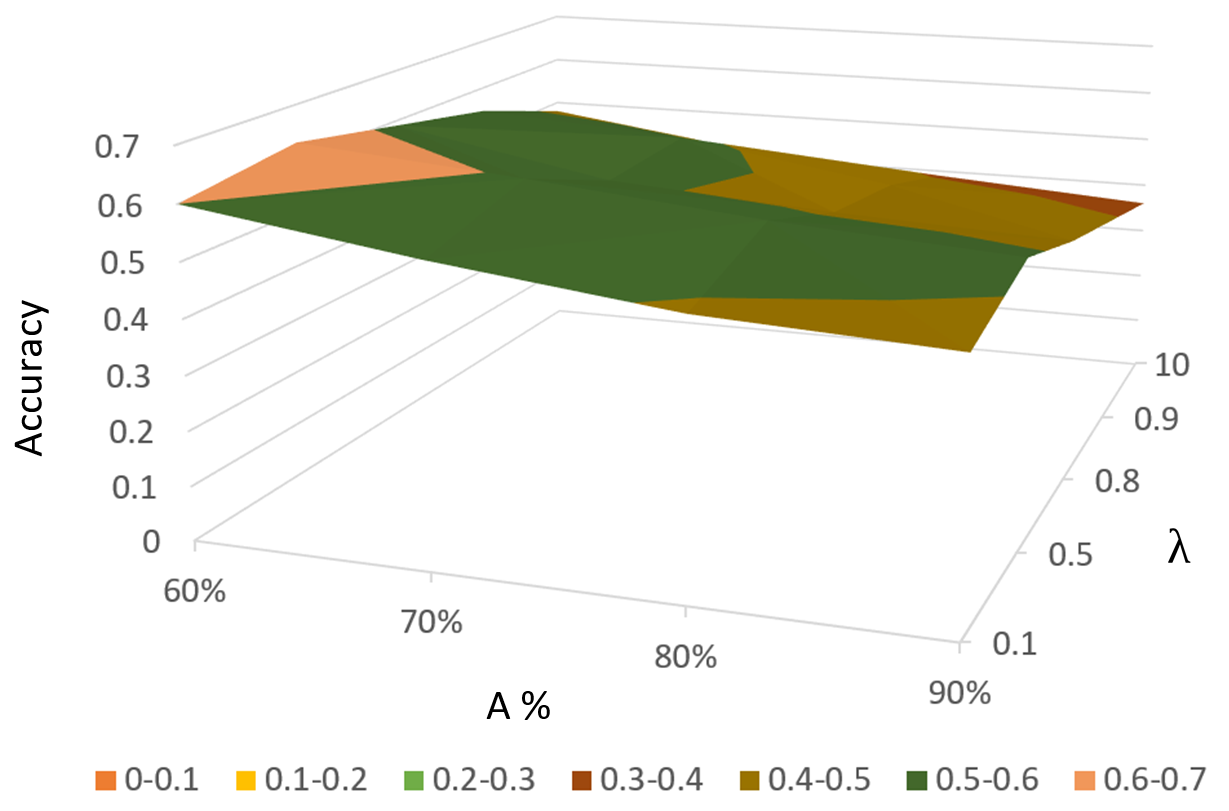}
  \caption{$TDTrust$ regularisation effects in the Epinions dataset}
   \label{fig:TDTrustregularization2}
\end{figure*}

\subsubsection{Epinions Dataset}

As Table~\ref{tab:TDTrustEpinionsRes} demonstrates, for the Epinions dataset, as $A$ increases, so does the prediction performance of the models. Moreover, in all experiments, $TDTrust$ gave better prediction performance than the other approaches. For example, when $A$ is 60\%, the prediction performance of $TDTrust$ is around 13\% and 36\% higher compared to $SETTrust$ and $Zheng$, respectively. Similarly, $TDTrust$'s prediction performance is twice that of $hTrust$, three times that of both $sTrust$ and $MF$ and significantly better than both $RS$ and $Random$. 
For other sizes of $A$ (70\%, 80\%, and 90\%), $TDTrust$ also has a better prediction performance compared to the other approaches.

\subsection{TDTrust Regularisation Effects}

We use $\lambda$ as a controlling parameter to control our proposed trust prediction model effects. In this section, we investigate the effects of this parameter on $TDTrust$'s prediction performance. To do so, for different sizes of $A$, we consider different values for $\lambda =\{0.1, 0.5, 0.8, 0.9, 10\}$. Our experiments demonstrate that:
(i)~the best performance of $TDTrust$ is achieved when $\lambda = 0.5$;
(ii)~$TDTrust$'s accuracy increases when $\lambda$ increases from 0 to 0.5; and
(iii)~for $\lambda > 0.5$, we see a gradual performance degradation in $TDTrust$, especially when $\lambda > 1$. Figures~\ref{fig:TDTrustregularization1} and ~\ref{fig:TDTrustregularization2} illustrate the prediction performance of $TDTrust$ when applying different values for $\lambda$.

\textbf{In summary,} in contrast to $SETTrust$, $TDTrust$ directly considers the context of trust in its mathematical model. $TDTrust$ employs a three-dimensional TD model for its trust prediction procedure. It also uses social context factors inspired by social psychology theories to capture the context of trust in OSNs.
The experimental results on two real-word datasets (Ciao and Epinions) demonstrate that $TDTrust$ has the highest prediction performance compared to other approaches.  
However, although $SETTrust$ and $TDTrust$ both have a high prediction performance, we further propose a context-aware trust prediction approach modelled on a deep network structure. Deep learning techniques are very popular in the big data era. These techniques allow for the incremental learning of high-level features, reducing the reliance on domain expertise and hard core feature extraction typical of traditional machine learning techniques~\footnote{https://towardsdatascience.com/why-deep-learning-is-needed-over-traditional-machine-learning-1b6a99177063}. Surprisingly, there has been almost no attempt in the pair-wise trust prediction area to employ deep neural network structures in predicting trust relations among users.




\section{DCAT: A Deep Context-Aware Trust Prediction Approach}
~\label{ch:chapt5sec6}
In this section we aim to propose a deep learning-based approach, $DCAT$, which considers users as nodes and their trust relations as the edges of a graph. Next, with the help of a graph convolutional network-based model, DCAT, as a deep classifier, tends to predict trust relations among users. 
\section{Background}
~\label{ch:chapt5sec7}
\subsection{Generalising Convolutions in the Graph Domain}
Studies in this domain can be classified into two approaches: spectral and non-spectral~\cite{GNN}. 

\subsubsection{Spectral Approaches.}
Spectral approaches focus on a spectral representation of graphs and have been used for node classification. Bruna et al.~\cite{JoanBruna} define convolution operation in the Fourier domain by `computing the eigen decomposition of the graph Laplacian'~\cite{GNN}. Defferrard et al.~\cite{Defferrard} introduced a model based on a Chebyshev expansion~\cite{Chebyshev} of the graph
Laplacian. Finally, Kipf and Welling~\cite{Kipf} limited filters to work in a one-step neighbourhood near nodes.
\subsubsection{Non-Spectral Approaches.}
Non-spectral approaches focus on groups of spatially close neighbours~\cite{GNN}. Monti et al.~\cite{Monti} developed a spatial approach called MoNet, that proposes a `unified generalization of CNN architectures to graphs'~\cite{GNN}. Hamilton et al.~\cite{Hamilton} proposed a model, named GraphSAGE, that focuses on `sampling a fixed-size neighborhood of each node and then performing a specific aggregator over it'~\cite{GNN}. GraphSAGE is this latter model that we use to design our deep learning-based structure. 

\subsubsection{GraphSAGE}

To have low-dimensional embeddings of nodes in large graphs, GraphSAGE, proposed by Hamilton et al.~\cite{Hamilton}, `allows embeddings to be efficiently generated for unseen nodes' for prediction tasks ~\cite{Hamilton}.
`GraphSAGE is an inductive variant of GCNs'~\cite{application1}.
Most existing approaches for creating low-dimensional embeddings of nodes either `require that all nodes in the
graph are presented during training of the embeddings' or `focus on embedding nodes from a single fixed graph'~\cite{Hamilton}. GraphSAGE does not have these shortcomings. It uses node-feature
information to generate node embeddings in the case of previously unseen data~\cite{Hamilton}.  
GraphSAGE applies a `graph-based loss function to the output representations and tune[s] the weight matrices and parameters of the aggregator functions via stochastic gradient descent'~\cite{Hamilton}. GraphSAGE investigates three
candidate aggregator functions:

\begin{itemize}
    \item Mean aggregator: this function takes the element-wise mean of the node's vectors~\cite{Hamilton}.
    \item LSTM aggregator: this function is based on LSTM architecture~\cite{LSTM} which may have a larger expressive capability~\cite{Hamilton}.
    \item Pooling aggregator: a pooling approach in which `each neighbor's vector is independently fed through a fully-connected neural network; following this transformation, an element-wise max-pooling operation is applied to aggregate information across the neighbor set'~\cite{Hamilton}.
\end{itemize}

 \subsubsection{Applications of GraphSAGE}
Despite GraphSAGE having been proposed only recently, it has attracted a great deal of attention and several studies have employed it in different domains. For instance, Ying~\cite{application1} proposed a large-scale deep recommendation engine based on GraphSAGE, while Haija~\cite{application2} based their multi-scale graph convolution
for semi-supervised node classification on GraphSAGE. Further, GraphSAGE's application for characterising and detecting hateful users on twitter has been investigated by Ribeiro et al.~\cite{application3}, and Shchur et al.~\cite{application4} employed GraphSAGE for semi-supervised node classification and GNNs evaluation.\\

\begin{figure*}[h]
\centering
  \includegraphics[width=0.99 \textwidth]{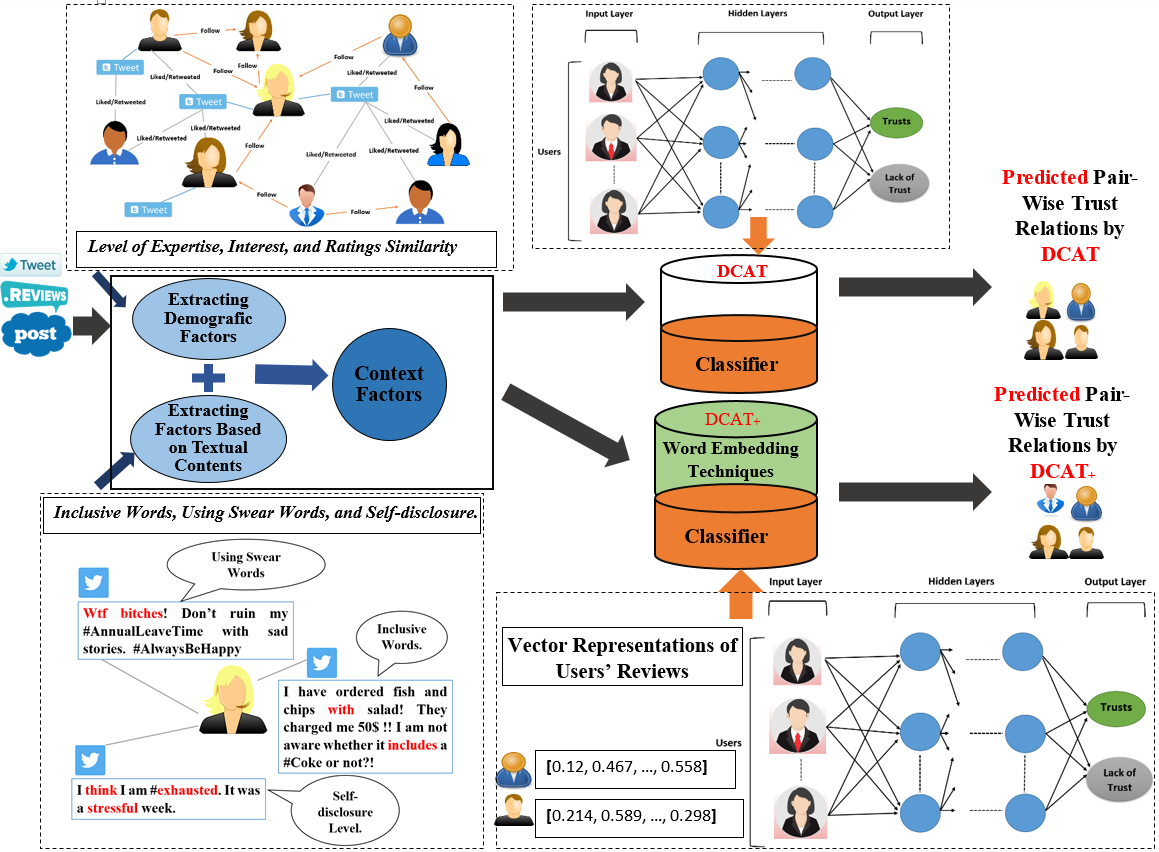}
 \caption{Our proposed framework: We first focus on extracting some context factors based on demographic features and textual contents. Then, with the help of a deep classifier, we predict the pair-wise trust relations and call this approach $DCAT$. Although we have some text analysis to extract textual content-based context factors, we further analyse the textual footprint of users by focusing on their reviews' textual content. We use word-embedding techniques and convert the reviews of users in a particular context to vectors of numbers. Next, we add these vectors to our designed classifier's feature set and call this version of the classifier $DCAT_{+}$. Finally, we compare the performance of $DCAT$ and $DCAT_{+}$ with other state-of-the-art trust prediction approaches.}
  \label{fig:DCATarch}
\end{figure*}

\section{Proposed Framework}
~\label{ch:chapt5sec8}
Figure~\ref{fig:DCATarch} illustrates our proposed framework. We first capture context factors based on demographic and textual contents (indicated in the blue circles). The demographic factors are based on level of expertise, interest and rating similarity. The textual content features use inclusive words, swear words and self-disclosure. Factors are then incorporated into classifiers (DCAT and $DCAT_{+}$) (indicated by the cylinders). Word embedding techniques (indicated in green) use vector representations from user reviews.
We describe the factors and classifier in the forthcoming sections. Assume that we have $n$ user $U=\{u_1,u_2,...,u_n\}$ and m contexts of trust $C=\{c_1,c_2,...,c_m\}$, and that $c_d$ indicates the $d^{th}$ context of trust.


\subsection{User Modelling}
\textbf{Classifier}. `Classification is a task to specify which of K categories some input belongs to'~\cite{Goodfellow}. To identify whether a pair of users trusts each other, we design a binary classifier. In this section, we present our deep learning-based classifier (Figure~\ref{fig:DCATarch}). This classifier uses Stellargraph\footnote{Stellargraph is a python API that implements machine learning for graph analytics, available at: \url{https://github.com/stellargraph/stellargraph}.}  library~\cite{StellarGraphReference}~\cite{StellarGraphReference} and employs GraphSAGE which can be used for inductive node embedding~\cite{Hamilton}. GraphSAGE is mainly based on MF and can be used to `learn an embedding function that generalizes to unseen nodes"~\cite{Inclusive}. GraphSAGE can be employed for node classification and link prediction in homogeneous networks. We consider the users as nodes of a graph, users' characteristics as the node attributes and trust relations as the edges of the graph (the absence of a link between two users means they do not trust each other). We divide users' characteristics into two sub-categories---demographic factors and textual contents-based factors---which can be used to capture the context of trust relations among users. We can further divide these characteristics based on whether they relate to individuals or pairs of users:

\textbf{Individual User's Features:}
These kinds of features are related to users' personal characteristics. Hence, we consider level of expertise, self-disclosure level, using swear words and using inclusive words as features of individual users. 

\textbf{Pair-wise Features:}
Pair-wise features are those features established between two users. For instance, we consider similar interest and rating similarity as pair-wise features. 

\subsection{Demographic Factors}
\textbf{Rating Similarity.}
Based on homophily theory, `similar people
have a higher likelihood to establish trust relations'~\cite{TangGHL13}. Here, we use the rating similarity (RS) metric~\cite{TangGHL13}, which is a measure for evaluating users' similarity. $RS$ can be used to `capture
different tastes from different users'~\cite{TangGHL13},
as well as users' ratings on similar items. In other words, it captures the tastes of different users on similar items, as follows~\cite{TangGHL13}:
\begin{equation}
RS_{i,j,c_{d}}= \frac{ \sum_q^T Rating_{iqc_{d}} \times Rating_{jqc_{d}}}{\sqrt{\sum_q^T Rating_{iqc_{d}}^2} \times \sqrt{\sum_q^T Rating_{jqc_{d}}^2}},
\end{equation}

where $RS_{i,j,c_{d}}$ is the cosine similarity between the ratings of $u_i$ and $u_j$ in the $d^{th}$ context of trust, $Rating_{iqc_{d}}$ is the rating that $u_i$ gave to the $q^{th}$ item in the $d^{th}$ context of trust. $Rating_{jqc_{d}}$ is the rating that $u_j$ gave to the $q^{th}$ item in the $d^{th}$ context of trust, and $T$ is the total amount of feedback that $u_i$ and $u_j$ gave to common items in the $d^{th}$ context of trust.  
We also consider the level of expertise of target users ($LevelofExpertise_{j,c_d}$) and the similarity in the interests of users ($Int_{i,j,c_d}$) as other demographic features for capturing the context of trust. These are defined as in Section~\ref{sec:levelofexpertise}.

\subsection{Textual Contents-Based Factors}

\textbf{Inclusive Words.} An inclusive environment, using inclusive language, can increase trust among people~\cite{Inclusive} and increase the diversity of a user's audience\footnote{Department of Education: https://documentcentre.education.tas.gov. \\
au/Documents/Guidelines-for-Inclusive-Language.pdf}. 
We propose a formula to capture the level of inclusive word usage, such as `with', `and', `include', `along' and `also'.
We use LIWC~\cite{LIWC} to detect inclusive words, and calculate the level of inclusive word usage as follows:
\begin{equation}
Inclusivec_{j,c_{d}}= \frac{Inc_{j,c_{d}}}{Total_{j,c_{d}}}, 
\end{equation} 

where $Inc_{j,c_{d}}$ represents the number of inclusive words that the $j^{th}$ user wrote in $d^{th}$ context of trust, $Total_{j,c_{d}}$ is the total number of words they wrote in the $d^{th}$ context of trust, and $Inclusivec_{j,c_{d}}$ is their inclusive word usage in the $d^{th}$ context of trust. In addition to the level of inclusive word usage, we also consider the level of self-disclosure ($Selflevel_{j,c_d}$) and of using swear words ($Swearlevel_{j,c_d}$). The definition of self-disclosure can be found in Section~\ref{sec: SelfdisclouserTDTrust}, while the definitions of bad language detection can be found in Section~\ref{sec:levelofexpertise}.

\subsection{$DCAT_{+}$: Improving $DCAT$ with User Embeddings}
 We use word-embedding techniques and convert the reviews of a user in a particular context to a vector of numbers. We call this vector $W2V_{j,c_d}$, which is the word representation of the reviews of a user in the context $c_d$.
 We employ Word2Vec~\cite{word2vec} to capture the vector representations of words, known as word-embedding. To generate users' embeddings, we first gather all the reviews of a user in a particular context. Then, using Word2Vec we create vectors of numbers from each review. Next, we average all these vectors and create a final vector for each user. 
Finally, we add this vector to the individual user's features. Our aim is to investigate whether using numeric vectors representing the reviews of users has a  positive effect on the performance of $DCAT$. We call this classifier $DCAT_{+}$.

\subsection{A Deep Classifier}
After modelling users' characteristics, we introduce our deep structure. We first create a user graph, inspired by~\cite{WSDMPersonality2vec}. For our deep trust prediction platform, we leverage GraphSage~\cite{Hamilton} which is a GNN model, capable of learning node embeddings in an inductive way. In GraphSage, each node is represented by the aggregation of its neighbourhoods. GraphSage can represent a node based on its neighbourhood even if that node was not part of the training set.

\subsubsection{Aggregator Architecture}

Gathering input from previous neurons is called the aggregation function. Inspired by \cite{Hamilton}, we assume that we have learned the parameters of $D$ (denoted as search depth) aggregator functions ($AGGREGATE_d$), where $d={1, 2, ..., D}$, which is neighbourhood nodes' aggregate information~\cite{Hamilton}:
\begin{equation}
\small
\begin{aligned}
h_{N(v)} ^d  \leftarrow	AGGREGATE_d (\{h_u ^{d-1} , \forall u \in N(v) \})
\end{aligned}
\end{equation}

We use Mean aggregator~\cite{Hamilton} for our aggregation function. This function calculates the average of the latent vectors of a node and its neighbourhood nodes: 
\begin{equation}
\small
\begin{aligned}
h_v ^d  \leftarrow	\sigma (W . MEAN(\{h_v ^{d-1}\} \cup \{h_u ^{d-1} , \forall u \in N(v)\}))
\end{aligned}
\end{equation}

where \textit{G=(V,E)} is our graph with $V$ as its nodes and $E$ as its edges, and $d$ denotes the depth of the search of Algorithm~\ref{alg:DCAT}. Moreover, $h_v ^d$ is the representation of a node, $W$ is a weight matrix and $h_v ^{d-1}$ denotes the node representation of the immediate neighbourhood of the node. $N(v)$ is a set of users that are different in each iteration of $d$. 

\subsubsection{Learning the Parameters}
To optimise our approach, we use a graph-based loss function, as follows~\cite{Hamilton}:
\begin{equation}
\small
\begin{aligned}
L_G(Z_u)=-log (\sigma  (Z_u^T Z_v)) -Q . E_{v_n \sim P_n (v)} log (\sigma (-Z_u^T Z_{v_n}))
\end{aligned}
\end{equation}

where $v$ is a node `that at co-occurs near u on fixed-length random walk'~\cite{Hamilton}, and $\sigma$ denotes the sigmoid function. Further, $P_n$ is a negative sample distribution and $Q$ is the number of these samples. $Z_u$ is the output representations. This loss function `encourages nearby nodes to have similar representations, while enforcing that the representations of disparate nodes are highly distinct'~\cite{Hamilton}. We also use the sigmoid function as the activation function at the output layer, as:
\begin{equation}
\small
\begin{aligned}
\sigma (x)=1 / (1+e_{-x})
\end{aligned}
\end{equation}

where $x$ is the input vector. 
Our model was implemented in TensorFlow~\cite{TensorFlow} with the Adam optimiser~\cite{ADAM}. Algorithm~\ref{alg:DCAT} shows our proposed approach.

\begin{algorithm}[t!]
\caption{ Trust Prediction with $DCAT$ and $DCAT_{+}$}
\begin{algorithmic}[1]
\State \textbf{Input}: n (number of users), m (number of contexts), users' reviews, depth D, E
\State \textbf{Output}: Vector representations $Z_v$ for all $v \in V$.
 \State Collect users' textual footprints
 \State \textbf{for} d=1 to m
 \State \quad \textbf{for} i=1 to n \textbf{ do} 
 \State  \quad \quad Evaluate $LevelofExpertise_{j,c_d}$, $Inclusivec_{j,c_d}$, $Selflevel_{j,c_d}$, $Swearlevel_{j,c_d}$ and $W2V_{j,c_d}$
 \State \quad \quad \textbf{for} j=1 to n \quad  \textbf{do }    
 \State  \quad \quad  \quad \textbf{if} i!=j  \quad \textbf {then}
 \State   \quad \quad \quad \quad Evaluate the $Int_{i,j,c_d}$
 \State   \quad \quad \quad \quad Evaluate the $RS_{i,j,c_d}$
  \State  \quad \quad  \quad \textbf{end if}
 \State  \quad \quad \textbf{end for}   
 \State  \quad \textbf{end for}       
 \State  \textbf{end for}  
 \State Having nodes as V (users), node attributes (user modelling), and trust relations as E, create the representation graph as G=(V,E).
  \State  \textbf{for} d=1...D \textbf{do}
 \State  \quad \textbf{for} $v \in V$ \textbf{do}
 \State  \quad \quad $h_{N(v)} ^d  \leftarrow	AGGREGATE_d (\{h_u ^{d-1} , \forall u \in N(v) \})$
 \State  \quad \quad  $h_v ^d  \leftarrow	\sigma (W . MEAN(\{h_v ^{d-1}\} \cup \{h_u ^{d-1} , \forall u \in N(v)\}))$
  \State  \quad \textbf{end} 
 \State  \textbf{end}
 \State $Z_v \leftarrow h_v^K, \forall v \in V$
 \State  Trust relations prediction with DCAT
 \State  Add $W2V_{i,c_d}$ to the feature set of $DCAT$ and predict trust relations with $DCAT_{+}$
  \State Return predicted trust relations in different contexts
 \end{algorithmic}
 \label{alg:DCAT}
\end{algorithm}

\section{Experiments}
~\label{ch:chapt5sec9}
In this section, we employ datasets (Ciao and Epinions), evaluation metrics (MAE and RMSE) and baseline methods, as presented in Chapter~\ref{chap:DataandMetrics}.

\subsection{Training Setup}
$DCAT$ has been trained and tested by cross-validation and random fraction, using p=0.4, p=0.3, p=0.2, p=0.1 samples from positive links (i.e., users who trust each other)---where p is the size of the sample validation data with respect to the dataset size---and the same number of samples from negative links (i.e., users who do not trust each other)~\cite{StellarGraphReference}, resulting in respectively splits of 60\%-40\%, 70\%-30\%, 80\%-20\% and 90\%-10\% between training and test data\footnote{This split could have been performed in terms of other parameters such as network density, however, in this dissertation, we have done it only in terms of the number of links and users.}. To configure our classifier, we used the default values applied in GraphSAGE and we employed a Mean aggregator for our experiments. Moreover, for $DCAT_{+}$, we initialised our word-embedding with pre-trained word vectors, like 300d Glove on 6B Wikipedia 2014 and Gigaword 5, 300d Glove on 42B Common Crawl and the 27B Twitter corpus. For a valid comparison, the training size and validation size for evaluating $DCAT_{+}$ were similar to in the $DCAT$ evaluation process.

\begin{table*}[t!]
\centering
\caption{Comparison of trust relation prediction performance on the Ciao and Epinions datasets based on the MAE metric; Lower is better.
  }
\begin{tabular}{llllllll}
\toprule
 \textbf{Dataset} & \textbf{sTrust} & \textbf{hTrust} & \textbf{Zheng} &  \textbf{SETTrust}&\textbf{TDTrust} & \textbf{DCAT} &\textbf{$DCAT_{+}$}\\  \midrule
Ciao  & 1.80& 1.17 &1.07 & 0.97& 0.89 &\textbf{0.37}& 0.44 \\
Epinions  &  1.47 & 1.36 & 1.30 & 1.29 & 1.20 &\textbf{0.42}& 0.51 \\
\bottomrule
\end{tabular}
   \label{tab:DCATMAE}
\end{table*}

\begin{table*}[t!]
\centering
\caption{Comparison of trust relation prediction performance on the Ciao and Epinions datasets based on the RMSE metric; Lower is better.
  }
\begin{tabular}{llllllll}
\toprule
 \textbf{Dataset} & \textbf{sTrust} & \textbf{hTrust} & \textbf{Zheng} &  \textbf{SETTrust}&\textbf{TDTrust} & \textbf{DCAT} &\textbf{$DCAT_{+}$}\\  \midrule
Ciao  & 1.93 & 1.32 & 1.19 & 1.10 & 1.04 &\textbf{0.46}& 0.55 \\
Epinions  &2.15 & 1.56 & 1.31 &1.22 &1.15 &\textbf{0.57}& 0.67 \\
\bottomrule
\end{tabular}
   \label{tab:DCATRMSE}
\end{table*}

\begin{table*}[t!]
\centering
\caption{Comparison of the performance of $DCAT$ and $DCAT_{+}$ on the Ciao and Epinions datasets based on the MAE and RMSE metrics; Lower is better.
  }
\begin{tabular}{llllllll}
\toprule
 \textbf{Approach} & \textbf{Dataset} &\textbf{Metric} & \textbf{60\%} & \textbf{70\%} &  \textbf{80\% } & \textbf{90\%}\\  \midrule
 $DCAT$  & Ciao& MAE &\textbf{0.38}&  \textbf{0.38}  & \textbf{0.37}   &\textbf{0.36} \\
$DCAT_{+}$ & Ciao & MAE & 0.48  & 0.47   &0.43& 0.41 \\
 $DCAT$  & Ciao& RMSE & \textbf{0.51}&  \textbf{0.49}     &\textbf{0.44}& \textbf{0.41}  \\
$DCAT_{+}$   &Ciao & RMSE & 0.62  &0.56 &0.53& 0.50 \\
 $DCAT$  & Epinions& MAE &\textbf{0.46}&\textbf{0.44}&\textbf{0.41}&\textbf{0.40} \\
$DCAT_{+}$   & Epinions&  MAE&  0.54 &0.52 &0.52& 0.49 \\
 $DCAT$  & Epinions&  RMSE& \textbf{0.69} & \textbf{0.62} &\textbf{0.51}& \textbf{0.48} \\
$DCAT_{+}$   & Epinions&  RMSE&  0.76 & 0.71 & 0.64 & 0.58 \\
\bottomrule
\end{tabular}
   \label{tab:DCATDCAT+}
\end{table*}

\subsection{Results}

Tables~\ref{tab:DCATMAE} and ~\ref{tab:DCATRMSE} illustrate the comparison results for the Ciao and Epinions datasets with respect to the MAE and RMSE metrics, respectively. The best values are indicated in bold. In all cases, $DCAT$ achieves better performance compared to the other approaches. 
For instance, in the Ciao dataset, it has an around 2 times, 3 times, 3 times, 3 times and 5 times lower MAE compared to $TDTrust$, $SETTrust$, $Zheng$, $hTrust$ and $sTrust$, respectively.  
Similarly, for the Epinions dataset, $DCAT$ has around 3 times, 3 times, 3 times, 3 times and 3.5 times higher prediction performance than $sTrust$, $hTrust$, $Zheng$, $SETTrust$ and $TDTrust$, respectively with regard to the MAE metric. With respect to the RMSE metric, $DCAT$ also outperforms the other approaches in both the Ciao and Epinions datasets.

In the case of $DCAT_{+}$, we obtain the best performance when we initialise our model with the 27B Twitter corpus pre-trained dataset~\cite{pennington2014glove}. However, as Tables~\ref{tab:DCATMAE} and ~\ref{tab:DCATRMSE} illustrate, $DCAT$ has a slightly better performance than $DCAT{+}$ in terms of the MAE and RMSE metrics in both datasets. Therefore, $DCAT_{+}$ which incorporates user embeddings does not improve the performance of $DCAT$. However, it still has a better performance compared to the other baselines.

Finally, Table~\ref{tab:DCATDCAT+} demonstrates the performance of $DCAT$ and $DCAT_{+}$ with different training data sizes. With an increase in the training data size, the MAE and RMSE of $DCAT$ decrease. In both datasets, the lowest MAE and RMSE for $DCAT$ corresponded to a training data size of 90\%, while a training data size of 60\% gave the highest MAE and RMSE. The situation was the same for $DCAT_{+}$.

\section{Summary}
~\label{ch:chapt5sec10}
In this chapter, we proposed three novel context-aware trust prediction approaches. First, we proposed SETTrust, which is based on MF and can capture the context of trust by proposing some social contextual factors. The proposed model is mainly based on SET and suggests that if the cost of a relation is less than its benefit, a trust relation will be established between two users.
Second, we proposed a new context-aware trust prediction model, $TDTrust$, to predict pair-wise trust relations among users. $TDTrust$ monitors the social actor's behaviour (supported by theories from social psychology) and directly considers the context of trust in its mathematical model.

Finally, we proposed a deep classifier, $DCAT$, that has a feature set that includes level of  expertise, similar interests, rating similarity, self-disclosure level, use of swear words and use of inclusive words. We also applied word-embedding techniques (using Word2Vec) in this model to convert users' reviews into vectors of numbers for consideration as features of our classifier. Experimental results demonstrate the superior performance of SETTrust,  $TDTrust$ and $DCAT$ over other state-of-the-art approaches; however, these approaches still have a limitation:
they assume trust relations are fixed over time. For example, they assume that if David trusts John in the context of $c_k$ and at time $t_1$, this trust relation should persist at time $t_1 + h$, where h is a fraction of time. However, this is not the case in real-world scenarios, in which people may lose their trust in someone at any point for many reasons (e.g., a change in the behaviour or attitude of the person they trusted, or a change in their own interests and opinions). Hence, developing a trust prediction approach capable of considering the time factor might improve performance and make it more applicable to real-world scenarios.

\chapter{A Dynamic Deep Trust Prediction Approach for Online Social Networks}
\label{chap:time}

Trust relations in OSNs may change over time.
This makes modelling (pair-wise) trust relations 
that account for temporal changes a challenging task. Most existing trust prediction approaches assume that trust relations are fixed over time, and thus they
may fail to capture the dynamic behaviour of users in OSNs. In this chapter,
we propose a deep trust prediction model that can predict trust relations dynamically considering different time windows. We propose a novel deep structure that incorporates users'
emotions---which psychology
studies show have a significant
impact on trust~\cite{Emotion1}---and the textual contents they provide in OSNs. We use
word-embedding techniques to represent the users and their self-descriptions based on
their online profile. To evaluate our approach,
we created a large Twitter dataset, with the results of the evaluation demonstrating the effectiveness of our approach compared to other state-of-the-art
approaches.

\section{Motivation}

\begin{figure*}[t!]
\centering
  \includegraphics[width=0.90 \textwidth]{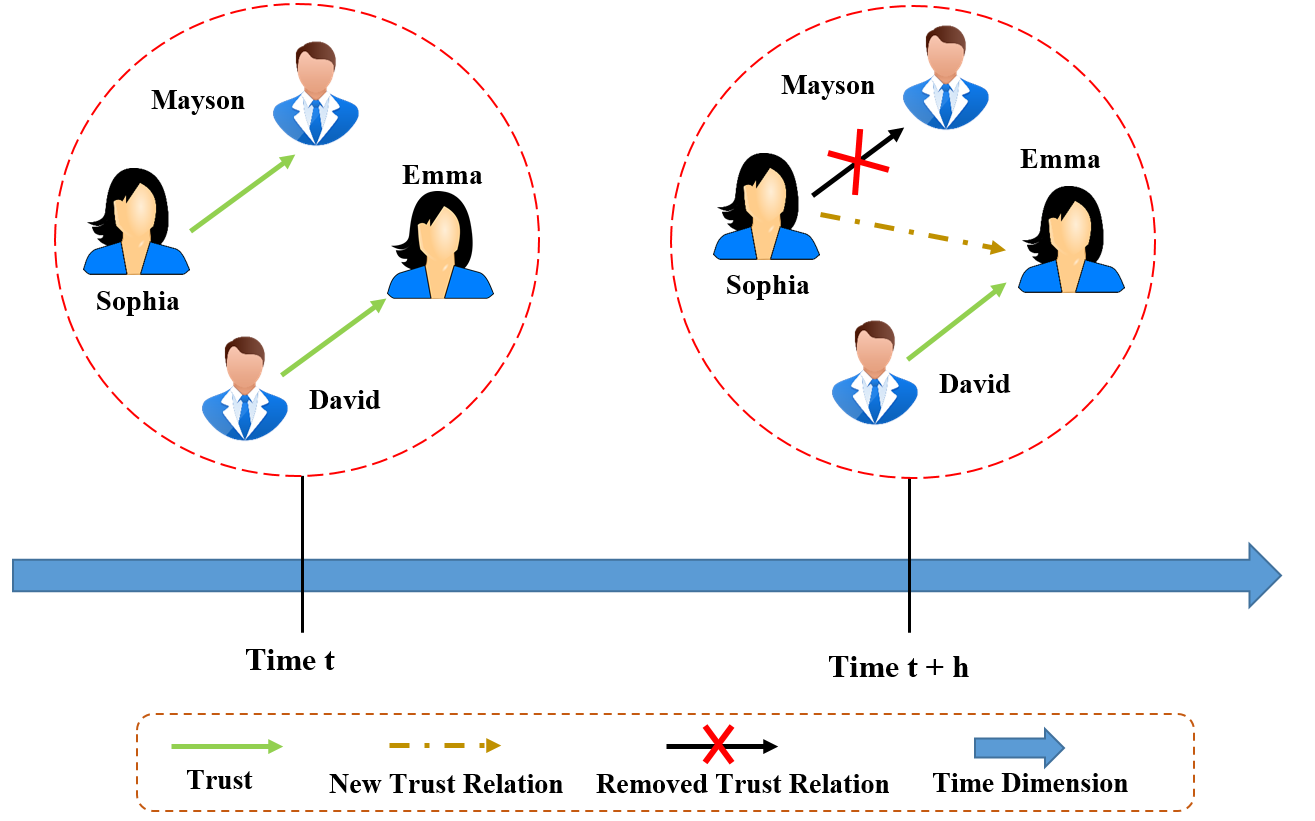}
  \caption{Pair-wise trust relations at time $t$ and $t+h$ (where $h$ is a fraction of time). At time $t$, there is a trust relation between Sophia and Mayson, but no trust relation between Sophia and Emma. However, at time $t+h$ Sophia no longer trusts Mayson, but does trust Emma.}
  \label{fig:timeexample1}
\end{figure*}

In the previous chapter, we proposed a series of trust prediction approaches to address the problem of pair-wise trust prediction; however, they are not time-aware. Trust is a time-dependent concept. For instance, as illustrated in Figure~\ref{fig:timeexample1}, if Sophia trusts Mayson at time $t$, this trust relation may change at time $t+h$ (where $h$ is a fraction of time) and it could be affected by many factors, such as Mayson's new online activity and behaviour or changes in Sophia's interests. As another example, Sophia may not trust Emma at time $t$, but she may trust her at time $t+h$ ($h$ is a fraction of time).
Hence, predicting pair-wise trust relations dynamically can be challenging.

Therefore, in this chapter, we investigate: (i) how we can dynamically monitor users' behaviour to predict their trust relations (ii) whether there is any relationship between users' emotions and their trust relations, and (iii) how this relationship can be used in a deep classifier to dynamically predict trust relations? Our answers to these questions result in a new dynamic trust prediction model, $DDTrust$. 
$DDTrust$ is a deep supervised model that evaluates the influence of users' emotions on trust relations by analysing the textual contents generated by users in OSNs.
By considering different time windows, $DDTrust$ is able to dynamically predict pair-wise trust relations. We evaluate $DDTrust$ on a large real-world Twitter dataset. 

The main contributions of this chapter are as follows:

\begin{itemize}
  \item We propose a novel deep trust prediction approach called $DDTrust$, which mainly focuses on analysing the textual contents provided by users. 
  \item We demonstrate the effects of users' emotions on pair-wise trust relations.
  \item To the best of our knowledge, $DDTrust$ is the first deep trust prediction approach that incorporates users' emotions and dynamically predicts pair-wise trust relations at different time points.
\end{itemize}

The rest of this chapter is organised as follows.
In Section~\ref{sec:dynamicproblem} we discuss our problem statement, before introducing our proposed approach in Section~\ref{sec:proposedDDTrust}. Our experimental results are presented and discussed in Section~\ref{sec:proposedDDTrustExperiments}, and we conclude this chapter in Section~\ref{sec:proposedDDTrustSummary}.

\section{Problem Statement}
\label{sec:dynamicproblem}

Let $U=\{u_1, u_2,..., u_n\}$ be a set of users and $n$ be the number of users. In this chapter, $G \in R^{n\times n \times t}$ demonstrates pair-wise trust relations in OSNs over different timestamps, where $t=\{1, 2, ..., m\}$ and $m$ is the number of considered timestamps. If $G_{i,j,h}=1$, it means that, $u_i$ trusts $u_j$ in the $h^{th}$ timestamp, while $G_{i,j,h}=0$ means there is no trust relation between these users at this timestamp.
For dynamically predicting a potential trust relation between $u_i$ and $u_j$ in the $h^{th}$ timestamp, it is necessary to monitor these users' activities between $t_h - \tau$ and $t_h$ (which we refer to as a time window, where $\tau$ is a timestamp and $\tau$ < $t_h$). However, determining the length of time windows can be challenging.
Therefore, our main research question in this chapter is:
\\\\
\noindent \textbf{RQ1.} Can prediction performance be improved by enabling the analysis of users' activities over time? 
\\
According to sociology and psychology studies,
people in a `grateful condition were significantly more trusting than were participants in other conditions, and participants in
the anger condition were significantly less trusting than were
participants in other conditions'~\cite{Emotion1}. In other words, a happy user is more
willing to trust other users. Conversely, users that feel sad trust other users less, while angry users are least likely to trust others. 
Therefore, the second question that we aim to address in this chapter is:
\\\\
\noindent \textbf{RQ2.} Does the emotional state of the source user affect pair-wise trust relations in OSNs?
\\\\
Although we investigated the impact of analysing users' activities using word-embedding techniques in $DCAT$ in Chapter~\ref{chap:context}, this was a static approach. In this chapter, we further exploit these techniques to improve the performance of our dynamic trust prediction approach. Therefore, the third research question addressed in this chapter is:
\\\\
\noindent \textbf{RQ3.} Does analysing users' activities through word-embedding improve the prediction performance of our dynamic trust prediction approach?
\\
\section{DDTrust: A Dynamic Deep Trust Prediction Approach}
\label{sec:proposedDDTrust}
\begin{figure}
\centering
  \includegraphics[width=0.99 \textwidth]{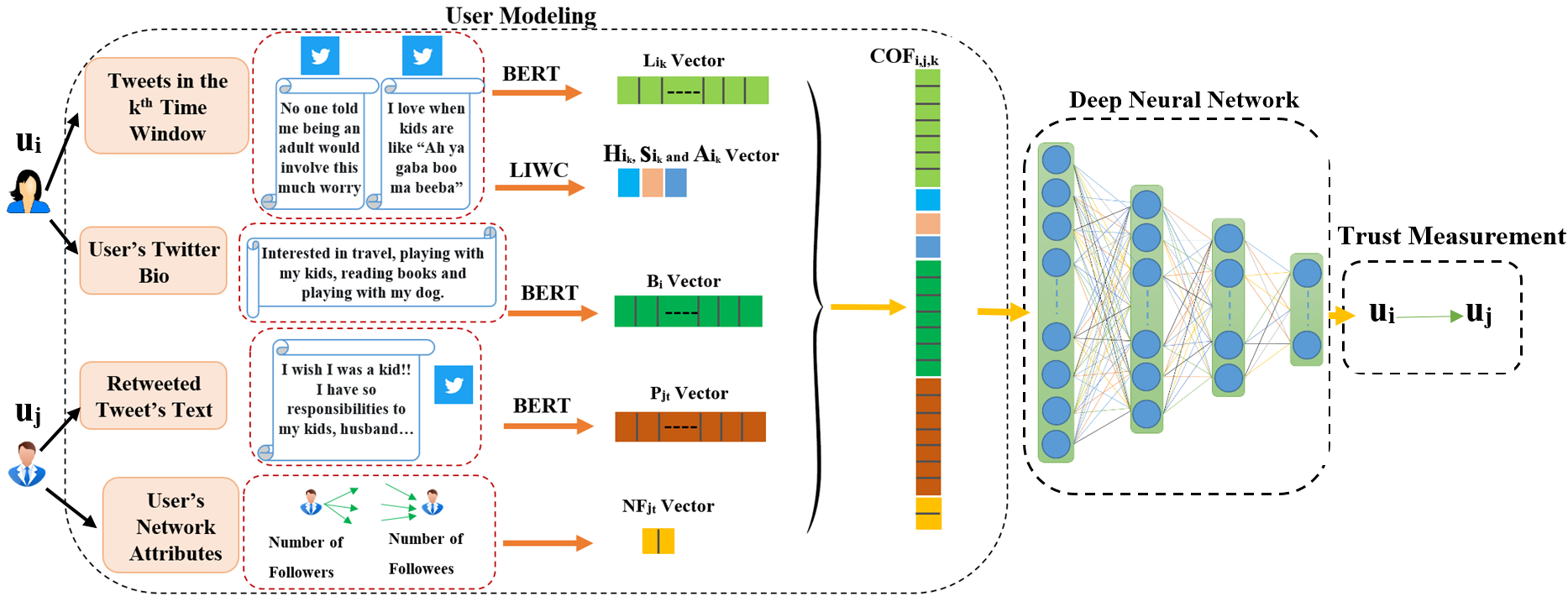}
  \caption{A snapshot of $DDTrust$ in the $k^{th}$ time window. DDTrust analyses the tweets and Twitter bio of the source user using LIWC and BERT to identify the users' emotional status and generate user embeddings. It also generates the user embeddings and  analyses the network attributes of the target users. Then, it concatenates the output vectors of these analyses and feeds them into a deep classifier.}
  \label{fig:timeexample}
\end{figure}

In this section, we introduce our proposed dynamic deep trust prediction approach (Figure~\ref{fig:timeexample}). It consists of three components: dynamic analysis of users' activities, user modelling and deep trust prediction. 
To the best of our knowledge, this is the first work to use
a dynamic deep learning-based model for trust prediction.

\subsection{Dynamic Analysis}

In this section, we discuss how to define the length of time windows for use in the trust prediction process. We define the length of the considered time windows according to the distribution of our test data (i.e., the Twitter dataset introduced in Chapter~\ref{chap:DataandMetrics}), and focus on both long and short time windows. Smaller time windows place greater emphasise on small changes in users' behaviour~\cite{BetaModel2} and help to investigate the concept of recency bias~\cite{Recency}. Based on this concept, we emphasise recent events, about which it is easier for people to access cognitive information (e.g., memory, pattern-matching and explanation), compared to events further in the past that require more thinking or analysis~\cite{Recency}.

In this chapter, the timestamp at which $u_i$ retweeted a tweet from $u_j$ is considered the \textit{current time}. Based on the distributions of timestamps of retweets in our dataset, the \textit{current time} is a value between August and September 2019. Then, we look at the timeline of $u_i$ and identify the timestamp of his or her first tweet and call that the \textit{oldest tweet in the timeline}. The difference between the \textit{current time} and the \textit{oldest tweet in the timeline} indicates the number of days that it took for $u_i$, after posting his or her first tweet, to retweet $u_j$'s tweet. The distribution of our test data (with respect to the \textit{oldest tweet in the timeline}) is from 2008 to 2019.
Over this period, for each year we select the month in which the highest number of users started their activities by posting their first tweet.

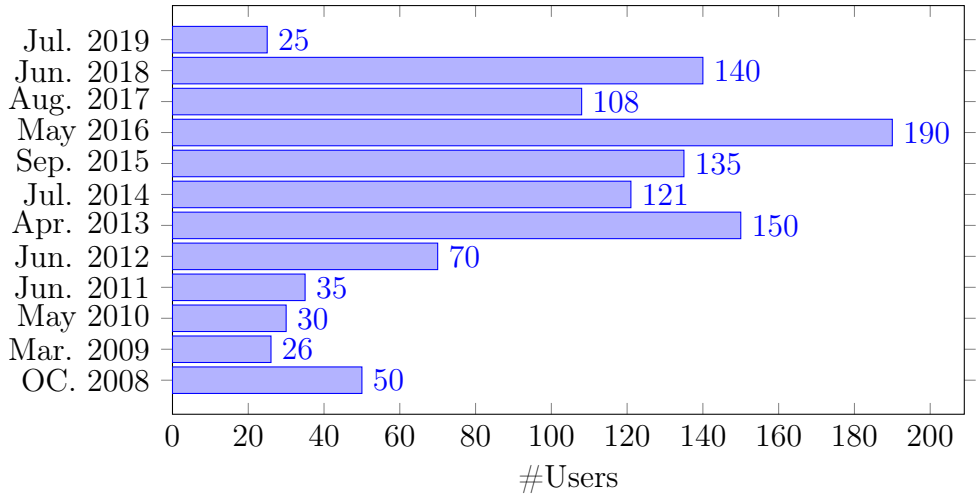
\begin{figure}[t!]
\centering
\begin{tikzpicture}
  \begin{axis}[
    xbar, xmin=0,
    width=12cm, height=7cm, enlarge y limits=0.1,
    xlabel={\#Users},
    symbolic y coords={OC. 2008, Mar. 2009, May 2010, Jun. 2011, Jun. 2012, Apr. 2013, Jul. 2014, Sep. 2015,May 2016, Aug. 2017,Jun. 2018,Jul. 2019},
    ytick=data,
    nodes near coords, nodes near coords align={horizontal},
    ]
    \addplot coordinates {(50,OC. 2008) (26,Mar. 2009) (30,May 2010) (35,Jun. 2011) (70,Jun. 2012) (150,Apr. 2013) (121,Jul. 2014) (135,Sep. 2015) (190,May 2016) (108,Aug. 2017) (140,Jun. 2018) (25,Jul. 2019)};
  \end{axis}
\end{tikzpicture}
   \caption{Distribution of users based on their starting date of activity (writing a tweet) on our Twitter dataset. This figure does not represent the total number of users as we only selected a peak month at each year that contains the most number of users.}
       \label{fig:dataditributionDDTRUST}
\end{figure}
Figure~\ref{fig:dataditributionDDTRUST} demonstrates the selected dates (as $TW_r$; $r=\{1, 2, ..., 12\}$). 
Based on this data distribution, the first time window ($TW_1$) is defined as `from October 2008 to the retweet time' (retweet time is the same as the \textit{current time}), the next time window ($TW_2$) is `from March 2009 onward' and the rest are as follows:

$TW_3$: from May 2010 onward, $TW_4$: from June 2011 onward, $TW_5$: from June 2012 onward, $TW_6$: from April 2013 onward, $TW_7$: from July 2014 onward, $TW_8$: from September 2015 onward, $TW_9$: from May 2016 onward, $TW_{10}$: from August 2017 onward, $TW_{11}$: from June 2018 onward and $TW_{12}$: from July 2019 onward. 

To investigate the recency bias further, we also consider time windows of one month ($TW_{13}$), one week ($TW_{14}$) and one day ($TW_{15}$). Our goal is to understand which time window more accurately captures the reasons behind establishing a trust relation. 



\subsection{User Modelling}  
In this section, we describe how we model a user's characteristics based on his/her provided textual contents in Twitter and his or her network attributes.

\subsubsection{Modelling Users' Tweet Behaviour:} On Twitter, a user's tweet behaviour is represented by the text of the tweets that he or she writes, retweets of others' tweets, and mentions or likes of tweets. In trust related studies in OSNs, retweet behaviour on the Twitter platform~\footnote{https://twitter.com} can be considered trust~\cite{retweet5}~\cite{retweet4}~\cite{retweet3}~\cite{retweet2}~\cite{ADALI}. In other words, we assume that if David retweets Sarah's tweet at time $t$, he trusts her at that timestamp. From now on, we consider the person who posted a tweet as the target user and the person who retweeted that tweet as the source user in a pair-wise trust relation. 

For analysing users' textual contents, we use two text analysis tools: LIWC and Bidirectional Encoder Representations from Transformers (BERT). LIWC~\cite{LIWC} is the gold standard text
analysis tool providing more than 88 linguistic categories
to which written words may belong (e.g., negative
emotion, positive emotion, and anger). BERT is a sate-of-the-art natural language processing (NLP) technique~\cite{BERT}.

\subsubsection{Users' Tweet Behaviour Features:}
In this section, we model the source and target users' tweet behaviour features in two phases: (A) source user modelling, including determining source users' emotional status and generating the word-embedding representations of  source users' tweets and their Twitter bios; and (B) target user modelling, including generating the word-embedding representation of the retweeted tweets belonging to the target users, and identifying these users' network attributes (i.e., number of followers and followees).
\\ \\ 

\noindent \textbf{\textit{A. Source User Modelling}}

\textbf{Emotional Status:}
Sociological and psychological studies have proven that people's emotions strongly influence their trust relations~\cite{Emotion1}. Our aim is to investigate the relation between trust relations and users' emotions in OSNs.  
The work in \cite{Emotion2} also points to a significant correlation between trust and users'
emotions. Unfortunately, it only considers users' given ratings as their emotions which can be too simplistic. 
Dunn and Schweitzer~\cite{Emotion1} found that 
`happy participants were more trusting than sad participants and that sad participants
were more trusting than angry participants'.

Let $H_{k}$, $S_{k}$, and $A_{k}$ be the happiness, sadness, and anger emotional representation vectors for users in $k^{th}$ time window, where $H_{k} \in R^{n\times 1 }$, $S_{k} \in R^{n\times 1 }$, and $A_{k} \in R^{n\times 1 }$.
To acquire users' emotions, we focus on the textual contents provided by them in their tweets. For analysing these textual contents, we first collect all tweets posted by the source users in the $k^{th}$ time window. Next, we analyse them using LIWC. Based on the LIWC report, and in particular the average percentage of a users' written text belonging to the negative emotion and negative feeling categories, we set $S_{k}$. We follow the same procedure for updating $H_{k}$ with the positive emotion and positive feeling categories. Finally, $A_{k}$ represents the degree of usage in a user's tweets of angry words, as per the anger category in LIWC.

\textbf{Word-Embedding Representation of Tweets:}
In most of the previous text-based analysis in this field, classical word-embedding methods were used to map a one-hot vector to a continuous vector. Recently, BERT has been proposed for assigning a vector to words after reading the whole sentence~\footnote{https://www.lexalytics.com/lexablog/bert-explained-next-level-nlp}. We first collect all tweets of a source user in the $k^{th}$ time window and then use BERT, to create a vector of numbers representing the tweets. To the best of our knowledge, this is the first time that a trust prediction approach, has used BERT in its text analysis step. $L_k$ represents a matrix containing the vectors of numbers related to the tweets of source users in the $k^{th}$ time window. For $u_i$, $L_{i_k}$ represents a vector containing his or her word-embedding representation of tweets in the $k^{th}$ time window. $L_k \in R^{q \times n}$, $L_{i_k} \in R^{q\times 1 }$ and $q$ is the length of the required numeric representation of the tweets of a user. 

\textbf{Word-Embedding Representation of Twitter Bio:}
On Twitter, as in many other OSNs, users can write a short description about themselves; that is, their bio. We use BERT to create a vector of numbers representing a user's Twitter bio. $B$ represents a matrix containing vectors of numbers related to the Twitter bios of users. For $u_i$, $B_i$ represents a vector containing his or her Twitter bio word-embedding. $B \in R^{v \times n}$, $B_i \in R^{v\times 1 }$ and $v$ is the length of the required numeric representation of the Twitter bio of a user. 

\noindent \textbf{\textit{B. Target Users Modelling}}

\textbf{Word-Embedding Representation of the Retweeted Tweets:}
We analyse the text of the retweeted tweet, originally posted at time $t$ by the target user. $P_t$ represents a matrix containing vectors of numbers related to the word-embedding of the retweeted tweets posted by the target users at time $t$. For $u_j$, $P_{j_t}$ represents a vector containing his or her tweet features at time $t$. $P_t \in R^{c \times n}$, $P_{j_t} \in R^{c\times 1 }$ and $c$ is the length of the required numeric representation of the retweeted tweet). 

\textbf{Network Attributes:}
A larger number of followers may indicate a user is well-known in the community and that the information provided by him or her is more credible~\cite{GhafariWise}~\cite{Liuwwwj}. Further, a high number of followees can indicate that a user has `more opportunities ... to have a social connection with others in the community'~\cite{Liuwwwj}. Twitter API provides us the number of followers and followees of Twitter accounts. As a result, we consider the number of followers and followees of a target user as his or her network attributes affecting his or her potential trust relations with source users.
$NF_t$ denotes a matrix that contains the number of followers and followees of users in time $t$. For $u_j$, $NF_{j_t}$ represents a vector of length two that contains his or her  number of followers and followees at time $t$. $NF_t \in R^{2 \times n}$, $NF{_{j_t}} \in R^{2\times 1 }$.

\subsubsection{Concatenation of Users' Characteristics}

After modelling the different users' characteristics in the $k^{th}$ time window, we concatenate them to create a single vector for each pair of users as follows:
\begin{equation}
\small
\begin{aligned}
CoF_{i,j,k}= H_{it_{k}} \oplus S_{it_{k}} \oplus A_{it_{k}} \oplus L{_{i_k}}\oplus P_{j_t}	\oplus B_i \oplus NF_{j_t}
\end{aligned}
\label{for: TimeCOF}
\end{equation}

where $CoF_{i,j,k}$ is a vector containing the features of $u_i$ and $u_j$ in the $k^{th}$ time window. 
Next, we use this concatenated vector as the input vector of our deep trust prediction model.

\subsection{Deep Trust Prediction}

Our deep neural network structure has a  multi-layer perception unit. Hence, for each pair of users ($u_i$ and $u_j$), we create $CoF_{i,j,k}$ and then feed this into our DNN (Figure~\ref{fig:timeexample}).
A DNN can have input and output vectors as $x$ and $y$, while it contains $Q$ number of hidden layers as $l_c$, a weight matrix as $W_c$ and $bias_c$, which denotes a value that can help to adjust the output, where $c=\{1,2, ..., Q\}$:
\begin{equation}
\small
\begin{aligned}
l_c=g(w_c \times l_c +b_c) \\
\label{for:1}
\end{aligned}
\end{equation}\\
\begin{equation}
\small
\begin{aligned}
 y= f(w_Q \times L_Q +bias_Q)
\label{for:2}
\end{aligned}
\end{equation}

where $g$ and $f$ are our activation functions: $g$ is the ReLU function and $f$ is the Sigmoid function. The activation function of our hidden layers can be presented as: 
\begin{equation}
\small
\begin{aligned}
g(z) = max(0,z)
\label{for:3}
\end{aligned}
\end{equation} 

We employ the Sigmoid function as the activation function of the output, to give a prediction probability in the range of `0' to `1' as follows:
\begin{equation}
\small
\begin{aligned}
f(z) = \dfrac{1}{1+e^-z}
\end{aligned}
\end{equation} 

\textbf{\textit{Learning the Parameters.}} We use a binary cross-entropy loss function for optimising our DNN, since it measures the performance of a classification model whose output is a value between `0' and `1':
\begin{equation}
\small
\begin{aligned}
loss(z,z') = (Z \times log(z') + (1-z) \times log (1-z'))
\end{aligned}
\label{formula:TimeLLLLOSSS}
\end{equation}

where $loss(z,z')$ is the binary cross-entropy loss function and $z'$ is the predicted value. Our model was implemented by TensorFlow~\cite{TensorFlow} with the Adam optimiser~\cite{ADAM}. The algorithm of our proposed approach can be found in Algorithm~\ref{alg:algorithm1}. To the best of our knowledge, this is the first dynamic pair-wise trust prediction approach that models users' features as the input of a deep neural network to dynamically predict trust in OSNs.

\begin{algorithm}[h]
\caption{Trust Prediction with DDTrust}
\begin{algorithmic}[1]
 \State Input: pairs of users, $t={t_1, t_2, ..., t_m}$, users' tweets, users' network attributes
 \State Output: predicted trust relations. 
 \State initialise the $H_{t_{k}}$, $S_{t_{k}}$, and $A_{t_{k}}$.
 \State Randomly, initialise the $AUk$.
 
\State \textbf{for} k=1...m \textbf{do}
\State \quad \textbf{for} 1 to the number of training iterations \textbf{do}
 \State \quad \quad  \textbf{for} each pair of users ($u_i, u_j$) \textbf{do}
 \State   \quad \quad \quad set $L_{i_k}$, $H_{i_k}$, $s_{i_k}$, $A_{i_k}$, $B_i$, $P_{j_t}$ and $NF_{j_t}$  
 \State \quad \quad \quad set $COF_{i,j,k}$ with Formula~\ref{for: TimeCOF}
\State \quad \quad \quad set $loss$ function with Formula~\ref{formula:TimeLLLLOSSS} 
\State \quad \quad \quad set $y$
\State \quad \quad \quad Use back propagation to optimise the parameters of our DNN
\State \quad \quad \textbf{end} 
\State \quad \textbf{end}
\State \textbf{end} 
\label{alg:algorithm1}
\end{algorithmic}
\end{algorithm}

\section{Experiments}
\label{sec:proposedDDTrustExperiments}

\subsection{Experimental Settings}

\textbf{DNN Setup.}
We empirically set our hidden layer to 4, as no significant improvement was captured by adding more layers.
We initialised the weights of our DNN according to~\cite{huang2013learning}~\cite{ICDM} in the following range:
\begin{equation}
\small
\begin{aligned}
-\sqrt{\dfrac{6}{Input Size+ Output Size}} < weights' range < \sqrt{\dfrac{6}{Input Size+ Output Size}}
\end{aligned}
\end{equation} 

where \textit{InputSize} and \textit{OutputSize} are the size of the input and output vectors, respectively.
The learning rate of our DNN's parameters is 0.002. The batch size of this model is 150. The performance values are reported as an average of five time experiments.

\subsection{Effectiveness of Our Model}
To demonstrate the effectiveness of our model and to answer \textbf{RQ1}, \textbf{RQ2} and \textbf{RQ3}, we compare the prediction performance of $DDTrust$ with several baselines, including the state-of-the-art trust prediction approaches. 
Table~\ref{tab:comparison22} illustrates the prediction performance of different trust prediction approaches. For the dynamic approaches (i.e., Liu and Datta [2011], Liu and Datta [2012] and all versions of $DDTrust$), we considered the average prediction accuracy of different time windows.

\begin{table}[t!]
\begin{center}
 \caption{Comparison of the performance of different trust prediction approaches with respect to the MAE and RMSE metrics; lower is better. L-D1 and L-D2 represents Liu and Datta [2011] and Liu and Datta [2012], respectively.
 }
 \begin{tabular}{||c | c|c|c|c|c|c|c|c|c ||} 
 \hline
 \hline
\small{Metric} & \small{$DDTrust_{Opt}$} & \small{DDTrust} &   \small{L-D1} & \small{L-D2} & \small{$DDTrust_{-Em}$}& \small{$DDTrust_{-text}$}&\small{sTrust}& \small{MF}\\
\hline
\hline
MAE  & \textbf{0.15} & 0.27& 0.32& 0.36 & 0.43& 0.51& 0.59&  0.67 \\
\hline
RMSE & \textbf{0.21} & 0.31& 0.39& 0.40 & 0.48& 0.54& 0.62& 0.69\\
\hline
\hline
\end{tabular}
   \label{tab:comparison22}
\end{center}
\end{table}

Table~\ref{tab:comparison22} demonstrates that $DDTrust$ (which in our experiments, represents the average prediction performance of $DDTrust$ in different time windows) has the lowest MAE and RMSE of all compared approaches. For $DDTrust$, optimal prediction performance with respect to both MAE and RMSE is achieved when the time window is $T_{13}$ (one month). This optimal performance is reported as $DDTrust_{Opt}$ in Table~\ref{tab:comparison22}. The mean of the distribution of performance prediction in different time windows is 0.272 and the standard deviation is 0.0851. For the model proposed in Liu and Datta [2012] and Liu and Datta [2011], optimal prediction performance is 0.29 and 0.30 with respect to the MAE metric, and 0.33 and 0.35 with respect to the RMSE metric, respectively. 

The superior performance of $DDTrust$ and the other two dynamic trust prediction approaches (Liu and Datta [2012], and Liu and Datta [2011]) compared to the static approaches ($sTrust$ and $MF$) suggest that \textbf{RQ1} can be answered in the affirmative. To answer \textbf{RQ2}, we tested $DDTrust$ without considering the emotional status of users, calling this version $DDTrust_{Em}$. The performance of
$DDTrust_{Em}$ was significantly lower than $DDTrust$, confirming the importance of considering the information about emotional status. Thus, \textbf{RQ2} can also be answered positively. Finally, to answer \textbf{RQ3}, we tested $DDTrust$ without analysing the textual contents provided by users and calling this version $DDTrust_{text}$. The experimental results demonstrate that $DDTrust_{text}$ has lower prediction performance compared to $DDTrust$. This positively answers \textbf{RQ3}. 

\begin{figure*}[t!]
\centering
  \includegraphics[width=\textwidth]{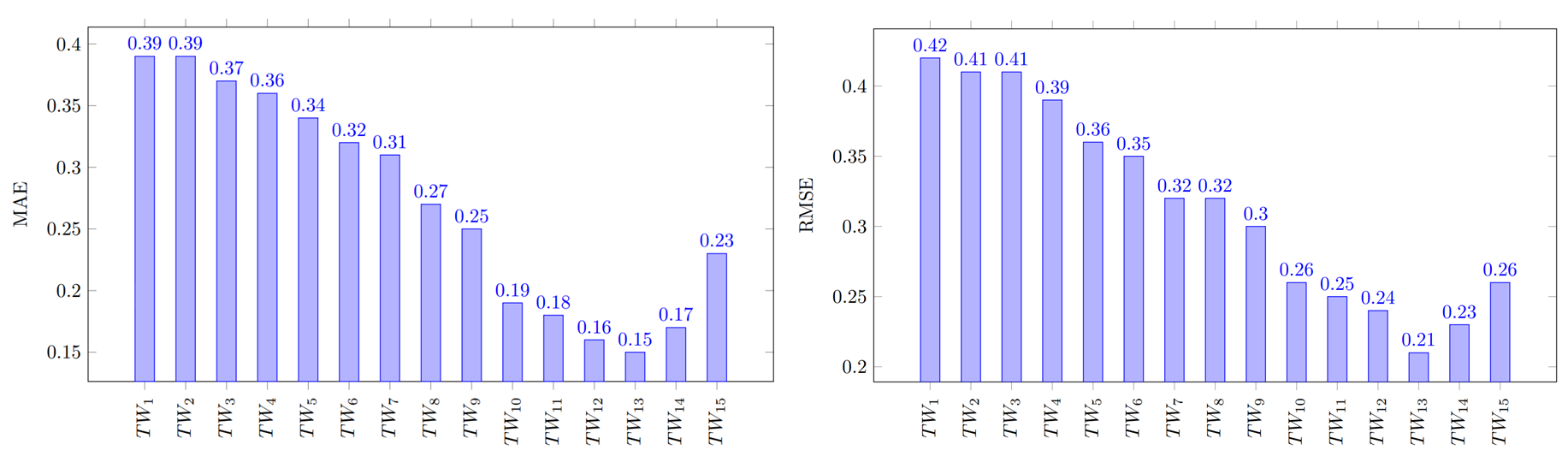}
  \caption{The prediction performance of $DDTrust$ with respect to the MAE (the left chart) and RMSE (the right chart) in different time windows.}
  \label{fig:timewindow}
\end{figure*}

\subsection{Impact of Time Window Length}

In this section, we investigate the impact of the length of the considered time windows on the prediction performance of $DDTrust$. Figure~\ref{fig:timewindow} illustrates that the prediction performance of $DDTrust$ improves when we focus on shorter time windows. When we consider very large time windows (e.g., $TW_1$, $TW_2$, $TW_3$, $TW_4$ or $TW_5$, which require analysing the timelines of users over the previous couple of years), the performance of $DDTrust$ is similar to the static trust prediction approaches. 
Trust is a complex concept, changes in which may reflect a variety of factors, including shifting interests, increased knowledge of a particular domain or the establishment of new personal relationship. Moreover, the emotional status of a person can change over time. As a result, considering shorter time windows can better capture the reasons behind establishing trust relations.   

The best performance of $DDTrust$ in terms of both MAE and RMSE is obtained when the size of the time window is between a couple of months to a couple of weeks ($TW_{10}$, $TW_{14}$). At $TW_{15}$ or when we consider only one day as our time window, the performance of $DDTrust$ drops significantly. We can also observe the same performance degradation between $TW_{13}$ and $TW_{14}$. This provides insight into the effect of recency bias on pair-wise trust relations in OSNs. Although analysing more recent events can improve the prediction performance of $DDTrust$, sufficient information about users' activities may not be available when using very short time windows.

\begin{figure*}[t!]
\centering
  \includegraphics[width=\textwidth]{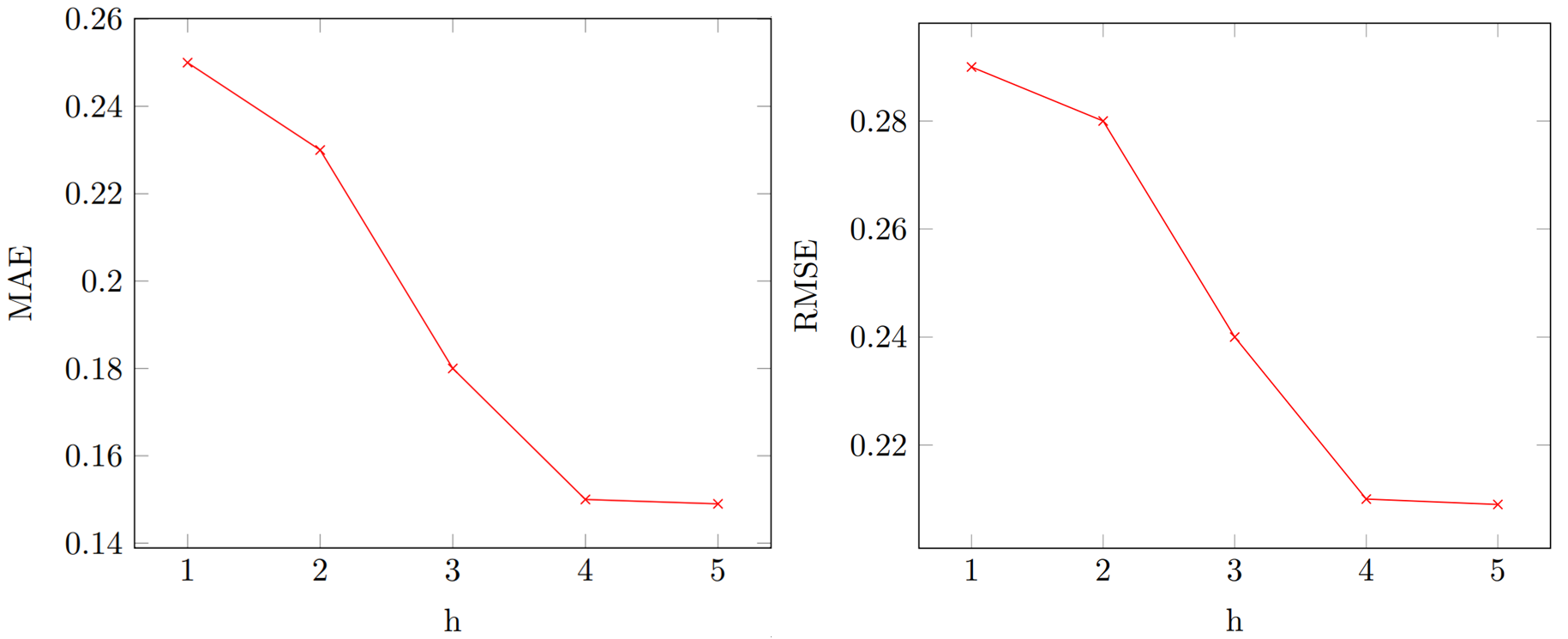}
  \caption{The impact of the number of hidden layers on the prediction accuracy of the $DDTrust_{opt}$}
  \label{fig:DepthDDTRUST}
\end{figure*}


\subsection{Impact of Depth of DNN}

In this section, we investigate the impact of the depth of the hidden layers of $DDTrust$ on its prediction performance. Assume $h$ denotes the number of hidden layers, where $h=\{1,2,...,5\}$. Figure~\ref{fig:DepthDDTRUST} demonstrates that as $h$ increases, from $h=1$ to $h=4$, the prediction performance of $DDTrust$ increases. This demonstrates that $DDTrust$ may better deal with the pair-wise trust prediction problem, if it has a deep neural network model. However, there is no significant prediction performance improvement when $h$ is greater than 4. Thus, the optimal depth of layers in $DDTrust$ is 4.

\section{Summary}
\label{sec:proposedDDTrustSummary}
Trust relations between users in OSNs can change over time; however, until now, only minor attempts have been made to capture trust relations dynamically. In this chapter, we proposed a dynamic deep trust prediction approach, called $DDTrust$. This approach defines several time windows and analyses users' online activities in those time windows. $DDTrust$ also investigates the impact of recency bias on pair-wise trust relations by considering several short time windows. The focus of $DDTrust$ is on analysing users' textual contents to identify their emotional status and generating word-embedding related to users' tweets and Twitter bios. After concatenating these features, they are feed into a DNN to predict trust relations dynamically and for different time windows. Our experimental results demonstrated that although recency bias is an important concept in accurately predicting trust relations, defining a very short time window can reduce the prediction performance of $DDTrust$. Moreover, we found that trust prediction approaches that dynamically predict trust relations have a higher prediction accuracy compared to static approaches.

\chapter{Conclusions and Future Work}
\label{chap:conclusion}

\section{Concluding Remarks}

OSNs enable users to connect with others, expand their social networks, share multimedia content and write reviews on specific items. Users in OSNs are bombarded with information and trust can play an important role in their decision making. Due to the lack of interactions between the majority of participants on OSNs, predicting pair-wise trust relations in this context is a daunting task. In this dissertation, we have made three major contributions to effectively and accurately predicting trust relations between unknown pairs of users. Below,
we summarise these contributions:

\begin{itemize}
  \item A novel trust prediction approach for different degrees of data sparsity (Chapter~\ref{chap:chapter4}). We proposed a new pair-wise trust prediction approach that can alleviate the effect of
the sparsity of trust relations. Particularly, it focuses on the personality traits of users as additional information in a trust prediction process. It also seeks a low-rank representation of users in OSNs and employs a three-dimensional TD model. We first analysed the homophily effects on trust relations, in particular similarity of the personality traits of users involved in a trust relation. We proposed a mathematical model for considering the personality traits of users and the level of similarity of the source and target users. Our experiments found this approach to be insensitive to changes in the degree of data sparsity. 
  \item Context-aware trust prediction approaches for taking the context of trust into account when making predictions (Chapter~\ref{chap:context}). First, we proposed a new SET-based context-aware trust prediction approach, mathematically modelled by an MF approach. We also proposed several social context factors to capture the context of trust relations. Secondly, to improve our proposed approach, we directly considered context of trust and proposed a three-dimensional TD model. Finally, we proposed a deep classifier capable of predicting trust relations in OSNs, while distinguishing them based on their context of trust. To the best of our knowledge, this is the first deep context-aware pair-wise trust prediction approach. 
  \item A time-aware trust prediction approach (Chapter~\ref{chap:time}). We proposed a new trust prediction approach that, instead of analysing the whole online behaviour history of users in OSNs, can accurately predict trust relations based on monitoring users' behaviours in different time windows of reasonable length. This approach uses the textual contents provided by users on Twitter to capture their emotional status, which is known to be related to their likelihood to trust. To the best of our knowledge, this is the first deep dynamic pair-wise trust prediction approach. 
\end{itemize}

The lessons that we learned from our experiments include: (i) Because the degree of sparsity of user-specified trust relations is very high in OSNs, without having a proper strategy to address this problem, the algorithm fails to predict trust relations well. Approaches that seek a low-rank representation of users or that incorporate additional information (e.g., users' similarity, users' social status or users' personality traits) may have better trust prediction performance when faced with the data sparsity problem. (ii) It is not correct to assume that trust is a time-independent concept. Many factors (e.g., emotional status) can affect trust either positively or negatively. Taking these factors into account can improve the performance of a trust prediction approach in dynamically capturing pair-wise trust relations.

\section{Future Directions}

In this dissertation, we have investigated the problem of predicting trust between two unknown users in OSNs. We believe this is an important research area that has important applications for business and government in understanding trends in the diffusion of misinformation on OSNs. For example, trust prediction on social media can help in understanding
important questions in political science and identifying campaigns of fake news and propaganda.
We believe this important research area, will attract a great deal of attention from the research community over the coming years. Below, we summarise some significant research directions in this area.

\textbf{Context and Data Curation}
Although several context-aware trust prediction approaches have been proposed in the literature, there remains room to study the factors that can accurately capture the context of trust relations. 
Two of our proposed trust prediction approaches in this dissertation have focused on 
analysing the textual contents provided by users in OSNs; however, it would be useful to investigate the use of textual contents in trust prediction approaches. As an almost unexplored research topic in trust prediction area, researchers need to use natural language processing techniques~\cite{NLP} to analyse textual contents as a rich source of information about users' activities and behaviour. Such analysis would enrich our available data about users and their relations, potentially helping to alleviate the data sparsity problem.

Accordingly, understanding the content and context of social data can help in understanding the trust relations among users in OSNs. For example, if a user retweets a tweet on Twitter, it would be helpful to understand the text of the tweet, whether it contains an image or URL, and the keywords or entities (e.g., people, organisations, locations and products) and topics mentioned.
In this context, data curation~\cite{curation1,Curation3,Curation4,intelKG} (i.e., the task of preparing the raw data for analytics) can help in turning raw data into contextualised data and knowledge. For example, curating a raw tweet from Twitter can tell us if the tweet contains a mention of a person named Barak Obama (using entity extraction and coreference resolution techniques~\cite{CDCR}) who was the 44th president of the United States (using linking techniques~\cite{linking} to link this entity to external knowledge sources such as Wikidata~\footnote{https://www.wikidata.org/}). We can also understand if the topic of the tweet is related to politics (using topic extraction~\cite{topic}) and if the tweet is discussing a social issue (using rule-based techniques~\cite{rule}).
A future direction would be to use data curation in OSNs to improve the accuracy of predicting the trust relation between two users. 

\textbf{Time and Business Processes}
In this dissertation, we take the first step towards introducing a novel time-aware trust prediction approach that can dynamically predict trust relations. However, the time complexity of these approaches in real-world scenarios must be critically examined.
In other words, the next trust prediction approaches should focus on decreasing the execution times. Many of the existing trust prediction models are based on a computationally complex model with a high execution time. By decreasing the execution time of trust prediction approaches, we make them more feasible for real-world applications.

An important application in this category is to understand customer's personality, behaviour and attitude in business processes~\cite{ProcessAnalytics,BPM} and to predict how their trust in companies and products may change over time. Business processes are a set of tasks and activities performed to accomplish a specific organisational goal~\cite{bpquery,iProcess}. For example, consider a bank customer who has decided to change their bank or a specific product offered by a bank. Analysing the time-aware activities of bank customers may allow the loss of a trust relation for an existing product to be predicted. 
Another interesting avenue for future work in this domain would be to use data provenance~\cite{prov1,prov2} to model and understand the evolution of social items over time. For example,
to help predict customers' personality, behaviour and attitude in business processes, their retweets, likes and views could be analysed over time~\cite{WSDMPersonality2vec}.

\textbf{Benchmarking Datasets}
Surprisingly, even after several years of research in the trust prediction area, researchers still suffer from an absence of test datasets that provide sufficient contextual information about users and the dynamic timestamp of their trust relations. As an urgent need in this domain, providing such a dataset for trust prediction related research could help to attract many more researchers to this research area. Future work in this domain would be to use crowdsourcing techniques~\cite{crowd0,Curation2,crowd1} to facilitate the labelling of such datasets.

%
%
%
 \bibliographystyle{splncs04}
 \bibliography{ms}



\backmatter



\end{document}